\documentclass[pdflatex,sn-mathphys-num]{sn-jnl}

\usepackage{hyperref}
\usepackage{lmodern}
\usepackage[T1]{fontenc}
\usepackage{graphicx}%
\usepackage{multirow}%
\usepackage{amsmath,amssymb,amsfonts}%
\usepackage{amsthm}%
\usepackage{mathrsfs}%
\usepackage[title]{appendix}%
\usepackage{xcolor}%
\usepackage{textcomp}%
\usepackage{manyfoot}%
\usepackage{booktabs}%
\usepackage{algorithm}%
\usepackage{algorithmicx}%
\usepackage{algpseudocode}%
\usepackage{listings}%

\usepackage{colortbl}
\usepackage{longtable}
\usepackage{caption}
\usepackage{subcaption}
\usepackage{array}
\usepackage{graphicx}
\definecolor{Gray}{rgb}{0.99, 0.88, 0.88}

\begin{document}

\title[Article Title]{A Systematic Review of Echo Chamber Research: Comparative Analysis of Conceptualizations, Operationalizations, and Varying Outcomes}

\author*[1,2]{\fnm{David} \sur{Hartmann}\footnotetext[1]{\textsuperscript{a}\url{https://scholar.google.com/citations?user=t0Aljh0AAAAJ&hl=en}}}\email{d.hartmann@tu-berlin.de}

\author[4]{\fnm{Sonja Mei} \sur{Wang}\footnotetext[2]{\textsuperscript{b}\url{https://scholar.google.com/citations?user=qJuTFFkAAAAJ&hl=en}}}\email{swang@uni-wuppertal.de}

\author[1,2]{\fnm{Lena} \sur{Pohlmann}}\email{l.pohmann@tu-berlin.de}

\author[1,2,3]{\fnm{Bettina} \sur{Berendt}\footnotetext[2]{\textsuperscript{c}\url{https://scholar.google.com/scholar?hl=en\&as_sdt=0\%2C5\&q=bettina+berendt}}}\email{berendt@tu-berlin.de}

\affil*[1]{\orgname{Technische Universität Berlin}, \country{Germany}}

\affil[2]{\orgname{Weizenbaum Institute for the Networked Society}, \country{Germany}}

\affil[3]{\orgname{KU Leuven}, \country{Belgium}}
\affil[4]{\orgname{University of Wuppertal}, \country{Germany}}


\abstract{This systematic review synthesizes research on echo chambers and filter bubbles to explore the reasons behind dissent regarding their existence, antecedents, and effects. It provides a taxonomy of conceptualizations and operationalizations, analyzing how measurement approaches and contextual factors influence outcomes.

The review of 129 studies identifies variations in measurement approaches, as well as regional, political, cultural, and platform-specific biases, as key factors contributing to the lack of consensus. Studies based on homophily and computational social science methods often support the echo chamber hypothesis, while research on content exposure and broader media environments, such as surveys, tends to challenge it. Group behavior, cultural influences, instant messaging platforms, and short video platforms remain underexplored. The strong geographic focus on the United States further highlights the need for studies in multi-party systems and regions beyond the Global North.

Future research should prioritize cross-platform studies, continuous algorithmic audits, and investigations into the causal links between polarization, fragmentation, and echo chambers to advance the field. This review also provides recommendations for using the EU’s Digital Services Act to enhance research in this area and conduct studies outside the US in multi-party systems. By addressing these gaps, this review contributes to a more comprehensive understanding of echo chambers, their measurement, and their societal impacts.
}

\keywords{Computational Social Science, Systematic Review, Echo Chamber, Filter Bubble, Measurement Modeling, DSA}



\maketitle

\section{Introduction}
The impact of social media on public discussions and democratic processes has been a hot topic for more than a decade. It remains relevant with elections taking place in more than 50 countries in the year 2024 (including the United States, India, and the EU), and the introduction of EU regulation on very large online platforms (see the Digital Services Act \cite{noauthor_regulation_2022}). The emerging shift to the political right, along with declining trust in democratic institutions and science, raises concerns that the advent of social media platforms may have contributed to these phenomena \cite{Bail.2018,quattrociocchi_social_2017, LorenzSpreen.2023}. 

Many believe that the constructs \textit{echo chamber} and \textit{filter bubble} can explain the decline in democratic exchange, as they cause social media users to become polarized (e.g., \citet{Bail.2018,minh_pham_effect_2020}) and radicalized (e.g., \citet{Torregrosa.2020}). \textit{Echo chambers} -- a term first introduced by \citet{sunstein2001republic} who repeatedly emphasized the dangers of echo chambers -- are frequently defined similarly as ``environments in which the opinion, political leaning, or belief of users about a topic gets reinforced due to repeated interactions with peers or sources having similar tendencies and attitudes'' \cite{Cinelli.2020}. In the context of social media, this involves the interchangeably used but differing concept of filter bubbles. \textit{Filter bubbles} -- coined by \citet{pariser_filter_2011} -- are created by personalized recommendation systems that expose users to content similar to their beliefs. These algorithms rank and moderate content to provide users with a personalized universe of information \cite{pariser_filter_2011}. Similarly to previous research, we will use the term echo chamber to describe both Sunstein's construct of an echo chamber and Pariser's filter bubble. This is due to the similarity in conceptualization and the fact that Sunstein included algorithms in the characterization of echo chambers in later work (see e.g., \citet{sunstein2017republic}).

The public discussion around both of these theoretical constructs \cite{van_aelst_political_2017, Dubois.2018} underscores the need for exposure to diverse viewpoints, which, according to \citet{habermas} and \citet{dahlgreen}, is crucial for the public ability to influence politics through a critical exchange of ideas and for consensus-building in democracies. The fear regarding echo chambers on social media arises from users having a vast range of media environments to choose from on the internet. Thus, users can choose to be exposed to conversations with like-minded users and content that reflects their existing preferences or beliefs  \cite{van_aelst_political_2017, Dubois.2018}. According to the \textit{echo chamber hypothesis} \cite{Terren.2021}\footnote{Also referred to as the “echo chamber argument,” \cite{Boulianne.2020}, or the “the echo chamber effect”  \cite{Cinelli.2021}.}, echo chambers in social networks lead to the fragmentation of increasingly polarized groups, which can profoundly impact public debate. There is a growing fear that echo chambers have led or could lead to an epistemological crisis and seriously threaten democratic societies \cite{benkler_network_2018}. 

However, there is conflicting evidence on the existence and effects of echo chambers. Thus, the motivation for conducting this review is the fact that empirical research -- especially regarding computation social science -- on echo chambers remains inconclusive. The review will address the following research questions:
\begin{enumerate}
\item How does academic literature characterize echo chambers in social networks, including their antecedents and effects?
 \item How are echo chambers measured (conceptualization and operationalization)? 
    \item How can the varying outcomes in echo chamber research be explained?
\end{enumerate}
The systematic review offers an overview of antecedents, properties (like existence), and effects of echo chambers in social networks based on research published until December 31st, 2023. We reviewed 129 peer-reviewed studies. From an initial set of 1,706 studies, we selected these 129 studies based on criteria explained in Section \ref{sec:inclusion}. We identify a taxonomy of conceptualizations and operationalizations of echo chambers and determine antecedents and effects that organize existing work and support future studies. This analysis aims to highlight certain assumptions and choices in the research design of echo chambers. We find that studies differ in (1) outcomes, (2) focus, (3) construct conceptualization and (4) operationalization, and (5) granularity.

Like \citet{jacobs} in the case of fairness, we propose to use measurement modeling from the quantitative social sciences \cite{measurement} to discuss the echo chamber construct and to compare conceptualizations and operationalizations of echo chambers. We argue that employing measurement modeling can help clarify how research defines and measures the concept of an ``echo chamber'' and how this is linked to the differences we observe in research outcomes.

Furthermore, we argue that a broader perspective on antecedents, properties, and effects of echo chambers as a concept is essential to understanding the sociotechnical structure and mechanisms of social media. We understand social media and recommendation technologies as sociotechnical systems embedded in a complex, dynamic political and social system that may have multiple feedback loops. A multi-perspective and multi-granular (individual level, structural level) approach to the topic is crucial to promote the understanding of these structures and their reproduction by such systems. This distinguishes this review from existing reviews on echo chambers \cite{Terren.2021, arguedas_echo_2022}, echo chamber detection \cite{mahmoudi}, filter bubbles \cite{michielsWhatAreFilter2022}, and social media \cite{LorenzSpreen.2023}. 

Our review identifies variations in measurement approaches, and regional, political, cultural, and platform-specific biases as key factors contributing to the lack of consensus. Studies based on homophily and computational social science methods often support the echo chamber hypothesis, while
research on content exposure that analyzes broader media environments tends to challenge it. 

\section{Related Work and Contribution}

In 2021, \citet{Terren.2021} conducted a systematic review to explore the existence of echo chambers in social networks. Similar to this systematic review, it aimed to provide a better overview of the current research (up to January 2020) and identify differences in research approaches. The authors examined 55 studies related to the existence of echo chambers in social networks. They found that differences in research design impacted whether the studies concluded that echo chambers exist in social networks. In particular, different data sets on which the research was conducted produced different results. Trace data clearly showed the existence of echo chambers, whereas self-reported data from survey studies did not \cite{Terren.2021}. While \citet{Terren.2021} sorted existing echo chamber literature by data set and foci, referring to foci such as interaction-centric studies and content-centered studies; they only provided a summary of studies that explicitly referred to the echo chamber hypothesis. In contrast, the present study provides a more comprehensive analysis of echo chambers and social media, including antecedents, properties, and effects of echo chambers and their measurement. 

Furthermore, non-systematic literature reviews on echo chambers, filter bubbles, and thematically related phenomena exist. \citet{arguedas_echo_2022} examined social science evidence of online echo chambers' existence, antecedents, and effects. They found no evidence for the echo chamber hypothesis and presented a mixed picture of how news and media affect polarization. \citet{Kitchens.2020} outlined several potential reasons for the divergent and conflicting findings related to echo chambers. However, their method was a trace data analysis, not a review study.

\citet{LorenzSpreen.2023} analyzed nearly 500 articles on the impact of digital media on democracy, including examining echo chambers. Findings are mixed: while social media platforms are shown to diversify news consumption, they also contribute to the formation and confirmation of ideologically similar social groups. They suggest a link between the increased knowledge through digital media and the rise of homophilic social structures, which may relate to the spread of hate speech and anti-outgroup sentiments. The systematic review of \citet{Interian} examined network polarization measures of 78 studies and identified the most used ones in research on online social data.

In summary, variations in research design have played a critical role in shaping conclusions about the existence of echo chambers in social networks. Differences in datasets used across studies have led to inconsistent results, with much of the evidence relying on data collected in 2020. Key open questions include the need for updated evidence on the antecedents, existence, and effects of echo chambers up to December 31, 2023, a comprehensive review of how echo chambers are measured, and an analysis of how measurement approaches influence research outcomes beyond the choice of datasets. Our work builds on these findings, incorporates evidence published in recent years, and provides novel insights into echo chamber research.

Furthermore, this review expands the scope of analysis by examining diverse conceptualizations and measurement methods of echo chamber phenomena, highlighting their impact on research conclusions. By advancing the theoretical understanding of echo chambers, refining measurement practices, and situating their study within broader methodological frameworks, this work contributes meaningfully to the field. Importantly, it offers actionable recommendations for both research and policy, underscoring the need for precision in measuring echo chambers to improve our understanding of their dynamics and implications.

\section{Method}
\label{sec:method}
We follow the \textit{PRISMA 2020} \cite{page_prisma_2021} guideline to present a transparent account of how we conducted our systematic review. A systematic review following the PRISMA 2020 guideline involves the following elements \cite{gough_weight_2007, grant_typology_2009}: (1) a definition of research questions (see below), a strategy regarding the search for relevant literature to answer these questions (Section \ref{sec:search}), explicit criteria for inclusion and exclusion of research literature (Section \ref{sec:criteria}), synthesis of evidence that can be derived from the literature (Appendix, Section \ref{sec:appendix}), and a summary of the results presented in a structured manner (Section \ref{sec:results}). 

To allow for reproducibility, we make all steps of the reviewing process explicit and describe them in detail. Our research questions are conceptual questions and methodology questions, as we are both interested in how echo chambers are conceptualized and what methods have been used to study them, to find how the varying outcomes in the literature emerge \cite{booth_systematic_2012}. 

\begin{figure}[t!]
    \centering
    \includegraphics[width=0.75\linewidth]{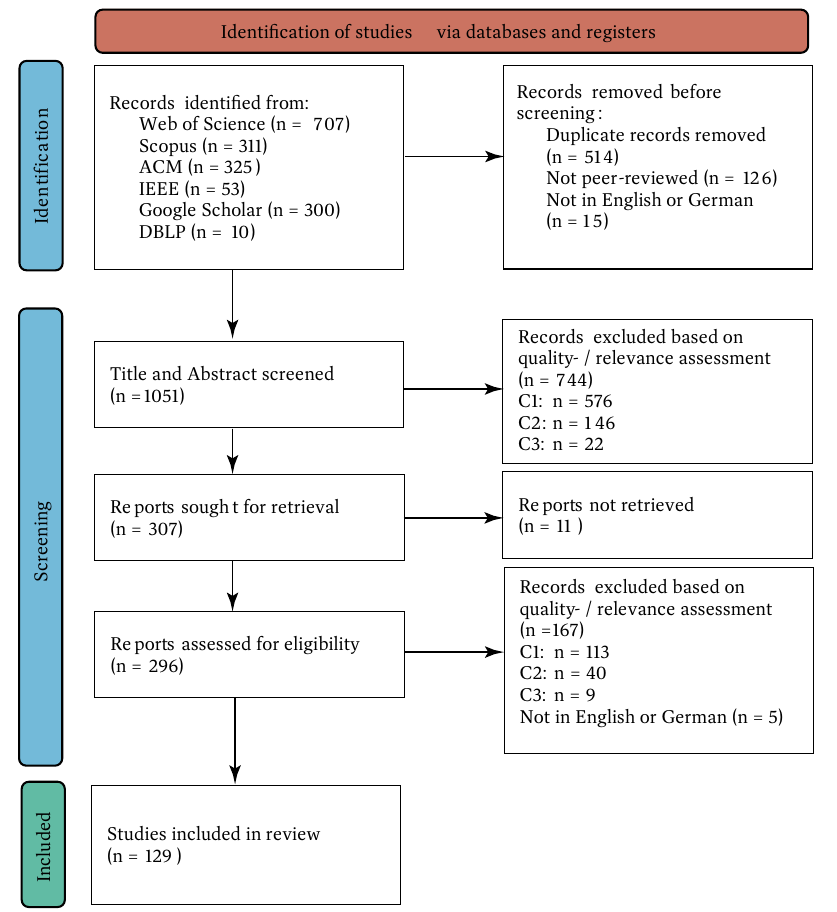}
    \caption{PRISMA 2020}
    \label{fig:prisma}
\end{figure}

We use standardized methods for search, evaluation, and selection of studies reduce subjective influences \cite{biondi-zoccai_rough_2011}. The search strategy is discussed in more detail in section \ref{sec:search}. It includes keywords and scientific databases we identified as relevant to the search, which we lay out in Section \ref{sec:database}. The criteria for inclusion and exclusion of papers are presented in Section \ref{sec:inclusion}. The PRISMA 2020 flow diagram is depicted in Figure \ref{fig:prisma}.

\subsection{Search Protocol}
\label{sec:search}

While formulating the search string, we aimed to reduce the number of irrelevant results. Some searches, such as ``echo chamber'' as a keyword, produced too many results. Other terms, such as ``echo chamber hypothesis'' or ``existence'' were too limiting and returned only a few results. Keywords were refined iteratively. Two main groups of keywords that identified broad and relevant research were merged, reducing irrelevant results. The first group included the keywords ``echo chamber'' and ``filter bubble''. The keywords of group one are presented in Figure \ref{fig:g1}. 

To keep the review focused, terms like ``selective exposure'', ``content exposure'', ``recommendation algorithms'' were not included. Using these keywords for searches resulted in more than 10,000 results. 

The second group of keywords, presented in Figure \ref{fig:g2}, exclusively contained all keywords related to social media platforms. After careful consideration regarding the inclusion of publications not written in English, we decided to remove these studies from the corpus. This was done to improve transparency of our analysis and synthesis, as not all readers can read German. Nonetheless, we want to emphasize the importance of publications written in other languages and have added citations for those studies excluded due to not being written in English. 
The second group included names of the most used social media platforms as of April 2024 \cite{wearesocial_2024}. This inclusion guaranteed that individual studies on specific platforms were taken into account. To avoid a search query that was too restricted, keywords were searched for in both titles and abstracts.\\

\subsection{Databases}
\label{sec:database}
After reviewing the most-used databases for computer science \cite{silva} and social science, the following databases were chosen: Web of Science, Scopus, ACM, IEEE, Google Scholar, and DBLP. 

Although Google Scholar does not have a main search system to compile research, it offers extensive coverage and serves as a comprehensive collection of scientific knowledge, including a large amount of grey literature - which refers to literature outside traditional publishing sources, such as dissertations, reports, and pre-prints \cite{mahood2014searching}. This can help counteract publication bias \cite{haddaway_role_2015}. While we only include peer-reviewed literature in our review, adding Google Scholar articles has expanded our sources.

Search precision for systematic searches in Google Scholar is much lower than one percent \cite{Boeker.2013}. For our query, Google Scholar returned 20,000 results. Therefore, Google Scholar was only used as a supplemental database, and its search query results were trimmed to 300 results (an amount deemed relevant for systematic reviews according to \citet{haddaway_role_2015}). 

\begin{figure}
    \centering
    \small\textbf{Echo Chamber* OR Filter Bubble*}
    \caption{Keywords for search protocol: Group 1}
    \label{fig:g1}
\end{figure}
\begin{figure}
    \centering
    \small\textbf{Social Media* OR Social Network* OR Social Platform* OR Youtube* OR  Twitter* OR X* OR Facebook* OR  Instagram* OR  Tik Tok* OR Reddit* OR YouTube* OR WeChat* OR Weibo* OR QQ* OR Kuaishou* OR Douyin* OR Snapchat* OR Pinterest* OR WhatsApp* OR Telegram* OR Signal*}
    \caption{Keywords for search protocol: Group 2}
    \label{fig:g2}
\end{figure}
\subsection{Search Results}
Database searches yielded a total of 1,706 results.
Each source's bibliographic information and abstract were downloaded into a single Citavi library\footnote{See \url{https://www.citavi.com/en}}. Citavi was used to remove duplicates during the inclusion and exclusion process and to code papers in the screening phase. 514 duplicates were removed. 
Furthermore, only peer-reviewed work was included as justified in Section \ref{sec:scope}. This selection resulted in 1,051 unique publications that were then screened for eligibility.

\subsection{Identification and Screening scope}
\label{sec:scope}
\subsubsection{Peer-reviewed Work and Publication Bias}
\label{sec:pubbias}
We want to briefly discuss potential biases that can occur in systematic reviews and how they apply to this one.  The likelihood of publication bias is one of the possible drawbacks of systematic reviews \cite{biondi-zoccai_rough_2011}. Publication bias refers to the tendency of researchers, reviewers, and editors to submit or accept articles for publication based on the direction or intensity of the study findings.

Consequently, journals may tend to prioritize publishing data with substantial or positive findings, skewing the balance in favor of such results \cite{publication}. This bias can influence systematic reviews, which often include only peer-reviewed papers to manage research volume and reduce potential small-study biases. However, retrieving unpublished data to address this issue is time-consuming, labor-intensive, and may not always yield reliable information.

While accessing unpublished data can supplement the review of published literature, it remains ineffective in fully mitigating publication bias \cite{publication}. Given this bias, it is plausible that studies presenting negative findings related to the echo chamber hypothesis or its effects face greater challenges in achieving publication compared to studies with positive results. In this context, our work makes a deliberate effort to emphasize and include negative results, contributing to a more balanced understanding of the topic.

\subsubsection{Other Potential Biases in Systematic Reviews}
\label{sec:bias}

Location bias refers to the publication of research findings in journals with varying accessibility to researchers and levels of indexing in standard databases. For this review, accessibility bias occured, since we found 11 studies that were inaccessible to us during the screening phase. While these studies have a low citation count, we recognize that their exclusion could still introduce bias and potentially overlook valuable perspectives.

Language bias occurs when research findings are published in a specific language based on the type and direction of the results. In this review, we only used English keywords and included studies written in English for transparency and reproducibility reasons. This means that studies relating to certain countries might be excluded from our corpus and there could be results relating to non-English speaking countries and the Global South that are not included in our analysis. 
Outcome reporting bias involves selectively reporting some outcomes while omitting others based on the kind and direction of the findings \cite{DRUCKER2016e109}. Due to the large amount of studies, we cannot discuss each in detail. By compiling the outcomes of studies in a table in the Appendix (see \ref{sec:appendix}) we aim to give an overview of the studies' diverse methodologies, results, and conceptualizations.

\subsubsection{Inclusion and Exclusion Criteria}
\label{sec:criteria}
Exclusion criteria were applied in the search for evidence; the inclusion criteria were the negations of the exclusion criteria. Our exclusion criteria (C1-C3) are presented in Table \ref{tab:criteria}. Criterion C1 is based on our research questions. Due to the difficulty of comparing qualitative and quantitative research, we only included papers that use quantitative methods with observational data, surveys, mixed methods, and experiments (see C2). While we discussed the existence of echo chambers in different media, such as television or newspapers, our focus is on echo chambers in social media to allow for comparisons between studies, as formulated in C3. 

\begin{table}[t]

\centering
\caption{Criteria for exclusion}
\label{tab:criteria}
\begin{tabular}[]{l p{11cm}}

\textbf{Criterion} & 
\textbf{Definition}\\
\hline
\textbf{C1} & Is not about the existence, antecedents, and effects nor any other properties of echo chambers\\
\textbf{C2} & Papers are excluded if their methodology does not employ quantitative research designs, such as surveys, experiments, CSS methods like social media data analysis, or if they solely use qualitative methods without incorporating quantitative elements. \\
\textbf{C3}&Is not about social media in relation to echo chambers \\

\end{tabular}
\end{table}%

\subsection{Screening of Search Results}
\label{sec:inclusion}
The identification of studies and the screening process in the PRISMA 2020 format can be seen in Figure \ref{fig:prisma}. During the screening process, the eligibility criteria were applied in a hierarchical cascade method, with each source being verified first against C1, then against C2, and then against C3. 
A total of 1,051 studies were screened based on title, excluding 744 studies that did not meet the three criteria. Most papers were excluded based on criterion C1. 
During the first screening round, we decided if an article was eligible based on its title. This step was conducted through one coder (the first author) who flagged articles to be discussed with a second coder (the second author) to discuss whether an exclusion is warranted. In this phase, only clearly out-of-scope articles, like articles on physical echo chambers, were excluded.
In the subsequent screening round, articles were excluded based on their abstract. To calculate Inter-coder reliability (ICR), 20 \% of articles were coded by a third coder (the third author) who was unfamiliar with the previous discussions \cite{lombard_intercoder_2006,oconnor_jaffe_2020}. The agreement between the coding conducted by the first author and the third author -- student researcher -- was moderate, and differences were then discussed. After the adjustment, the value for Cohens' $\kappa$ was $0.62$ \cite{cohen1960coefficient}, meaning a substantial agreement.

\subsection{Analysis and Synthesis}
\label{sec:coding}
The final full-text coding followed an inductive approach to content analysis, as research on echo chambers is dispersed and inconclusive \cite{elo2008qualitative}. Codes were established without a preconceived theoretical framework and then grouped into categories concerning conceptualizations and operationalizations of echo chambers, as well as findings concerning antecedents, properties, and effects \cite{hsieh2005three}. An example of a code that was inductively formed is ``Focus and Granularity'' (see \ref{tab:coding_subcoding}) for all codes and subcodes). During the coding process, it became apparent that there are significant differences between studies in regards to these two aspects. Some studies look at specific groups, others focus on individuals and their behaviourial patterns. Some studies are focused on one platform, others focus on multiple platforms. We wanted to highlight these differences since platform design can influence the behavior of users \cite{olteanu2019social}. Other codes were also formed based on the intention to highlight similarities and differences (e.g. type of data used). A short version of the codebook is presented in Table \ref{tab:coding_subcoding}.

After the codebook was established through an initial coding round and discussions between the first author and second author, the third coder independently coded 19 random studies (roughly a 15 \% sample \cite{oconnor_jaffe_2020}). Here, ICR was substantial (Cohen’s $\kappa = 0.66$\footnote{For explanation and categories see \citet{cohen1960coefficient}.}). We used MAXQDA for coding\footnote{See \url{https://www.maxqda.com/}}.

Cohen's $\kappa$ is a statistic used to measure the level of agreement between two coders while accounting for the agreement that could occur by chance. For our study, it was calculated by comparing the primary codes assigned by the first and third coder to a subset of the articles, focusing on the categories of antecedents, existence, and effects. The observed agreement was then adjusted to reflect the likelihood of random agreement, resulting in a Cohen's $\kappa$ value of 0.66, indicating substantial agreement.

Conducting a meta-analysis would not have been possible on the included papers, as research designs across identified studies are not similar \cite{siddaway2019systematic}. To organize the results of our synthesis, we instead follow a narrative synthesis approach with bibliometric features, which is suitable to show relationships in and between quantitative studies using diverse methodologies and different conceptualizations to identify factors that can explain differences \cite{popay2006guidance, siddaway2019systematic}. 

We discuss these relationships by giving an overview of the type of data and datasets used and the context of the studies, such as the country. We group studies by conceptualization, methodology, and outcome.
\let\cline\cmidrule
\newcommand{\tabitem}{~~\llap{\textbullet}~~}
\begin{small}
\begin{table}[]
    \centering

\begin{tabular}{p{6.5cm}p{6.5cm}}
\rowcolor{black}
\multicolumn{2}{c}{\color{white} \textbf{CODEBOOK}} \\ \hline
\multicolumn{1}{c}{\textbf{Code}} &
\multicolumn{1}{c}{\textbf{Subcodes}} \\ 
\hline 

\multirow{4}{6cm}
{FOCUS AND GRANULARITY: What are the foci of the publication and which factors / conceptualizations of echo chambers are used?} 
& Focus  \\  
& Group \\ 
& Platform  \\ 
& Cross-Platform \\ 
& Holistic \\
\hline
\multirow{4}{6cm}
{EC CONSTRUCT AND CONCEPTUALIZATION: Which factors and conceptualizations of echo chambers are used to analyze echo chambers?} 
& Well-defined conceptualization? \\ 
& Homophily \\  
& Content Exposure \\ 
& User Behaviour \\ 
& Group Behaviour \\ 
& Recommender Systems \\
\hline
\multirow{2}{6cm}
{SOCIAL MEDIA DATA: 
What type of social media data was used?} 
& Platform(s) \\
& Data set\\
& Countries\\
& Number of Users \\
& Number of Data Points\\
\hline
\multirow{2}{6cm}{METHOD: Which methods, operationalization, and metrics are used?} 
& Method(s) \\
& Metrics and Measures \\
\hline
\multirow{2}{6cm}{RESULTS ANTECEDENTS: Which mechanisms or antecedents are associated with echo chamber formation?} 
& Political Institutions\\
& Populism \\
& Fragmented Society  \\  
& Content Moderation \& Recommender Systems \\ 
& User Behaviour \\
\hline
\multirow{2}{6cm}{RESULTS ATTRIBUTES: Which attributes are associated with echo chamber formation?} 
& Evidence EC\\
& Mixed-Results\\
& Counter-Evidence EC\\
& Active Users \\ 
& Around Events? \\ 
& Mitigation? \\  
\hline
\multirow{2}{6cm}{RESULTS EFFECTS: Which effects are associated with echo chamber formation?} 
& Toxicity \\ 
& Emotional Contagion \\ 
& No Effect \\ 
& Polarization \\  
& Extremism \\ 
& Trust-loss \\ 
& Misinformation \\
\hline 
    \end{tabular}
   \caption{Table of the codes and subcodes from the detailed analysis\label{long}. The more detailed codebook with descriptions is shown in Appendix \ref{sec:appendix}.}
\label{tab:coding_subcoding}
\end{table}
\end{small}

\section{Results}
\label{sec:results}
In this section, we start by giving a brief corpus overview and then present the results of our analysis in sections corresponding to the research questions. 

\begin{figure}[t]
    \centering
    \begin{subfigure}{0.49 \textwidth}
        \includegraphics[width=\linewidth]{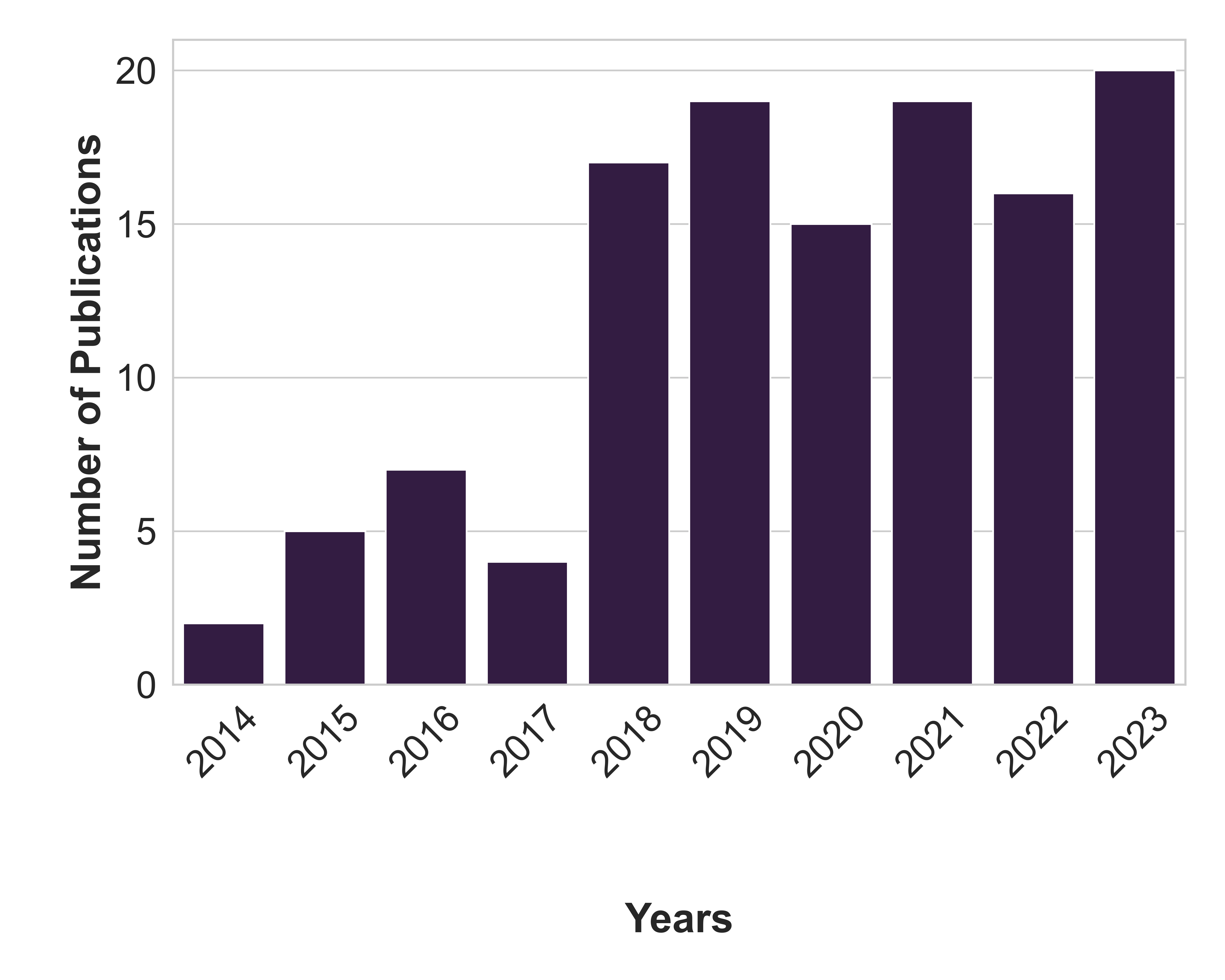}
        \subcaption{Publication years of studies included in the systematic review.}
        \label{fig:years}
    \end{subfigure}
    \hfill
    \begin{subfigure}{0.49 \textwidth}
        \includegraphics[width=\linewidth]{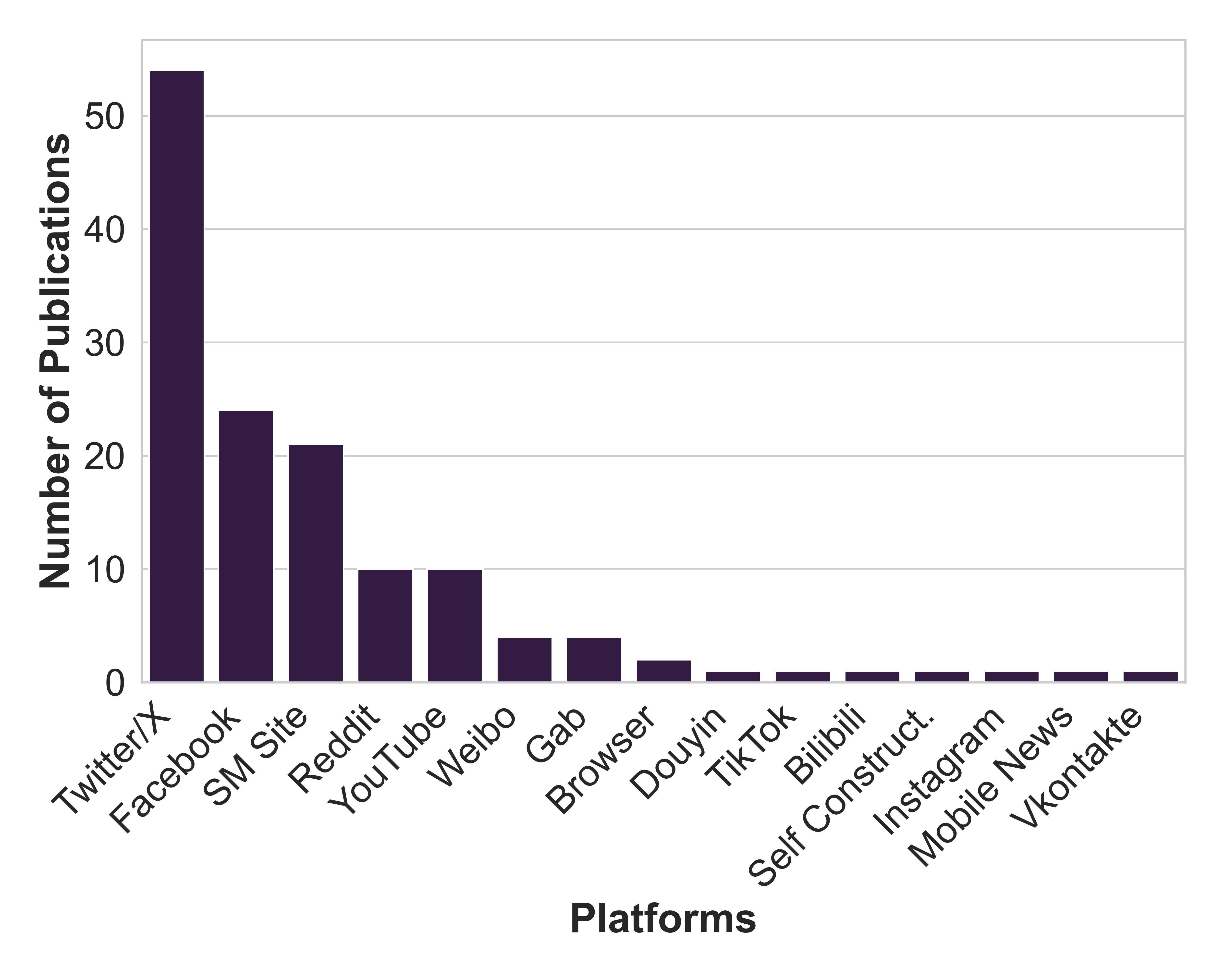}
        \subcaption{The distribution of platforms that were analyzed in the corpus.}
        \label{fig:publication}
    \end{subfigure}
    \caption{Overview of publication years and studied platforms}
\end{figure}

The 129 studies included in the systematic review were all published between 2014 and 2024, as Figure \ref{fig:years} illustrates. The year of publication was not restricted in the search query for this review. The distribution of the identified studies across different publication sources demonstrates the interdisciplinarity of this topic but also represents the need to systemize operationalizations (see Table \ref{tab:pub} in the Appendix). 

In this review, conceptualizations refer to the theoretical frameworks that define and explain the phenomenon. These include homophily, which describes the tendency of individuals to associate with like-minded others, content exposure, which examines the diversity or narrowness of information individuals encounter, and user behavior versus group behavior, which distinguishes between individual-level dynamics and collective patterns within online communities. 

Operationalizations, meanwhile, are the methods employed to measure and analyze these conceptual frameworks. Common approaches include computational social science (CSS) methods, such as network analyses and algorithmic modeling, which leverage large-scale digital data; surveys, which capture self-reported behaviors and attitudes; experiments, which create controlled environments to test causal effects; and mixed-method studies, which combine qualitative and quantitative approaches to provide a more comprehensive understanding.

\subsection{Evidence Supporting the Existence, Antecedents, and Effects of Echo Chambers in Social Networks (RQ1)}
\begin{figure}[t]
\centering
    \includegraphics[width=0.6\linewidth]{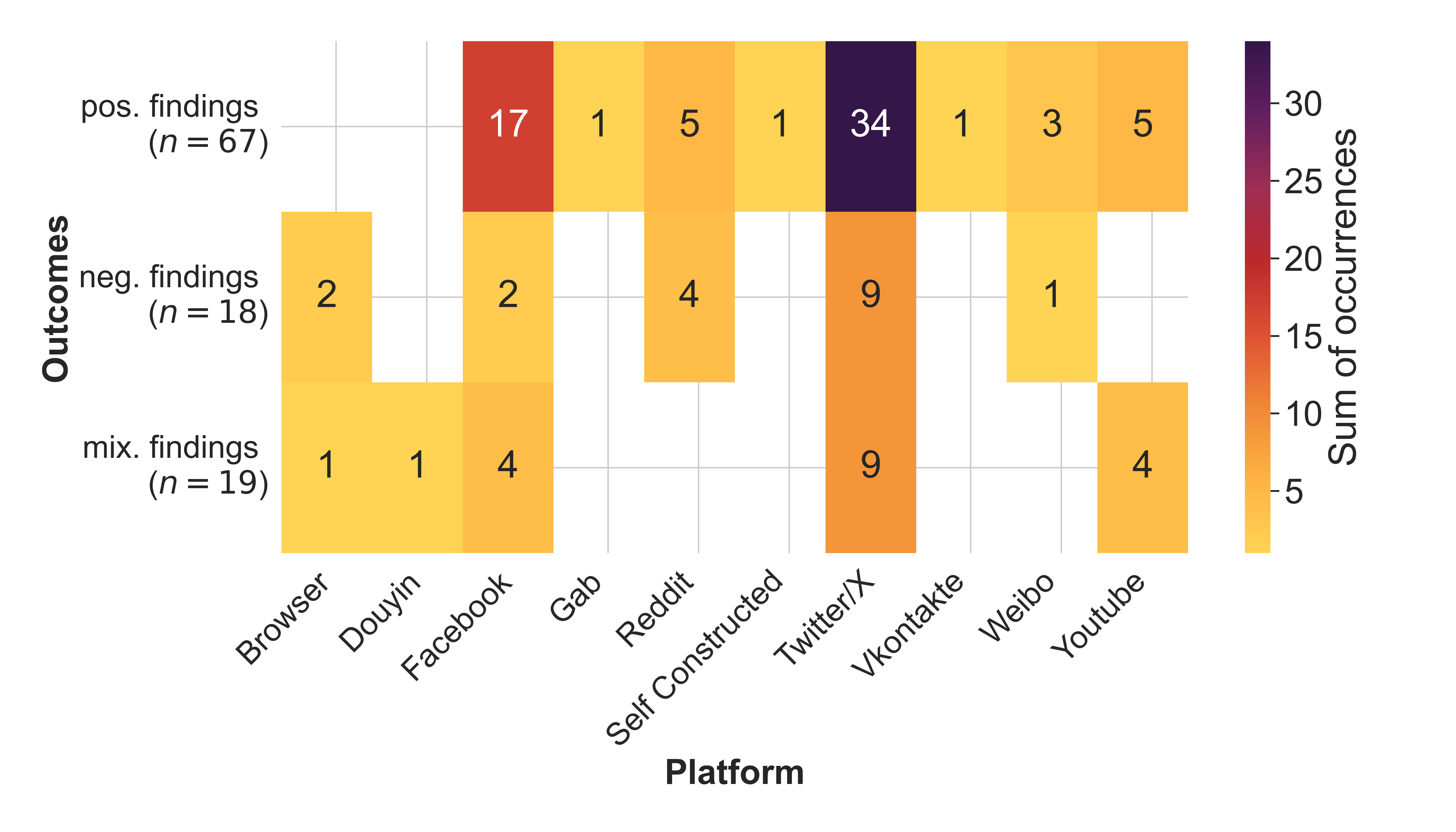}
    
\captionof{figure}{Heatmap of coded outcomes in echo chamber studies per researched platform. The colors indicate the total number of occurrences. It is important to note that most survey studies and experiments did not research one platform in particular and are therefore not considered in this plot.}
\label{fig:platform_res}
\end{figure} 
\subsubsection{Existence of Echo Chambers}
In the following, we discuss evidence for the echo chamber hypothesis, mixed results, and counter-evidence. We will include findings on specific platforms. Figure \ref{fig:publication} presents an overview of platforms when specific platforms were researched in a study. Looking at which platforms are studied shows that some platforms are studied in many publications, while other platforms appear only in a few. This is related to available APIs -- we will come back to this in the discussion -- and leads to the fact that not the most popular platforms \cite{most} were studied; instead most studies analyzed Twitter/X or Facebook, although Facebook is only in the top 10 of platforms with the most users on social media. 

Overall, the range of outcomes in echo chamber research points to a divergence of outcomes regarding the existence of echo chambers, how pronounced the phenomenon is in various communities or platforms, and what consequences its presence has. Despite contesting results, the overall findings support the echo chamber hypothesis. 

\paragraph{Evidence Supporting Echo Chamber Hypothesis}
A significant body of research ($n= 76$) demonstrates attitude-based homophily in social media. This means that in these studies, clusters of users around shared values, attributes, or beliefs were identified. This phenomenon is particularly pronounced in segregated networks, such as ``scientific'' vs. ``conspiracy'' communities \cite{Bessi.2016,DelVicario.2016,Schmidt.2018,Zollo.2017}, and among groups divided by \textit{political affiliation}, such as Democrats and Republicans. Studies examining major political events, including the 2015 Polish elections \cite{Batorski.2018}, the impeachment of Brazilian President Dilma Rousseff \cite{Cota.2019}, and the 2016 U.S. elections \cite{Guo.2020}, also support the echo chamber hypothesis, showing pronounced clustering in user interactions. 

On Facebook, research highlights the prevalence of homophilic news consumption patterns. Users often consume partisan content, reinforcing existing beliefs \cite{Kitchens.2020, Nikolov.2019}. For example, \citet{Cinelli.2021} found that Facebook facilitates the formation of segregated communities with limited exposure to cross-cutting content, especially compared to Reddit.

Twitter/X studies similarly document echo chambers, particularly in retweet networks. Polarization often arises during contentious events, such as elections, where users amplify content from ideologically aligned sources \cite{Masip.2020, Guarino.2020}.  Figure \ref{fig:platform_res} displays the outcomes for each platform and demonstrates that most research for Facebook and Twitter/X supports the echo chamber hypothesis on these platforms.    

\paragraph{Mixed Results}
 Several studies ($n = 27$) documented mixed results, meaning they found segregated communities or tendencies for echo chambers, but only under certain circumstances , assumptions, or partially in specific communities. Such studies found segregated communities in the Catalan parliamentary relations network on Twitter/X but not in the retweet network \cite{DelValle.2018}, on specific political events where users tend to group themselves around authority hubs but only for these events\cite{Guarino.2020}, in the Hungarian Twitter/X follower-followee network but not in the Polish one \cite{Matuszewski.2019b}, and in the Subreddit on the Men's Rights Movement Online \cite{Rafail.2019} but not in all Subreddits \cite{Whittaker.2021, monti2023, Kitchens.2020, treen2022}. 

\textit{Platform design} plays a significant role in shaping echo chamber tendencies. For example, Reddit's customizable recommendation algorithms encourage exposure to diverse viewpoints, reducing echo chamber effects compared to Facebook \cite{Cinelli.2021, Morales.2021, monti2023}. However, echo chambers were partially observed in specific Subreddits \cite{Rafail.2019}, as mentioned previously.

Comparative studies, such as \citet{Kitchens.2020}, found that Facebook users tend to consume more partisan news than Reddit users, who encounter greater content diversity. Twitter/X occupies a middle ground, with users showing limited changes in exposure to opposing views over time.
\citet{Nikolov.2019} also observe that content bias varies by platform, suggesting the impact of platform design on echo chamber formation.   

In our corpus, only a few studies examined short video platforms like Instagram and TikTok, likely due to barriers such as limited access to APIs and the complexities involved in analyzing video content. However, a study by \citet{Gao2023} analyzed platforms such as Douyin, TikTok, and Bilibili, demonstrating how feed algorithms foster selective exposure and homophily, ultimately contributing to echo chamber formation, while also highlighting that cultural differences can correlate with the development of echo chambers and their influence is not equivalently likely for each community and culture.

Another important dimension is \textit{user behavior}, particularly how engagement and media consumption habits influence the formation of echo chambers. User engagement has been found to correlate positively with echo chambers and political affiliation. The survey by \citet{Dubois.2020} found that political involvement and media variety reduce the risk of being in an echo chamber. They find that the amount of media a person consumes is anti-proportional to their chance of being trapped in an echo chamber, which contradicts findings by CSS studies \cite{DelVicario.2016, Schmidt.2020, Zollo.2017, Koivula.2019} that find that engagement and echo chambers are positively correlated. Greater media diversity reduces the likelihood of echo chamber formation, as studies such as \citet{Dubois.2020, jones2022can, burnett2022selfcensoring, Boulianne.2020} found only minimal indication of echo chambers when examining the complete multi-media environment. Many studies found a small echo chamber effect for Republicans, but none for Democrats before and in the aftermath of the 2016 U.S. election \cite{Justwan.2018, Neely.2021, Eady.2019, Boutyline.2017}.

\paragraph{Counter-Evidence}
In contrast, 26 studies provide little to no support for the echo chamber hypothesis. On Facebook, \citet{Beam.2018} found that users exposed to newsfeeds experienced depolarization over time, suggesting that social media could reduce rather than reinforce ideological segregation. Similarly, \citet{Eady.2019} and \citet{Barbera.2015} observed significant ideological overlap among the most liberal and conservative users, contradicting the assumption of rigid polarization. 

In general, studies examining the broader media environment including newspapers and TV do not find substantial evidence for echo chambers. For instance, \citet{Nordbrandt., Nyhan2023} observed that while social media usage does not directly influence affective polarization -- the individuals’ sympathy or antipathy toward specific political parties -- prior patterns of social media engagement can shape this relationship. Similarly, \citet{Dubois.2018} demonstrated that users with more diverse media consumption habits are less likely to become trapped in echo chambers, as varied exposure to information promotes openness to differing viewpoints and, again, politically engaged users are particularly resilient to echo chambers, with their media choices further mitigating risks of ideological isolation. Supporting this, \citet{Bail.2018} further challenge the polarization assumption, showing that social media use can increase exposure to non-partisan content.

Reddit and Weibo present unique counter-evidence. On Reddit, studies find minimal evidence of echo chambers due to platform features encouraging cross-cutting exposure \cite{Cinelli.2021, monti2023, Morales.2021}. On Weibo, \citet{Wang.2020} observed that retweeting mechanisms promote polarization, while commenting fosters consensus, highlighting the importance of interaction types in shaping polarization dynamics. Cross-platform comparisons further highlight the influence of platform characteristics on echo chamber formation. For example, \citet{Masip.2020} showed that Twitter/X users are exposed to more diverse viewpoints than Facebook users.

\subsubsection{Antecedents}
This section examines the antecedents of echo chambers, i.e. what leads to the formation of echo chambers, focusing on group and user behavior, content moderation and recommender systems, as well as polarization and fragmentation. These factors were most present in the corpus as potential contributors to the formation and persistence of echo chambers.

\paragraph{Group Behavior}
\label{sec:group}
Certain group behaviors and attributes contribute to the emergence of echo chambers. Behaviors such as avoidance, unfriending, and discreditation are mechanisms that reinforce group cohesion by excluding opposing views. Studies operationalize these behaviors through concepts like the spiral of silence (i.e., the theory that people fear isolation and, therefore, fear speaking out against mass media opinions) and affective polarization (i.e., people's emotions become more positive towards members of their own party or own community than opposing groups). Although group behaviors play a role in echo chamber formation, few studies have investigated them comprehensively or as antecedents of the construct.

\textit{Avoidance and Discreditation}: \citet{Neely.2021} found that political leanings and perceived credibility influence avoidance behaviors, particularly among Republicans, during the 2020 U.S. elections. Avoidance often manifested in politically motivated unfriending, driven by discreditation of opposing views. Similarly, \citet{Lin2023}  demonstrated that cross-cutting discussions can paradoxically heighten affective polarization through unfriending, with this effect exacerbated by exposure to incivility on social media.

\textit{Self-Censorship:} Some studies such as \citet{burnett2022selfcensoring, Powers.2019} applied the spiral of silence theory to social media, suggesting that platform dynamics might invert traditional silencing mechanisms. This inversion allows a vocal minority to overshadow the silent majority, particularly in ideologically polarized contexts. \citet{burnett2022selfcensoring} highlights significant ideological differences in self-censorship practices and fears of isolation, showing how political identity influences individuals' willingness to engage in public discourse which is as we have demonstrated correlated to echo chamber formation. Accordingly, \citet{Powers.2019} observed that college students avoid expressing political opinions online to protect their civic identity, reflecting a strategic approach to maintain social harmony in ideologically congruent networks.

\textit{Counter-Evidence:} Contrarily, \citet{Beam.2018} reported that exposure to counter-attitudinal news on Facebook could promote depolarization, challenging assumptions about affective polarization. 

Although group behaviors play a role in echo chamber formation, few studies have investigated them comprehensively or as antecedents of the phenomenon.

\paragraph{User Attributes and Behavior}
Selective exposure, demographics, emotions, and personality traits have been identified as key factors in echo chamber formation that we summarize under the category of user attributes and behaviour. Although selective exposure is the most studied and affirmed contributing factor among these user behaviours, it remains unclear how selective exposure and recommendation systems interact with each other, given that previous studies have focused on one or the other. Interestingly, despite the tendency for selective exposure, many users still consume a variety of content through diverse media channels \cite{Dubois.2018}. This can even amplify the echo chamber effect since exposure to counter-opinions may reinforce one's opinion when combined with avoidance and discreditation as described before \cite{bail2018exposure}. Consequently, some studies analyzed demographics, emotions, and personality traits as antecedents of echo chambers.

\textit{Selective Exposure:} Users often choose content that aligns with their existing opinions, reinforcing confirmation biases. However, as \citet{Dubois.2020} notes, exposure to diverse media and political engagement can counteract this tendency, reducing echo chamber risks.

\textit{Demographics and Personalization:} \citet{Bodo.2019} highlighted that younger and less educated users are more susceptible to personalized content, potentially limiting exposure to diverse viewpoints. Conversely, political interest and efficacy are associated with a reduced likelihood of being in echo chambers \cite{Chan.2019}.

\textit{Personality Traits:} Studies found that traits such as openness to experience and extraversion correlate with greater exposure to diverse viewpoints, while introversion and conscientiousness are linked to like-minded discussions \cite{matz2021personal, Boulianne.2022}.

These findings suggest that while user behavior and individual attributes play significant roles in echo chamber formation, they also offer pathways to mitigate such effects through active engagement and media diversity.
\paragraph{Content Moderation and Recommendation Systems}

recommendation systems and content moderation are frequently cited as primary drivers of echo chambers due to their role in shaping user content exposure.

\textit{Platform-Specific Differences:} Studies comparing Reddit and YouTube demonstrate stark differences. Findings show that Reddit’s customizable algorithms reduce echo chambers by encouraging diverse exposure \cite{Morales.2021}, while YouTube's recommendations amplify far-right content, fostering echo chambers \cite{Whittaker.2021, Kaiser.2020}.
    
\textit{Algorithmic Effects: }\citet{Hilbert.2018} emphasized the influence of initial network creation by recommendation systems, which can embed users in echo chambers depending on their initial interactions.
    
\textit{Limited Research on Moderation:} Despite the critical role of content moderation, few studies have explored its relationship with echo chambers due to restricted access to platform data about how content moderation works and how content is deleted and recommended \cite{hartmann2024addressing}.

While evidence suggests that recommendation algorithms significantly influence echo chambers, gaps remain in understanding the interplay between moderation practices and echo chamber dynamics. 

\paragraph{Polarization and Fragmentation}
The relationship between societal polarization, fragmentation, and echo chambers is complex and often correlational rather than causal.

\textit{Offline-Online Interaction:} \citet{Bright.2018} found that offline political success correlates with detachment (isolation of political actors) in online networks, suggesting that online fragmentation is not solely a product of digital interactions but also reflects offline political dynamics.

\textit{Geographic and Community Influences:} \citet{Bastos.2018} demonstrated that Brexit-related echo chambers had strong geographic and offline community influences, highlighting the interplay between online and offline environments. Aligning with \citet{Mahrt2019} that demanded more research on cultural context and polarization, novel research \cite{Gao2023, Grusauskaite2023, monti2023} finds that culture or demographics drive polarization and echo chambers.

These findings emphasize the need to account for offline contexts when studying echo chambers and suggest that future research should integrate online and offline dynamics to better understand these relationships.
\subsubsection{Effects of Echo Chambers}

This section synthesizes the documented effects of echo chambers, focusing on their role in polarization, the spread of misinformation, and their potential to amplify extremism.

\paragraph{Polarization and Fragmentation of the Public Sphere}
Nearly a third of the included studies in the corpus identify polarization as an effect associated with echo chambers. Polarization, a phenomenon where individuals with similar beliefs converge more strongly while those with opposing views diverge further, has been central to these studies \cite{Interian}. Researchers predominantly operationalized polarization through political leaning metrics, calculating the distance between the mean political leanings of opposing communities. Additional measures include:

\textit{Affective polarization:} which captures the degree of emotional divides between groups, such as hostility or negative sentiments toward opposing ideologies \cite{LorenzSpreen.2023} was studied by many studies that focused on group or user behavior.

\textit{Bimodality coefficients and homophily:} assessing the clustering of users based on shared characteristics or beliefs. Many studies demonstrated that shared partisanship increases the likelihood of social tie formation on platforms like Twitter, reinforcing homophily (e.g., \citet{Mosleh.2021, Cann.2021, GuerreroSole.2018}). Similarly, studies of vaccine debates on Facebook reveal how polarized communities emerge around pro- and anti-vaccine content, driven by selective exposure \cite{Schmidt.2018}, which is similar in climate change debates \cite{Cann.2021} and in ``science'' and ``conspiracy'' communities \cite{DelVicario.2017}. However, these communities might have been polarized from the beginning.

\textit{Challenge avoidance and reinforcement seeking:} Both are cognitive mechanisms that explain the persistence of polarized behavior. Studies (e.g., \citet{Brugnoli.2019, delvicarion2016emotional} found that these mechanisms, paired with peer influence, promote the formation of tightly clustered subgroups within echo chambers. Reinforcement-seeking amplifies group coherence by encouraging users to engage with content that aligns with their preexisting beliefs, while challenge avoidance leads users to ignore dissenting information, deepening polarization.

\textit{Counter-Evidence:} \citet{Nordbrandt.} challenged the assumption that social media drives societal polarization, instead proposing that existing polarization shapes social media usage patterns. This finding aligns with findings by \citet{monti2023}, \citet{Gao2023}, and \citet{Grusauskaite2023} who find that culture or demographics drive polarization. Furthermore, \citet{Nyhan2023} found that although exposure to like-minded content is prevalent on Facebook, reducing such exposure during the 2020 US presidential election had no measurable effect on polarization in beliefs or attitudes.

On the other hand, fragmentation, measured through the size and density of communities across entire platforms, has been comparatively understudied. While fragmentation reflects the structural segmentation of networks into isolated clusters, it has not received the same analytical attention as polarization.

Despite diverse operationalizations, the majority of results are consistent with the echo chamber hypothesis: echo chambers contribute to both polarization and societal fragmentation. However, novel studies from 2023 also highlight that polarization may stem from other factors, such as traditional media, cultural differences, demographics, or other offline influences, complicating the attribution of causality solely to social media.

While the relationship between polarization and echo chambers has been widely explored, the causal direction of this association remains unclear, as most studies focus on echo chambers increasing polarization and fragmentation -- aligning with the echo chamber hypothesis -- and use correlational methods. However, the study by \citet{Nyhan2023} demonstrates using a randomized controlled experiment that reducing exposure to like-minded sources may not significantly alter polarization, further questioning the causal role of echo chambers in driving societal divides.  Strategies to mitigate polarization, such as algorithmic interventions or promoting diverse interactions, are discussed by \citet{Interian} as potential recommendations for addressing these issues \cite{Interian}.

\paragraph{Spread of and Belief in Misinformation}
Echo chambers are closely tied to the belief in and spread of misinformation. Twenty-three studies in the corpus address this connection, with seven focusing explicitly on misinformation in the context of echo chambers.

\textit{Homophily and Content Distribution:}
\citet{DelVicario.2016} found that users on Facebook cluster around two types of pages, "conspiracy" and "scientific," selectively consuming and distributing content aligned with a single narrative while ignoring conflicting information. The authors argue that homophily is a key driver of content dissemination and that polarization within these communities exacerbates the spread of misinformation. Notably, the first two hours of an information cascade are critical for developing opinion clusters.

\textit{Influential Users and Information Cascades:}
Specific hub nodes play a pivotal role in spreading misinformation within networks and shaping echo chambers \cite{Asatani.2021}. For example, \citet{choi2020rumor} found that a small number of highly connected users are responsible for a significant portion of misinformation spread, with 10\% of hub communities generating 36\% of retweets of false information. These findings highlight the importance of early cascades and densely connected nodes in amplifying misinformation.

\textit{Polarized Consumption:}
\citet{Schmidt.2020} analyzed anti-vaccination communities and found that users tend to exclusively consume and share polarized content, either supporting or opposing vaccinations. This behavior has gained heightened attention since the COVID-19 pandemic, generating substantial research on echo chambers and misinformation.

In summary, misinformation is intricately linked to echo chambers, with highly connected and influential users driving both phenomena. Research emphasizes the importance of focusing on these nodes when studying the spread of false information.
\paragraph{Potential for Increasing Populism and Extremism}
14 studies in the corpus investigate the relationship between echo chambers and extremism, examining how homophily and algorithmic amplification contribute to radicalization.

\textit{Algorithmic Influence and Weak Ties:}
\citet{Wolfowicz.} analyzed trace data from ego-centric networks on Twitter/X and found that recommendation algorithms often suggest accounts aligned with users' ideological profiles. While these weak ties expose users to radical ideas, the lack of reciprocal connections may limit deeper engagement. Similarly, \citet{Torregrosa.2020} observed that network relevance is associated with extremist content, highlighting the role of user connections in spreading radical ideologies.

\textit{Homophily and Political Ideologies:}
\citet{Boutyline.2017} found that conservatives in the U.S. are more homophilic than liberals on Twitter/X, with conservative homophily rates 3.8 times higher. This structural advantage makes conservative groups more prone to echo chambers and potential radicalization. Homophily appears to facilitate mobilization within politically extreme groups.

\textit{Selective Communication:}
\citet{Bright.2018} found that individuals with radical beliefs tend to limit their interactions to like-minded individuals, avoiding discussions with those holding opposing or more moderate views. This selective communication reinforces echo chambers by deepening ideological isolation.

\textit{Platform-Specific Differences:}
\citet{Whittaker.2021} used sock puppet accounts to study recommendation systems across platforms. They found that YouTube amplifies extreme and fringe content through algorithmic recommendations, while Reddit and Gab do not exhibit the same patterns. The findings underscore the need for policy interventions to address algorithmic amplification of extremism.

\textit{Counter-evidence:} Not all studies support this connection. \citet{Boulianne.2020} found no evidence that social media increases support for right-wing populism, suggesting that online behavior often reflects existing offline communities rather than driving polarization or radicalization.

\subsection{Measurement of Echo Chambers (RQ2)}
From the perspective of measurement modeling, “echo chamber” is a theoretical construct. It describes a phenomenon observed in online and offline communities that cannot be directly measured. Instead, researchers rely on measurable proxies that serve as indirect indicators of echo chambers. Conceptualization involves defining the theoretical construct of an echo chamber in terms of measurable concepts, while operationalization is the process of translating these concepts into concrete, observable measures \cite{jacobs}. 

In the following sections, we provide an overview of the various conceptualizations and operationalizations of echo chambers in the literature. We distinguish operationalization into two components: the method of measurement (e.g., network analysis, sentiment analysis, or surveys) and the granularity of measurement (e.g., individual behaviors, group dynamics, or platform-wide patterns). Granularity acts as a link between conceptualization and operationalization, shaping how abstract concepts about echo chambers are transformed into specific, observable phenomena. 
\subsubsection{Conceptualizations}
We identified four key concepts integral to echo chamber conceptualizations: (1) \textit{homophily}, (2) \textit{content exposure}, (3) \textit{individual user behaviors}, such as selective exposure, confirmation bias or fear, and (4) \textit{group behaviors}, including exclusion of outsiders and collective unfriending.
Another dimension in the definition and then measurement of echo chambers is the granularity of the echo chamber, meaning how granular we define and measure echo chambers. We will explore how these concepts intersect with theoretical frameworks, providing an overview of the predominant conceptualizations within the corpus. 

Although both echo chamber and filter bubble as concepts were established due to concern that social media and other information-gathering platforms could influence people’s decisions about what they consume, what they think, and how they interact, the study field has conceptualized them in different ways. According to \citet{pariser_filter_2011}, personalization technology is the underlying mechanism of echo chambers, and he contends that this technology will show users information that confirms their own opinions at the expense of information that challenges them. The isolation of the individual has negative effects, as it leads to epistemic bubbles where personal ideas go unchallenged and untested \cite{nguyen_2020}. This is why many studies put more focus on the content that an individual sees through their news feed curated by recommendation systems. We call this conceptualization \textit{content exposure}. 

Both in 2001 \cite{sunstein2001republic} and then again in 2017 \cite{sunstein2017republic}, Sunstein emphasized  that technology has the potential to increase fragmentation on a larger scale, with people no longer living separately but rather forming groups where those with similar ideological preferences associate exclusively with one another.  Studies that focus on groups, interactions, and communication between users relate to the phenomenon of individuals selecting like-minded people as communication partners and information sources \cite{mcpherson}. This conceptualization of echo chambers that focuses more on the social structures than on media and information diet is captured by the term \textit{homophily}. 
A term frequently used in the context of echo chambers is selective exposure, which is said to be one of the main antecedents and mechanisms driving echo chambers. This is based on the many choices of media environments on the Internet. Users of social media platforms can choose to be exposed to conversations with like-minded users and content that reflects their thinking, thus reinforcing their existing preferences or beliefs  \cite{van_aelst_political_2017, Dubois.2018}. This tendency is accompanied by the avoidance of cognitive dissonance through avoiding challenging information and is also known under the term \textit{confirmation bias}. Selective exposure and other user-specific behaviors or attributes partially conceptualized by the corpus literature. These user attributes that are linked to echo chambers include -- besides selective exposure -- personality traits (e.g., \citet{Boulianne.2020}), fear of isolation and anger (e.g, \citet{Wollebaek.2019}), openness (e.g., \citet{matz2021personal}), and reflectiveness (e.g., \citet{Mosleh.2021}). 

Recent philosophical debates on social epistemology have built on the work of \citet{jamieson2008echo} and approach the understanding of echo chambers differently, as they strictly separate them from epistemic bubbles. \citet{nguyen_2020} argues that an echo chamber is “a social epistemic structure in which other relevant voices have been actively discredited'', meaning that people who do not share the opinions of that group are discredited as sources of information in general. The overall structure of echo chambers in his account is built on creating a substantial trust imbalance between members and non-members so outsiders are no longer taken seriously \cite{nguyen_2020}. However, such an echo chamber conceptualization is hardly reflected in the corpus, although there is research that examines particular \textit{group behavior} and presents it as echo chamber-specific.

\subsubsection{Granularity of Measurement}


Granularity is an essential dimension of an echo chamber operationalization. How granular are echo chambers analyzed, and from what perspective? After analysis of studies, we  differentiate between \textit{single-user}, \textit{group}, \textit{platform}, \textit{cross-platform}, and \textit{holistic}.
While single-user studies examine whether echo chambers originate from single individuals and how the individual influences echo chambers, group studies examine specific groups on specific platforms. Platform studies generally examine an entire platform to see whether it has an affinity for echo chambers or whether echo chambers occur there. In contrast, cross-platform studies compare platforms regarding their tendency for echo chambers. Holistic studies look at the entire media diet of users, i.e., not only certain social media but also newspapers, television, and other media simultaneously.

Most studies examined specific groups or communities on platforms, such as climate change deniers (e.g., \citet{Cann.2021}), vaccination opponents and advocates (e.g., \citet{Schmidt.2018}), and Trump supporters (e.g., \citet{Boulianne.2020}). However, few papers looked at the specific group behavior of these groups but mainly examined homophily over time and information flow to the group as opposed to outside the group. Some research conceptualizing specific user behavior also looked at specific users on platforms and how their behavior can create echo chambers. Some papers examined entire platforms and whether they correlate with echo chambers. 

Only seven studies compared platforms in cross-platform studies with CSS methods. However, if specific platforms were researched, CSS was usually utilized. This is the case as surveys mostly observed the whole media diet but not one particular platform. 
    \subsubsection{Operationalization and Methods}
\paragraph{Data-driven Computational Social Science}
\label{sec:homophily}
In the following, the general procedure of echo chamber detection and evaluation via data-driven CSS methods or Social Data Analysis (SDA)\footnote{Data-driven CSS methods and SDA will be used interchangeably in the following.}, used in 85 studies, will be presented. 
\subparagraph{General Procedure}
In general, most CSS studies used a form of social network analysis and followed this procedure: 
\begin{enumerate}
    \item Gathering social media data through an API or scraping 
    \item Building an undirected graph from this data
    \item Transforming the undirected graph into a bipartite graph
    \item Calculating network attributes from this bipartite graph
    \item Using community detection algorithms to detect clusters of homophilic relationships or use latent space models to find higher order dependencies between users
    \item Using a specific metric to determine a threshold $t$ which indicates whether there is an echo chamber/multiple echo chambers
\end{enumerate}

\subparagraph{Content-considering and Interaction Data Sets}
The underlying data sets can be classified into API and scraping data, crowdsourced or donated data, and tracking data. The latter also includes sock puppet studies -- bots that impersonate users to analyze platforms or algorithmic systems from the user's perspective \cite{sandvig_auditing_2014}. They also differ in their central research focus and granularity. API and scraping studies are predominantly platform-centered, while data donations and tracking data are mainly user-centered \cite{ohme}. In the case of this corpus of echo chamber research, most of the data sets were API-gathered data sets (Facebook API, and the Twitter/X API back when it still offered free academic access). Some used scraping or tracking via plugins (e.g., Bing toolbar, YouGov). Most data sets were collected from digital traces of users collected on specific platforms, not donated by users. 

We find another distinction between the underlying observational trace data sets for echo chamber research: purely structural data (interaction-based) and content-considering data sets. The former data sets consist of interactions between users such as `likes', `posts', `comments' ,`friendship', `shares', etc.; the latter use the content of `posts' and `comments' to gain insights into the users political leaning, polarization, radicalism or other attributes.

\label{sec:api}

The size of the data sets varied by platform and by study as Figure \ref{fig:usersAndData} in the Appendix shows. Data consisted of posts, comments, likes, and other digital traces. Different amounts of users and data were used for the analysis, with interaction-based methods having the advantage that larger data sets could often be examined, whereas content-considering data sets either had to be labeled, cleaned or had to apply resource-intensive Machine Learning (ML) algorithms to make sense of their data. The differences by platform could be related to the platforms' API access restrictions. Studies that included data donated by users included additional information to the traces, but were smaller in size\footnote{This could make them more prone to selection bias \cite{ohme}.}. Data gathering periods differ among studies as well, and most of the studies look at a certain event, such as an election to investigate emerging echo chambers around that event. There were also works that examined longer time spans. An overview of interaction-based data sets and content considering data sets can be seen in Table \ref{tab:datasets}. 
\begin{table}[t]
\small
    \centering
\resizebox{\textwidth}{!}{
    \begin{tabular}{p{2cm}p{6cm}p{6cm}}
    
         source & interaction-based data sets & content-considering data sets \\
         \hline
    \hline
         Twitter/X API& \small retweet graph \cite{Bastos.2018, Barbera.2015, Aruguete.2023, Torregrosa.2020, Wieringa.2018, rusche2022, Kratzke, Hagen.2022}, retweet cascades on rumors \cite{choi2020rumor,Asatani.2021}, follower-followee network \cite{Boutyline.2017,Colleoni.2014,bruns2017echo, Matuszewski.2019}, followers and mentions of political parties \cite{Bright.2018, DelValle.2022, DelValle.2018, Enjolras.2022}, hashtags and mentions on specific events with two or more opposite opinions \cite{Cota.2019, Furman.2020, Guarino.2020, Radicioni.2021b}, followers \cite{Gaumont.2018}, hashtags \cite{Radicioni.2021} or mentions \cite{Guo.2020} of candidates of an election process & \small evaluated shared content \cite{Cann.2021}, shared keywords on political talk shows \cite{Ceron.2019}, term frequency in political tweets for a specific event like an election or topic (e.g., climate change, vaccines) \cite{Asatani.2021,Colleoni.2014, Samantray.2019}, shared news URLs with news outlets leaning scores \cite{Shore.2018, Wieringa.2018, Flamino.2023}, sentiments of postsin a specific group \cite{Torregrosa.2020} or for a specific topic  \cite{Tyagi.2021}, following specific affiliated accounts of elites and non-elites \cite{Barbera.2015}, hashtags for political leaning estimation \cite{Bastos.2018}\\
         \rowcolor{Gray} Facebook API& \small activity of users on specific Facebook pages \cite{Batorski.2018, Schmidt.2018, Matuszewski.2019b}, likes, posts, and comments in two labeled groups of Facebook pages 'scientific' and 'conspiracy' \cite{Bessi.2015, Bessi.2016,delvicarion2016emotional, DelVicario.2016,Brugnoli.2019} and their relation to misinformation \cite{Zollo.2015} debunking posts \cite{Zollo.2017} &\small sentiments of tweets and their personalities \cite{Bessi.2016}, shared news URLs with news outlets leaning scores \cite{Bakshy.2015, Cinelli.2020, DelVicario.2017}, term / keyword clustered groups \cite{Bessi.2015}, users’ reactions to ad-hoc articles published on the Corriere della Sera Facebook page \cite{Schmidt.2020}\\
         Cross-platform comparison through API trace data& \small posts, comments, and retweets from Facebook, Twitter/X, Reddit, and Gab \cite{Cinelli.2021}& \small classified tweets (reliable vs. questionable) on Twitter/X and Gab \cite{etta}, prediction of users polarization on Facebook with YouTube content exposure \cite{Bessi.2016b}, hatespeech scored posts on Reddit, Twitter/X and Gab \cite{goel} \\
         \rowcolor{Gray} Tracking and sock puppet data sets& \small desktop and laptop web tracking data (collected by YouGov) \cite{fletcher}&  \small web-browsing records collected via Bing toolbar \cite{Flaxman.2016}, comparison between TV panel and a second-level laptop/desktop Web browsing panel \cite{muise2022}\\
         VKontakte scraping& \small friendship ties and page \small followers \cite{Urman.2019}  &\\
         \rowcolor{Gray} Reddit API& \small subreddits of Clinton and Trump supporters \cite{Morales.2021}  & \small posts and upvotes in the Men's Rights Movement \cite{Rafail.2019}, climate change related subreddit posts \cite{treen2022}, subreddit r/news communities interaction, sentiment and demographic data \cite{monti2023}, 101 popular subreddits on politics and non-politics \cite{bond2022}\\
         YouTube API& \small node-centric analysis of recommendations from specific topics \cite{Roth.2020} &\small vaccine video comments \\
         \rowcolor{Gray} Gab API&  &\small shared news URLs with news outlets leaning scores \cite{Lima.2018}\\
         Weibo API&  &\small Coded posts addressing genetically modified organisms \cite{Wang.2021}, COVID-19 related content\cite{Wang.2020,wang2022}, \\
         
    \end{tabular}}
    \caption{Trace data sets included in the corpus clustered by interaction-based or content-based and data source}
    \label{tab:datasets}
\end{table}

Based on these two types of data sets, the general procedure was to build a graph for every time-dependent sample as social networks are dynamic networks \cite{oxford1} and use interaction (structural data sets) or semantic networks (content-considering data sets) \footnote{Graph is represented by $G= (V,E)$ with a node set $V = \{v_1,\dots,v_n \}$ and edge set $E \subseteq V \times V$.}. In the identified studies, nodes often represent users and edges denote interactions between users and interaction of users with content such as posts or hyperlinks to news pages (e.g., \citet{Aruguete.2023}). Thus, users can be connected in a graph through edges if they are friends (e.g., \citet{Urman.2019}) or have shared content. At the same time, the user is connected to posts or news pages that they have shared, thus also creating a semantic network. Each user can be characterized by a vector of node attributes\footnote{For a user $u$ in $V$ this vector is denoted by $u_i = (x_{1,\dots, j})^T$}, such as demographic information, interests, or activity patterns. The edges can be weighted by the strength of the social tie between users, such as the frequency, intensity of interactions, or a friendship tie.

\subparagraph{User-Attribute Estimation} Some of the content-considering data sets first transformed the network into a weighted, undirected network by computing the pairwise similarities between nodes of the same type (e.g., similarity between users based on their shared interactions with content or friendship ties). They estimated the user attributes like political leaning by calculating ideology scores or political preference \footnote{Calculated through: $f: (x_{1,\dots, j})^T \rightarrow \{-1, \dots, 1\}$} based on the users attributes with an ideology score ranging from -1 (conservative) to 1 (liberal) (e.g., \citet{Boutyline.2017}), by shared news pages which were sorted by their political affiliation (e.g., \citet{Aruguete.2023}, \citet{Bakshy.2015}), averaging over user's produced content scores with the number of posts ranging as well from $-1$ to $1$ (e.g., \citet{Cinelli.2021}) or by their shared content by using sentiment analysis (e.g., \citet{Bessi.2015}). 

These user-attribute estimation techniques are, as their name suggests, used to estimate user attributes, such as political leaning, based on the content they consume, share, or interact with. They enable the analysis of interactions between specific groups of attributes. User leaning was also estimated by activity in specific groups (e.g., \citet{Cinelli.2021b, Rafail.2019, Bessi.2015}), by semantics in the posts (e.g., \citet{Cann.2021, Asatani.2021, Colleoni.2014}), calculation of political slant of visited websites by the user \cite{Shore.2018}, shared news URLs with news outlets leaning score (e.g., \citet{Bakshy.2015, Cinelli.2020, DelVicario.2017}).

\subparagraph{Two-Mode and Bipartite Graphs} Some content-considering studies incorporate a second set of nodes to represent content items (e.g., posts, articles), creating bipartite graphs where users are connected to the content they interact with. From these, unipartite projections or multiple bipartite graphs are derived to study user-to-user connections or user-content dynamics \cite{DelVicario.2017, Cann.2021}. Polarization and homophily measures are frequently calculated on these networks to assess ideological clustering.

\subparagraph{Community Detection} The next methodological step for most of the social data methods was to use a community detection algorithm. The Louvain algorithm\footnote{The Louvain algorithm is a greedy agglomerative cluster algorithm that optimizes a modularity score by iteratively merging nodes into communities that maximize the increase in modularity. The modularity score measures the degree to which nodes within a community are more densely connected than nodes outside the community \cite{traag_louvain_2019}.}, Leiden algorithm\footnote{The Leiden algorithm is an extension of the Louvain algorithm that improves its community quality and scalability by employing a refinement step that further partitions communities into sub-communities \cite{traag_louvain_2019}.} (e.g., \citet{Asatani.2021}), random walk modelling (e.g., \citet{Aruguete.2023}, \cite{Bessi.2015}), greedy algorithms (e.g., \citet{Cann.2021}), latent space models  \footnote{Assumes that nodes in the network belong to one of K latent communities and that the probability of an edge between two nodes depends on their community membership \cite{latentspace}. In latent space models, the goal is to learn a low-dimensional embedding of the nodes in the network, where nodes that are close in the latent space are likely to have similar connectivity patterns.}(e.g., \citet{choi2020rumor}, \citet{Cinelli.2020}), flow stability (e.g., \citet{bovet2022}) or hierarchical clustering algorithms (e.g., \citet{Bastos.2018}) are used in most cases of the included studies to detect communities in the bipartite graph\footnote{Let $A = \{A_1,\dots,A_q\}$ be a set of node groups defined over $V$, that is, each $A_i \subseteq V$ for any $i = 1,\dots, q$. Then $A_i$ is a community of $G$ based on some node and edge attributes.}.
 
\subparagraph{CSS Metrics} In the context of the bipartite graph $B$, these algorithms are used to detect communities of users who are more likely to interact with content items within their community than with content items outside their community. These communities can be interpreted as echo chambers, where users are exposed to information that reinforces their existing beliefs and opinions, if these communities are communities of like-minded people with similar attributes. To asses the similarity between different users in communities, different metrics are used. 

To detect and analyze echo chambers, a variety of metrics are employed, categorized into interaction-based metrics, content-considering metrics, homophily and polarization measures, and modularity measures. These metrics differ by data type, granularity, and approach (connectionist vs. positional) \cite{oxford1}. 

\subparagraph{Interaction-based methods} These metrics rely heavily on centrality measures to identify influential users and assess the structural dynamics of networks:

\textit{Degree Centrality:} Nodes with a high in-degree, such as accounts receiving frequent retweets, signify prominence within the network \cite{Guarino.2020, Torregrosa.2020}.

\textit{Eigenvector Centrality:} Highlights nodes connected to influential accounts, portraying their authority within the network \cite{Kratzke, Torregrosa.2020}.

\textit{Closeness Centrality:} Identifies nodes with the shortest paths to all others, indicating their ability to disseminate information effectively \cite{Torregrosa.2020}.

\textit{Betweenness Centrality:} Measures nodes acting as bridges, facilitating the flow of information and potentially mitigating misinformation \cite{Torregrosa.2020}.

Centrality measures identify influential nodes, such as disseminators (high closeness centrality) or bridges (high betweenness centrality), offering insights into individual roles within echo chambers \cite{Torregrosa.2020, Kratzke}. While these metrics provide user-level insights, they do not directly capture broader interaction patterns or group dynamics. 

\subparagraph{Content-Considering Metrics} They estimate user attributes, such as political leanings, or analyze cross-cutting content exposure to understand how individuals engage with ideologically similar or diverse content. Metrics like polarity scores measure the distance between users based on their attributes when group membership is unavailable (e.g., \cite{choi2020rumor, Boutyline.2017}). For example: \citet{Bakshy.2015}, \citet{choi2020rumor}, and \citet{Brugnoli.2019} used polarity scores to assess user ideological alignment. \citet{DelVicario.2016} and \citet{Flaxman.2016} demonstrated that content-considering methods effectively detect echo chambers formed through selective content exposure.

\subparagraph{Homophily and Polarization Metrics} These are used interchangeably in many studies to evaluate the similarity between users within echo chambers.

        \textit{Node-Level Homophily:} Focuses on individual tendencies to connect with similar users (e.g., \cite{Boutyline.2017, Asatani.2021}).
        
        \textit{Group-Level Homophily:} Aggregates the homophily of individual group members to assess group polarization \cite{Interian}.
        
        \textit{Network-Level Homophily:} Measures the overall homophily of a network to understand the systemic tendencies for like-minded clustering \cite{Asatani.2021, Enjolras.2022}.
        
        \textit{EI-Index:} Calculates relative homophily, capturing the balance between in-group and out-group ties \cite{bruns2017echo, Kaiser.2020}.

While the measures can also indirectly represent structural trends in a network, homophily studies the propensity for
people to connect with those who share their qualities or traits and does not directly capture the structural organization
of these groups. This is why it is often combined with community detection. Homophily measures highlight user preferences and shared characteristics, directly linking echo chambers to ideological or demographic alignment \cite{Boutyline.2017, Interian}. Homophily measures, particularly ideological homophily (e.g., \cite{Boutyline.2017, Enjolras.2022}) and demographic homophily (e.g., \cite{srba2023}), are crucial for analyzing echo chambers with CSS research. The strong association of homophily and
polarization measures used in the corpus, especially in connection to conceptualizations of echo chambers as homophily
demonstrates the importance and advantages of using homophily measures in CSS echo chamber research. 

\subparagraph{Modularity Measures}
Modularity measures the organization and segregation of communities within a network, providing insights into the structural presence of echo chambers. Modularity captures larger structural trends, assessing how communities form and segregate \cite{Kratzke}. For example, \citet{Kratzke} used modularity to analyze community dynamics, complementing user-level centrality measures.

\paragraph{Surveys}
Out of all the analyzed studies, 20 were surveys. Most of these surveys were longitudinal, meaning they collected data over a period of time, whereas a few were cross-sectional, meaning data was collected at one point in time. About half of the surveys were conducted in the USA (a total of 8), and only 5 of these studies addressed the echo chamber hypothesis directly. The rest of the studies examined elective exposure of social media users, news consumption and sharing behavior, extremism, and voting patterns. As Figure \ref{fig:conc-ope} demonstrates, most survey studies conceptualized echo chambers with content exposure. These studies mostly researched how much cross-cutting content users self-report via their media usage. User behavior survey studies researched specific user behavior that is associated with echo chambers. 

 In both cross-sectional and longitudinal studies, some surveys were conducted around election periods. This means that panels were usually surveyed before and after the elections or a three-wave panel was used. In cross-sectional studies, participants were surveyed either shortly before or after an election. The longitudinal studies are all panels that were either built around elections or drawn from existing panels such as household panels (e.g., \citet{burnett2022selfcensoring}), panels about migration (e.g., \citet{zerback}) or social media and internet panels (e.g., \citet{Nordbrandt.}). 

Out of all survey studies, 17 looked at general social media usage or the whole media diet, while only two looked at specific platforms. \citet{Masip.2020} analyzed social media usage on Facebook, Instagram, and Twitter/X, focusing on news-sharing behavior in Spain. Similarly, \citet{Beam.2018} examined news polarization on Facebook before and after the 2016 U.S. elections. Interestingly, both studies found no echo chambers or polarization through social media, contrary to homophily studies.

18 surveys that focused on content exposure have examined the entire media landscape and compared the content exposure of social media with other media outlets. \citet{Dubois.2018}, for instance, studied the use of other media and found that social media is only one part of the users' media diet and argued that echo chamber research has to include a wide variety of media usage.

Some survey studies ($n = 14$) examine the role of user attitudes in shaping online behavior and mitigating the effects of echo chambers. For example, the study by \citet{Dubois.2020} examines the effects of fact-checking, political interest, and opinion leadership on individuals' exposure to different viewpoints and their susceptibility to echo chamber effects. Similarly, studies by \citet{Koivula.2019} and \citet{zerback} examine the role of political activity, extreme attitudes, and interpersonal communication in reinforcing or counteracting echo chambers in online communities. Studies by \citet{Chan.2019}, \citet{Boulianne.2020}, and \citet{Neely.2021} explore how factors like internal political efficacy, personality, and social network structures influence individuals' interactions with political content and their susceptibility to echo chamber effects. The focus is placed on user behavior and thus also conceptualized and operationalized by investigating how specific user attributes, such as political interest, change content exposure. There is little focus here on the social environment of the users but instead on what content they are exposed to with certain behaviors or specific attributes, such as fear of isolation or personality traits.

Survey research involves a broad set of measures like political engagement, media consumption, and specific topics like climate change, misinformation, COVID-19, vaccines, news dissemination, and other forms of media use such as TV and newspapers. The studies use between 5 and 41 items to gather data from a participant pool of 198 to 11,052 individuals. 

Control variables are mainly demographics and political ideology. To evaluate the broader societal implications of echo chamber behavior, researchers use dependent variables that we group into four categories: (1) political engagement and behavior, (2) media consumption and exposure, (3) homophily and polarization measures, (4) trust in specific information and misinformation sharing. 
Political engagement and behavior measures include political engagement (e.g., following political news, expressing political views), social media behavior and impact (e.g., reliance on Facebook, unfriending/unfollowing due to political posts) polarization, satisfaction with democracy, political ideology, party affiliation, and affective polarization, news consumption habits (e.g., interest in hard news, news trust).

Media consumption and exposure metrics include exposure to various news sources (e.g., news websites, TV news) and engagement with news, social media usage and attitudes, perception of the public sphere, universal news access, and privacy concerns.
Homophily and other echo chamber measures include like-minded discussion, perceived viewpoint diversity exposure, and cognitive attitude extremity \footnote{Meaning an attitude that is cognitively explainable and is not less emotionally based as affective attitudes \citet{zerback}.}.

Trust in information and misinformation measures include trust in news sources, traditional media use for COVID-19, and vaccine hesitancy.
 
\paragraph{Experiments}
Ten experiments are part of the corpus. Such experiments include observing test subjects while they follow a bot that shares certain types of political content, experiments with their own newsfeeds or self-constructed platforms, or even sock puppet studies, in which bots generate online content and collect what content they encounter. Most experiments focus on user behavior (e.g., \citet{Mosleh.2021}), recommendation systems' influence on echo chamber creation (mostly through sock puppet studies like e.g., \citet{minh_pham_effect_2020, Whittaker.2021}), on extremism and misinformation (e.g., \citet{Wolfowicz.}). Like surveys, these experiments used operationalizations that isolated the effects of echo chambers by controlling for variables like demographics and political ideology. 

Figure \ref{fig:outcome1} demonstrates that most experimental studies find echo chambers. These findings point mostly to individual behavior like selective exposure, extremism, and anger. Data sets for experiments vary from 102 to 1,652 users. They can be categorized into (1) sock puppet data, (2) user-centric data, and (3) trace data experiments. Sock puppet experiments (\citet{minh_pham_effect_2020, Whittaker.2021,Mosleh.2021}) used bots that mimic user behavior on Twitter/X, YouTube, and Weibo. In contrast to trace data, bots enable real-world experiments on platforms which comes with the advantage of controlled experiments but is limited to a certain number of bots and has to be ethically evaluated. Sock puppet studies used measures like the directed clustering coefficient, connection density, strong and weak co-partisanship and counter-partisanship\footnote{This means that users with ties on social media share partisanship for a party or not \cite{Mosleh.2021b}. Particularly important and tied to two-party systems.}, social attentiveness, attitude consistency, algorithmic polarization, and extremist media index. User-centric data sets were used, e.g., in an eye-tracking study by \citet{Suelflow.2019} which used measures like political affiliation, affective polarization, perceived polarization, contact with immigrants, and political interest. Trace data experiments like the experiment by \citet{bail2018exposure} measure the change of affiliation after the experiment through a survey.

\paragraph{Mixed Methods}

Six of the included studies used a mixed methods approach. The main idea of mixed methods approaches is to have different perspectives and granularities in echo chamber research and platform research. Most mixed methods approaches studied a particular platform; some did a cross-platform analysis (e.g., \citet{Kitchens.2020}). Most mixed method studies focused either on user behavior, recommendation systems' influence on echo chamber creation (mainly through sock puppet studies), or extremism and misinformation. 

Most link survey data with trace data, and some link experiments or surveys with trace data. As Figure \ref{fig:outcome1} suggests, mixed methods approaches have mixed results. Here again, studies using homophily as a conceptualization tend to have affirming results on the echo chamber hypothesis, and studies using context exposure tend to have negative findings. All but two mixed method studies used trace data and analyzed user groups ranging from 2,000 to 42,600 users (e.g., \citet{matz2021personal}). The other two studies mixed impression data with surveys (e.g., \citet{Hilbert.2018}). 

The survey and trace data linking approaches used categories of accounts: media elite, political elite, and non-elite (\citet{Eady.2019}), distinct news sites, slant dispersion, reverse Gini index, audience variety, mean slan, cross-cutting proportion (\citet{Kitchens.2020}), Big Five Personality Traits (\citet{matz2021personal}), and a combination of cognitive reflection and co-follower network (\citet{Mosleh.2021}). The two studies that combined impression data and surveys used attitude polarization, anger, and emotional 
valence (\citet{Hilbert.2018}).
\subsection{Varying Outcomes in Echo Chamber Research (RQ3)}

This section aims to answer RQ3, namely: ``How can the varying outcomes in echo chamber research be explained?''. Synthesizing the results from our first two research questions yields five dimensions where echo chamber studies diverge, providing a framework to examine the causes of conflicting outcomes in this field systematically. Specifically, echo chamber studies vary across dimensions of  (1) focus, (2) construct conceptualization, (3) operationalization, (4) granularity, and finally in (5) outcomes. 

In the following, we will give an extended overview of how studies vary across these dimensions and outcomes, analyze the relationship between outcomes and measurement, and discuss how geographical and contextual factors may influence research approaches and variations in findings.

\subsubsection{Varying Research, Varying Outcomes}

Firstly, the echo chamber construct is controversial, and researchers claim that it was ill-defined from the beginning and that researchers should either abandon or conceptualize it properly \cite{Mahrt2019}. This is largely because the concept does not represent any observable outcome since ``echo chambers'' describe a state of absence, namely the absence of an idealized model of a deliberative public sphere such as the Habermasian democratic model \cite{Kitchens.2020, bruns2017echo}. 

In addition, studies on echo chambers in social networks have been carried out by researchers from various fields -- such as computer science, sociology, political science, law, media, and communication science -- using differing methods and thus different \textit{operationalizations} of the echo chamber construct. Some studies have conducted a network analysis based on data from social networks \cite{Cinelli.2021}, while others have interviewed Twitter/X users after they were exposed to the contents of a social media bot \cite{Bail.2018}.  %

The difficulties in operationalizing extend to the \textit{granularity} of echo chamber detection as they can occur on individual users, groups, or entire platforms \cite{Kratzke} and related to the research decision to analyze specific groups or platforms, data utilized in the empirical research is subject to sampling bias even if there are attempts to counteract these differences \cite{Kitchens.2020,Dubois.2018}. 

Previous studies may have analyzed mainly active users and used scarce or incomplete data sets to support their conclusions. This fact may have led to misleading results, as the periphery of networks appears to be essential for the average behavior of social networks \cite{Shore.2018}. Studies analyzed online platforms either independently (one specific platform) or in aggregate (all of social media), which may make it difficult to determine the link between social aspects, technological characteristics, and information-limiting contexts \cite{Kitchens.2020}. 

If only one medium (e.g., one platform or one online newspaper) is analyzed, this may not provide relevant information on how political information moves across offline and online media \cite{Dubois.2018}. Moreover, demographics differ across platforms \cite{Mellon.2017,pewresearch}. 

Echo chamber research in our corpus also differs in \textit{focus}. Some research studies are primarily concerned with the echo chamber hypothesis, while others focus on the antecedents, properties, and effects of echo chambers. 

Finally, studies differ in \textit{outcomes}. Although the majority of studies has found evidence for echo chambers on platforms such as Facebook (e.g., \citet{Schmidt.2020}), there is still a substantial amount of studies that had mixed findings, and one-fourth of studies that focused on the echo chamber hypothesis have not identified evidence for echo chambers (e.g., \citet{Beam.2018,Bakshy.2015}). 

Echo chambers are also repeatedly associated with the spread of misinformation, which poses the risk of poor political decisions or the formation of opinions based on falsehoods \cite{DelVicario.2016, choi2020rumor,jasny_empirical_2015}. However, scientific findings have also been contrasting with the echo chamber hypothesis. Some studies indicate that social media networks are similar to real-life interpersonal networks (e.g., \citet{Dubois.2018}, \citet{Bastos.2018}), thus downplaying the asserted impact of social media or that social networks do not change the political discourse excessively (e.g., \citet{Barbera.2015,bruns2017echo,avoiding}). 

Some studies claim that echo chambers have no significant impact  (e.g., \citet{Dubois.2018}), do not contribute to polarization (e.g., \citet{Nyhan2023}), or do not demonstrate increased exposure to dissimilar opinions through social media (e.g., \citet{bail2018exposure}). According to some studies, societal polarization is not a phenomenon attributable exclusively to social media, but to other forms of media like radio and TV as well (e.g, \citet{jamieson2008echo, muise2022}). Other studies find that cultural and demographic factors are more important than structural (e.g., \citet{monti2023, Grusauskaite2023}). 

\begin{figure}[t]
    \centering
    \begin{subfigure}{0.49\textwidth}
    \centering
    \includegraphics[width=1\linewidth]{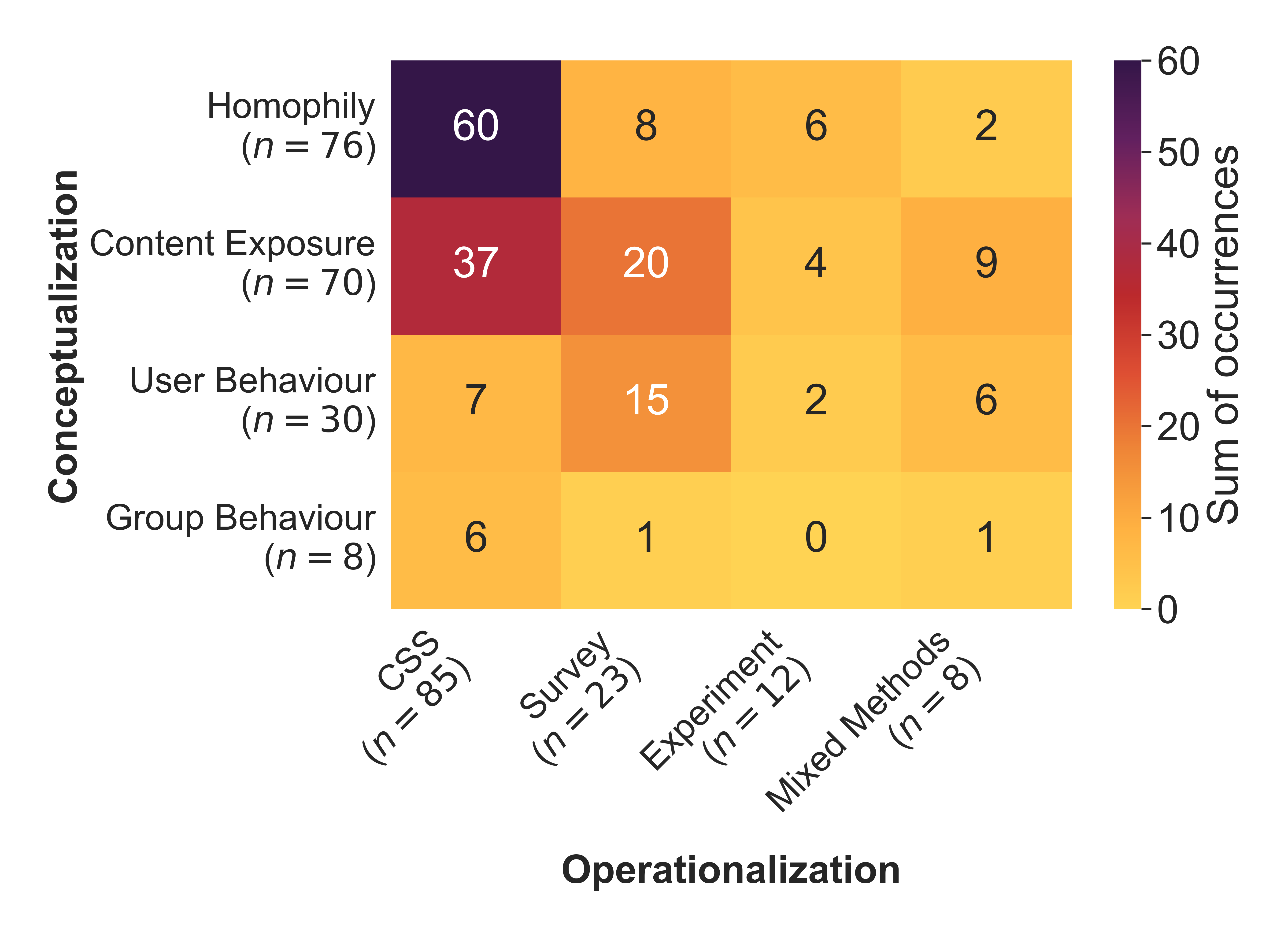}
    \caption{Contingency table of coded conceptualizations and operationalizations of the corpus literature.}
    \label{fig:conc-ope}
    \end{subfigure}
    \hfill
    \begin{subfigure}{0.49\textwidth}
    \centering
    \includegraphics[width=1\linewidth]{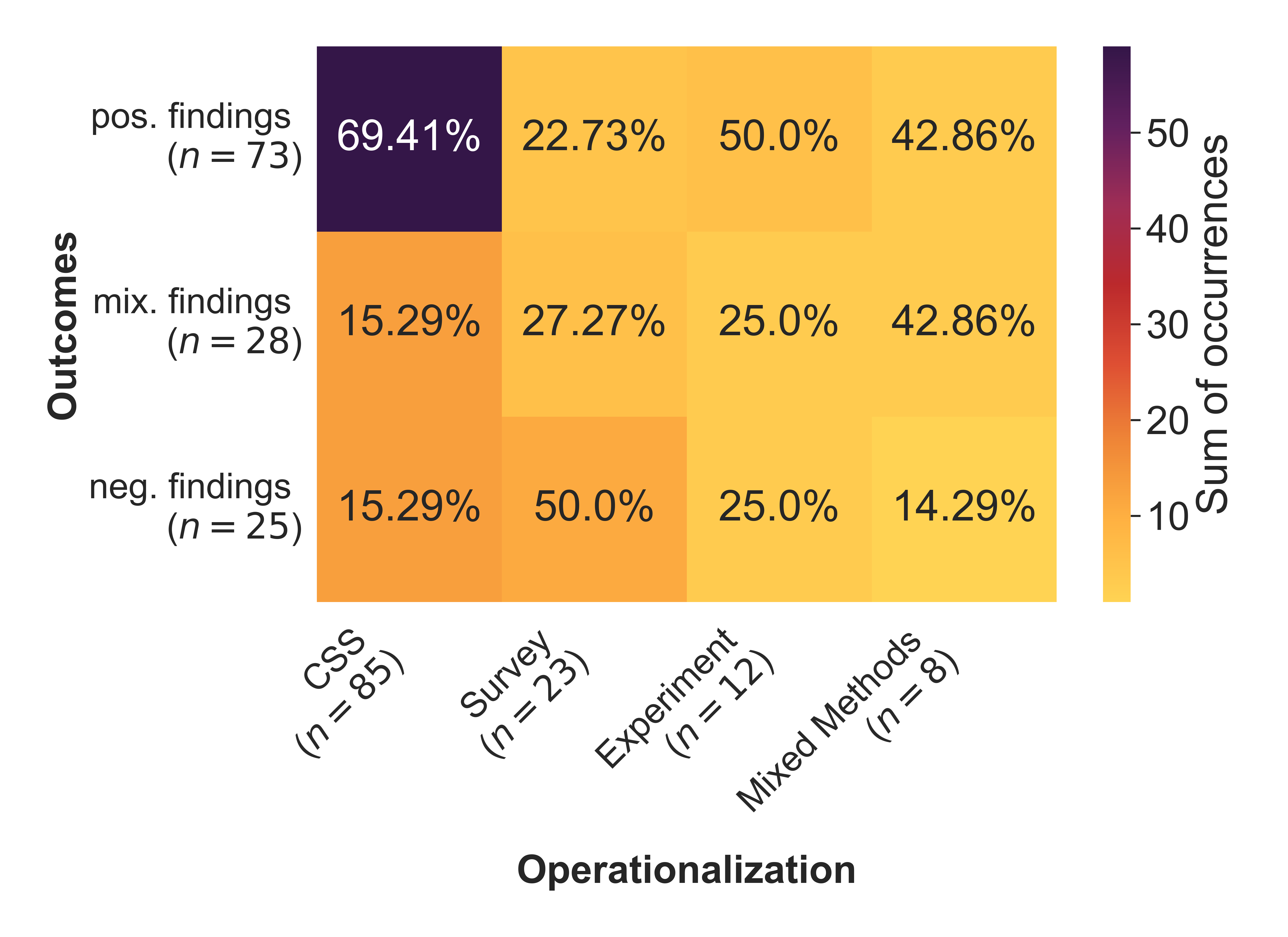}
    \caption{Overview of positive, negative, and mixed findings from echo chamber research, categorized by operationalization.}
    \label{fig:outcome1}
    \end{subfigure}
\end{figure}
\subsubsection{Relationship Between Outcomes and Measurement}

Our analysis culminates in the analysis of factors contributing to the varying outcomes in echo chamber research, with measurement playing a central role. Outcomes are closely linked to how echo chambers are conceptualized and operationalized. We start by examining the relationships between conceptualization, method, and granularity, followed by an analysis of how these elements influence outcomes. Conceptualization and operationalization are interconnected. Figure \ref{fig:conc-ope} shows that conceptualizations often align with specific methods. For instance, studies using CSS methods typically conceptualize echo chambers as attribute-based homophilic structures, whereas surveys and experimental studies focus on content exposure and user behavior.

The granularity of measurement varies depending on the method. Survey studies often assess exposure across broader media environments, encompassing both social media and traditional media. In contrast, CSS studies focus more narrowly on structural aspects of specific communities or platforms. These methodological differences shape the outcomes of the studies.

Differences in definitions of echo chambers—whether based on homophily, content exposure, or selective exposure—translate into distinct methodological choices, which in turn shape findings. Figure \ref{fig:outcome1} illustrates a clear relationship between methods and outcomes. Studies employing CSS methods are more likely to identify echo chambers, while surveys and experiments tend to produce mixed or less conclusive results.

Studies that challenge the echo chamber hypothesis often conceptualize echo chambers in terms of content exposure and emphasize the availability of cross-cutting content. These studies argue that users are exposed to diverse media sources beyond social networks, challenging the notion of isolated filter bubbles. However, they also find that echo chambers can still emerge under specific conditions, such as heightened political interest, selective exposure tendencies, or high user activity. Surveys and experiments in this area focus on individual-level behaviors and demographics, often reporting modest or mixed effects. These findings suggest that individual traits mediate the formation of echo chambers.

CSS studies, by contrast, frequently provide evidence supporting the echo chamber hypothesis. These studies focus on network analysis and trace data to examine homophily, showing how like-minded individuals cluster within networks and how these clusters contribute to polarization. The methodological emphasis on interaction patterns within platforms highlights structural aspects of echo chambers, such as clustering and information reinforcement. However, these approaches may overlook broader user experiences, leaving gaps in understanding how individuals engage with the wider media environment.

The relationship between methodology and outcomes is consistent. CSS studies often support the existence of echo chambers, while surveys and experiments, with their emphasis on content exposure and user perceptions, are more likely to report mixed results. This reflects how methodological choices shape the specific aspects of echo chambers that are studied. Furthermore, conceptualization and methodology are closely aligned. Studies that focus on network-based homophily tend to use CSS methods, while those examining user-level exposure typically rely on surveys or experiments. These interconnections demonstrate how conceptual frameworks and methodological approaches influence the outcomes of echo chamber research.

\subsubsection{Measurement Factors and Challenges}

The diverse measurements used in echo chamber research reveal different approaches' strengths and limitations. Each conceptualization and operationalization offers advantages for measurement while also presenting specific challenges. This section explores these factors, focusing on their implications for measuring and understanding echo chambers. 

\paragraph{Computational Social Science}

CSS methods allow researchers to unobtrusively observe behavioral choices in natural social media environments. This approach minimizes social desirability bias and measurement error \cite{Barbera.2015} and is relatively accessible, particularly for European researchers under the upcoming Digital Services Act (DSA). However, API restrictions have made access more challenging in recent years.

CSS methods are particularly effective at analyzing large-scale networks and identifying patterns of homophily, polarization, and content exposure. For example, researchers can use network metrics like degree centrality and betweenness centrality to analyze the structural components of social networks. Techniques like latent space models enable the estimation of ideological positions from behavioral data, although these approaches are not without limitations. Latent models can be biased by parameter choices and assumptions and may fall into circular reasoning, such as inferring political positions from network structures and then using those positions to explain the same structures \cite{barbera2013}. While content-based methods offer higher accuracy \cite{DelVicario.2016}, they require substantial labeled training data, which is often unavailable or costly to generate. These methods are also less effective for novel situations, such as the Russia-Ukraine war, where labeled data is scarce \cite{Kratzke}.

Bots pose another significant challenge for CSS studies. While \citet{Gallwitz.2021} argue that the role of social bots is overstated, other studies suggest that bots disproportionately benefit certain political groups, such as right-wing parties \cite{neudert2017junk}. For instance, when Twitter/X deleted millions of bots in 2017, radical right European MPs experienced the greatest loss of followers \cite{silva_proksch_2021}. Machine learning-based bot detection, as proposed by \citet{rusche2022}, and implemented by \citet{choi2020rumor} and \citet{Gaumont.2018}, offers a way to mitigate this issue.

Causal inference remains one of the most significant challenges in CSS. Observational data can identify associations but cannot establish causality. For example, polarization on social media may reflect broader societal polarization rather than causing it \cite{Morgan.2014}. Longitudinal data can help address this issue by disentangling network influence from selection effects, and some studies have shown that longitudinal approaches can approximate results from randomized experiments \cite{causal}.

\paragraph{Surveys}

Survey methods provide insights into users’ broader media diets and individual-level behaviors, capturing information beyond the scope of social media platforms. These methods are particularly useful for studying exposure to cross-cutting content. Longitudinal surveys, in particular, can track changes over time and help distinguish network effects from selection biases.

However, surveys face limitations in measuring homophily. Self-reported data often captures perceived rather than actual homophily, which can lead to underestimation of polarization \cite{Boutyline.2017}. Survey studies also suffer from smaller sample sizes, measurement errors, and social desirability biases, where respondents may overreport ``positive'' behaviors, such as exposure to diverse viewpoints, and underreport politically charged or ``negative'' activities \cite{Terren.2021}. Additionally, survey respondents may struggle to recall specific instances of content exposure, further complicating data collection \cite{Dubois.2018}.

\paragraph{Experiments}

Experiments provide a robust framework for identifying causality by allowing researchers to manipulate network structures or content exposure. For example, randomized experiments can isolate the effects of polarization and information flow on individual behavior \cite{causalexp}. Crowdsourcing platforms and sock puppet studies offer tools for quasi-experimental research. These methods can be cost-effective and scalable, but ethical concerns and the complexity of real-world social networks limit their use in echo chamber studies. However, more recent studies as conducted by \citet{Nyhan2023} are positive examples of how experiments can significantly contribute to echo chamber research. They conducted a large-scale, randomized field experiment with estimated treatment effects to directly evaluate whether reducing exposure to like-minded content can causally impact polarization.

\paragraph{Common Measurement Challenges}

Several challenges cut across all methodological approaches. Selection bias is a persistent issue, particularly in CSS studies. Highly active and ideologically driven users are often overrepresented in trace data, which can skew findings \cite{rusche2022}. For instance, \citet{Aruguete.2023} show that ideological congruence and issue salience correlate with political engagement, leading users on both ends of the political spectrum to disproportionately share ideologically aligned content. This overrepresentation of ideologues may amplify observed polarization and limit the generalizability of findings.

Another common issue is the “no interference assumption” in causal inference frameworks, which assumes that one individual's treatment does not affect another's outcome \cite{Morgan.2014}. This assumption often fails in highly interconnected social networks, where individuals influence one another across various ties. Although recent advancements have sought to extend causal frameworks to account for interference, they remain underutilized. Cluster-based approaches offer a partial solution but are difficult to implement in highly connected networks like social media.

Lastly, there are inconsistencies in how echo chambers are operationalized and measured across studies. Dependent variables, such as exposure to opposing viewpoints or clustering within networks, vary widely. These inconsistencies complicate comparisons across methods and highlight the importance of aligning conceptual frameworks with empirical approaches.

\subsubsection{Geographical and Contextual Factors}
In the following, we briefly analyze other factors such as regional factors, political system factors, demographics, and platform characteristics that could influence research outcomes. We want to compare studies' demographics and political environments, as \citet{Kitchens.2020} claimed that most research that confirms the echo chamber hypothesis has been conducted in the United States.

Table \ref{tab:countries} presents the distribution of countries in the corpus. This summarization includes studies with participants from these countries or CSS studies that used trace data from these populations. The table demonstrates that the majority of studies, namely 54, were conducted in the US. The other most studied countries are all European countries. Only four studies were conducted in China, two studies in Japan, and another two in Brazil. There is a lack of research, particularly about countries outside the Global North. This may be due to non-English publications, limited funding, or limited data access. This will probably only change with publicly accessible data and increased research funding.  

Concurrent with previous research, we find that most studies affirming the echo chamber hypothesis are conducted in the United States or use data concerning the United States population and are not representative of other regions. We find that more studies from the US and UK affirm the echo chamber hypothesis and less from countries such as Germany, France, and the Netherlands. 

This presents a tendency for echo chambers to be mostly demonstrated in countries or discourse spaces that are already highly fragmented. That is, in countries that have two-party systems and are considered divided or in polarized groups, such as conspiracy and science or vaccination and anti-vaccination. It is necessary to determine whether there is a reverse causality fallacy present in this context and whether fragmentation in society ultimately leads to the formation of echo chambers in social networks.

Figure \ref{fig:publication} illustrates the social media platforms analyzed in the reviewed studies, where specific platforms were identified. Importantly, this distribution does not reflect the actual usage patterns of social media platforms globally. Among the top 10 most popular platforms, only Facebook and YouTube were addressed in the included studies \cite{most}. Research on platforms such as Instagram, WhatsApp, Telegram, TikTok, and WeChat is notably absent, largely due to challenges in data accessibility. These platforms lack open APIs, introducing a significant data accessibility bias.

This gap is particularly concerning given the varying user demographics and interaction patterns across platforms. For instance, TikTok, which relies on algorithm-driven content delivery, features interactions that are fundamentally different from platforms like Twitter/X, where users can follow one another back. Such differences not only shape how information spreads but also influence user engagement and community dynamics. Neglecting these variations risks oversimplifying the complexity of online ecosystems.

Moreover, certain platforms like Telegram have emerged as critical spaces for misinformation, including content related to COVID-19 \cite{telegram} and disinformation about the Russian invasion of Ukraine \cite{STOKELWALKER20228}. Similarly, TikTok’s growing role in shaping public discourse and its distinct interaction model remain significantly underexplored.

Additionally, while instant messaging services like WhatsApp and Telegram are often excluded from the definition of social networks, their importance in information sharing is undeniable and growing. A significant portion of content sharing occurs through private channels such as instant messengers or email rather than public social networks \cite{madrigal2012dark}.

    \begin{table}[t]
    
    \caption{Countries that were covered in the corpus of the review}
    \centering
    \small
    \begin{tabular}{lrlrlrlr}
    
    \midrule
          US &            54 &       Italy &            10 &        UK &             8 &   Germany &           6 \\
      France &             4 & Netherlands &             4 &     China &             4 &        EU &           4 \\
      Poland &             3 &       Spain &             3 &     Japan &             2 &    Norway &           2\\
      Brasil &             2 &      Israel &             1 & Argentina &             1 & Palestine &           1 \\
      Russia &             1 &      Taiwan &             1 &   Hungary &             1 &   Finland &           1 \\
       South Korea &             1 &   Australia &             1 &     Swiss &             1 &       \\
    \bottomrule
    \end{tabular}
        
        \label{tab:countries}
    \end{table}   

\section{Discussion and Limitations}
Our systematic review demonstrates that differing conceptualizations and operationalizations of echo chambers significantly shape research outcomes. In the following, we outline research recommendations for echo chamber studies based on our findings and the limitations of our review.

Future research should focus on (1) proposing unified conceptual and operational frameworks to standardize the study of
echo chambers, (2) expanding research on group behavior and finding ways to integrate social and cultural context into measurements of echo chambers, (3) addressing gaps in causal evidence through longitudinal studies, experiments, or quasi-
experimental designs, (4) investigating cross-platform dynamics, including less-studied platforms like Instagram, TikTok, instant messengers, and decentralized networks, and (5) researching the relationship between content moderation and echo chambers, not solely focusing on the recommendation systems in the moderation process.

The field urgently needs standardized frameworks for both conceptualizations and operationalizations. The current diversity in how echo chambers are defined -- whether as homophily, selective exposure, or content exposure -- and the wide range of methodologies used make it challenging to synthesize evidence. This review contributes to a more nuanced understanding of the echo chamber construct by categorizing and systematically measuring these varying approaches. 

However, to achieve meaningful progress in understanding phenomena such as polarization, it is crucial to develop a unified perspective that integrates these differing views while distinguishing between structural and behavioral (individual or group) elements. To bridge this gap, researchers should adopt measurement modeling techniques to systematically link theoretical constructs with empirical data, ensuring coherence between what is being studied and how it is being measured.

While the corpus predominantly reflects these four established conceptualizations, some approaches remain underexplored. For instance, trust imbalances as the foundational structure of echo chambers \cite{nguyen_2020} have yet to be extensively operationalized in empirical studies. Similarly, \citet{Mahrt2019} argues for broader integration of social and cultural contexts into echo chamber research, emphasizing that fragmentation is often context-specific rather than universal. These perspectives call for re-conceptualizing echo chambers as embedded within wider societal dynamics rather than solely as digital phenomena. We recommend expanding operationalizations on group behavior and integrating context-specific measurement into research.

Researchers should establish transparency in reporting such measurement decisions and potential limitations. This will allow others to replicate findings and assess the robustness of conclusions, which should be done regularly. Validations of operationalization should accompany this. By addressing these issues and promoting interdisciplinary collaboration, the field can progress toward a standardized framework that captures the complex dynamics of echo chambers while accommodating their multifaceted nature.

The current limitations in echo chamber research design highlight the need for more comprehensive and interdisciplinary approaches. One issue is the focus on single platforms, which restricts the generalizability of findings and overlooks cross-platform dynamics. Future research should prioritize cross-platform studies that examine how users navigate diverse media ecosystems, integrating insights from platforms like Facebook, Twitter/X, Reddit, TikTok, and instant messengers. These studies could explore how platform-specific affordances, such as recommendation systems, influence echo chamber formation and how cross-cutting content from one platform may counteract clustering effects on another.

Another critical gap lies in capturing broader contextual factors. CSS methods are effective for identifying patterns and relationships within networks but often fail to account for the social, cultural, and political contexts that shape these networks. Combining CSS with surveys can address this gap, providing both behavioral data and self-reported insights into user motivations and experiences. For instance, surveys could explore how individuals perceive their exposure to diverse viewpoints, while CSS could validate these perceptions through trace data. Mixed-methods approaches, which combine trace data, surveys, and experiments, offer a promising pathway to uncovering the nuanced mechanisms driving echo chambers.

Randomized experiments, while challenging due to ethical and logistical constraints, are important methods for establishing causality. Innovative approaches, such as ethically designed simulations or controlled interventions (e.g., ``sock puppet'' accounts), should be explored to assess the impact of recommendation systems on polarization, misinformation, and selective exposure. Such experiments would help understand the interplay between homophily and selective exposure, which may be more critical than the overemphasized issue of content exposure or filter bubbles.

Causal inference remains a significant challenge in this field. While many studies identify associations, few establish causal relationships, such as whether echo chambers amplify polarization or whether pre-existing polarization fosters echo chambers. Future research should use longitudinal designs, instrumental variable approaches, and quasi-experimental methods to address this gap. 

Finally, addressing granularity requires designing studies that account for different levels of analysis, from individual users to platform-wide phenomena. Researchers should integrate multi-level data to examine how individual behaviors aggregate into group dynamics and how these, in turn, influence platform-wide trends. By explicitly linking findings across these levels, studies can provide a more holistic understanding of echo chambers and their societal impacts.

There is a lack of research about countries outside the US and outside the Global North. This may be due to the limits of literature searches in systematic reviews or the difficult data situation. This will probably only change with publicly accessible data. The DSA \cite{noauthor_regulation_2022} could lead to more studies focusing on EU countries, but similar regulatory efforts are needed to ensure broader applicability beyond the Global North.


In the following, we will briefly lay out the limitations of our work, although most limitations of systematic reviews have already been addressed in this study. For instance, ICR was discussed in Section \ref{sec:method}. Although the ICR falls within a reasonable range, there is room for improvement. The initial agreement between the first and third coders was moderate. While considered substantial by some standards, this highlights areas for enhancing the clarity and consistency of the coding process. This moderate agreement could stem from differences in the interpretation of the codebook or the inherent complexity of the coding task. To address this, disagreements were thoroughly discussed, and the codebook was refined during the process. These efforts improved consistency, resulting in a Cohen's $\kappa$ value of 0.66. Nonetheless, future studies could benefit from additional coder training or iterative testing of the codebook before formal coding begins.

It is worth noting that most systematic reviews do not measure or report ICR, nor do they provide transparency in how their coding processes are executed \cite{belur}. This study has aimed to counteract this common issue by employing measures such as publishing the codebook, detailing the derivation of code variables, and calculating ICR. These steps enhance the reproducibility and reliability of the findings.

The possible biases that can occur in systematic reviews are the already mentioned publication bias, time lag bias, multiple (duplicate) publication bias, location bias, citation bias, language bias, and the outcome reporting bias 
\cite{DRUCKER2016e109} in Section \ref{sec:pubbias} and Section \ref{sec:bias}, these biases and their potential to occur in this work were discussed.



\section{Policy Recommendations}
Based on our findings, the following section will explore implications and recommendations for policies.

The findings highlight the necessity of platform-level interventions that empower users with greater customization options and control over recommendations. Adjustable recommender systems like those tested on Reddit allow users to customize content diversity, reducing algorithmic reinforcement of homogeneous content and introducing contrasting perspectives. Regulatory policies should mandate their implementation to ensure users can meaningfully adjust their content exposure.

One major criticism of social media studies is their limited attention to algorithmic filtering changes. Although a few studies have conducted continuous audits to examine how algorithms evolve (e.g., \citet{Tomlein.2021, srba2023}), most research fails to address these dynamic shifts adequately. The inaccessibility of platform algorithms and their inherent complexity -- often requiring live observation to understand their real-world effects -- presents significant challenges.  

This reinforces the need for continuous algorithmic auditing, as the EU DSA emphasizes, which grants vetted researchers access to platform data and enables recommendation system audits \cite{hartmann2024addressing}. Institutionalizing live observation and ongoing audits in various regional and political contexts are necessary to create transparency and accountability and provide deeper insights into how recommender systems reinforce or mitigate echo chambers in these contexts. 

Policies should enable standardized researcher data access across platforms to create the possibility for cross-platform research. Integrating user data from multiple platforms, combined with survey, trace, and experimental data, would offer a more comprehensive understanding of how algorithmic filtering influences exposure to content across the digital ecosystem. Policies and standardization of social media data and data access should enable this. Whether the DSA will shape such a standardized data ecosystem for vetted researchers remains to be seen. 

Our findings demonstrated that we not only need more European research on echo chambers but also more data from outside the Global North. Thus, policies such as the DSA Art. 40 and systematic audits should be enabled outside the Global North to understand better how regional and political contexts shape the echo chamber phenomenon and its dynamics with polarization, misinformation, and extremism.

\section{Conclusion}
\label{sec:discussion}
We conducted a systematic review of 129 studies investigating the antecedents, characteristics, and effects of echo chambers on social media. Our analysis focused on how echo chambers are conceptualized and operationalized, using a measurement modeling perspective. The review demonstrated that varying conceptualizations and operationalizations often lead to divergent findings.

Specifically, studies that conceptualized echo chambers through the lens of homophily and employed data-driven CSS methods tended to support the echo chamber hypothesis and its link to polarization effects on social media. In contrast, studies utilizing content exposure analyses and surveys, which examined the broader spectrum of media exposure, often refuted the echo chamber hypothesis.

Notably, most of these studies were conducted in the context of the United States, highlighting a critical gap in understanding echo chambers within non-two-party political systems. To advance this field, future research should prioritize cross-platform analyses, incorporate continuous audits of algorithmic filtering, and examine the causal dynamics between polarization, fragmentation, and the formation of online echo chambers.

Additionally, research should strive for greater transparency in conceptualizing and operationalizing echo chambers. A more nuanced and granular approach to defining echo chambers is essential for advancing our understanding of this complex phenomenon.

\section*{Data Availability Statement}
The data that was produced during screening and analysis and synthesis of coding can be found under the following link\footnote{Anonymized for peer review.}: \url{https://anonymous.4open.science/r/sysreview_ec-E925/}.
\section*{Acknowledgements}
Funded by the German Federal Ministry of Education and Research (BMBF)—Nr. 16DII113f. Special thanks to PD Dr. Merja Mahrt for her valuable feedback, which was above and beyond the usual. We also sincerely thank the anonymous reviewers for their constructive feedback, which has significantly improved this paper and extend our gratitude to Dr. Milagros Miceli, and the entire research group \textit{Data, Algorithmic Systems, and Ethics} at the Weizenbaum Institute.

\section*{Conflict of Interest}
The authors declare that there is no conflict of interest regarding the publication of this paper.

\bibliography{ec-bib}


\begin{thebibliography}{188}
\ifx \bisbn   \undefined \def \bisbn  #1{ISBN #1}\fi
\ifx \binits  \undefined \def \binits#1{#1}\fi
\ifx \bauthor  \undefined \def \bauthor#1{#1}\fi
\ifx \batitle  \undefined \def \batitle#1{#1}\fi
\ifx \bjtitle  \undefined \def \bjtitle#1{#1}\fi
\ifx \bvolume  \undefined \def \bvolume#1{\textbf{#1}}\fi
\ifx \byear  \undefined \def \byear#1{#1}\fi
\ifx \bissue  \undefined \def \bissue#1{#1}\fi
\ifx \bfpage  \undefined \def \bfpage#1{#1}\fi
\ifx \blpage  \undefined \def \blpage #1{#1}\fi
\ifx \burl  \undefined \def \burl#1{\textsf{#1}}\fi
\ifx \doiurl  \undefined \def \doiurl#1{\url{https://doi.org/#1}}\fi
\ifx \betal  \undefined \def \betal{\textit{et al.}}\fi
\ifx \binstitute  \undefined \def \binstitute#1{#1}\fi
\ifx \binstitutionaled  \undefined \def \binstitutionaled#1{#1}\fi
\ifx \bctitle  \undefined \def \bctitle#1{#1}\fi
\ifx \beditor  \undefined \def \beditor#1{#1}\fi
\ifx \bpublisher  \undefined \def \bpublisher#1{#1}\fi
\ifx \bbtitle  \undefined \def \bbtitle#1{#1}\fi
\ifx \bedition  \undefined \def \bedition#1{#1}\fi
\ifx \bseriesno  \undefined \def \bseriesno#1{#1}\fi
\ifx \blocation  \undefined \def \blocation#1{#1}\fi
\ifx \bsertitle  \undefined \def \bsertitle#1{#1}\fi
\ifx \bsnm \undefined \def \bsnm#1{#1}\fi
\ifx \bsuffix \undefined \def \bsuffix#1{#1}\fi
\ifx \bparticle \undefined \def \bparticle#1{#1}\fi
\ifx \barticle \undefined \def \barticle#1{#1}\fi
\bibcommenthead
\ifx \bconfdate \undefined \def \bconfdate #1{#1}\fi
\ifx \botherref \undefined \def \botherref #1{#1}\fi
\ifx \url \undefined \def \url#1{\textsf{#1}}\fi
\ifx \bchapter \undefined \def \bchapter#1{#1}\fi
\ifx \bbook \undefined \def \bbook#1{#1}\fi
\ifx \bcomment \undefined \def \bcomment#1{#1}\fi
\ifx \oauthor \undefined \def \oauthor#1{#1}\fi
\ifx \citeauthoryear \undefined \def \citeauthoryear#1{#1}\fi
\ifx \endbibitem  \undefined \def \endbibitem {}\fi
\ifx \bconflocation  \undefined \def \bconflocation#1{#1}\fi
\ifx \arxivurl  \undefined \def \arxivurl#1{\textsf{#1}}\fi
\csname PreBibitemsHook\endcsname

\bibitem[\protect\citeauthoryear{{European Commission}}{2022}]{noauthor_regulation_2022}
\begin{botherref}
\oauthor{\bsnm{{European Commission}}}:
Regulation ({EU}) 2022/2065 of the {European} {Parliament} and of the {Council} of 19 {October} 2022 on a {Single} {Market} {For} {Digital} {Services} and amending {Directive} 2000/31/{EC} ({Digital} {Services} {Act}) ({Text} with {EEA} relevance).
Legislative Body: EP, CONSIL
(2022).
\url{http://data.europa.eu/eli/reg/2022/2065/oj/eng}
Accessed 2024-06-19
\end{botherref}
\endbibitem

\bibitem[\protect\citeauthoryear{Bail et~al.}{2018}]{Bail.2018}
\begin{barticle}
\bauthor{\bsnm{Bail}, \binits{C.A.}},
\bauthor{\bsnm{Argyle}, \binits{L.P.}},
\bauthor{\bsnm{Brown}, \binits{T.W.}},
\bauthor{\bsnm{Bumpus}, \binits{J.P.}},
\bauthor{\bsnm{Chen}, \binits{H.}},
\bauthor{\bsnm{Hunzaker}, \binits{M.B.F.}},
\bauthor{\bsnm{Lee}, \binits{J.}},
\bauthor{\bsnm{Mann}, \binits{M.}},
\bauthor{\bsnm{Merhout}, \binits{F.}},
\bauthor{\bsnm{Volfovsky}, \binits{A.}}:
\batitle{Exposure to opposing views on social media can increase political polarization}.
\bjtitle{Proceedings of the National Academy of Sciences of the United States of America}
\bvolume{115}(\bissue{37}),
\bfpage{9216}--\blpage{9221}
(\byear{2018})
\end{barticle}
\endbibitem

\bibitem[\protect\citeauthoryear{Quattrociocchi}{2017}]{quattrociocchi_social_2017}
\begin{bchapter}
\bauthor{\bsnm{Quattrociocchi}, \binits{W.}}:
\bctitle{Social and political challenges: Western democracy in crisis?}
In: \bbtitle{Global Risks Report 2017}
(\byear{2017})
\end{bchapter}
\endbibitem

\bibitem[\protect\citeauthoryear{Lorenz-Spreen et~al.}{2023}]{LorenzSpreen.2023}
\begin{barticle}
\bauthor{\bsnm{Lorenz-Spreen}, \binits{P.}},
\bauthor{\bsnm{Oswald}, \binits{L.}},
\bauthor{\bsnm{Lewandowsky}, \binits{S.}},
\bauthor{\bsnm{Hertwig}, \binits{R.}}:
\batitle{A systematic review of worldwide causal and correlational evidence on digital media and democracy}.
\bjtitle{Nature Human Behaviour}
\bvolume{7}(\bissue{1}),
\bfpage{74}--\blpage{101}
(\byear{2023})
\doiurl{10.1038/s41562-022-01460-1}
\end{barticle}
\endbibitem

\bibitem[\protect\citeauthoryear{Minh~Pham et~al.}{2020}]{minh_pham_effect_2020}
\begin{botherref}
\oauthor{\bsnm{Minh~Pham}, \binits{T.}},
\oauthor{\bsnm{Kondor}, \binits{I.}},
\oauthor{\bsnm{Hanel}, \binits{R.}},
\oauthor{\bsnm{Thurner}, \binits{S.}}:
The effect of social balance on social fragmentation
\textbf{17}(172),
7--52
(2020)
\end{botherref}
\endbibitem

\bibitem[\protect\citeauthoryear{Torregrosa et~al.}{2020}]{Torregrosa.2020}
\begin{botherref}
\oauthor{\bsnm{Torregrosa}, \binits{J.}},
\oauthor{\bsnm{Panizo-Lledot}, \binits{A.}},
\oauthor{\bsnm{Bello-Orgaz}, \binits{G.}},
\oauthor{\bsnm{Camacho}, \binits{D.}}:
Analyzing the relationship between relevance and extremist discourse in an alt-right network on twitter.
Social Network Analysis and Mining
\textbf{10}(1)
(2020)
\end{botherref}
\endbibitem

\bibitem[\protect\citeauthoryear{Sunstein}{2001}]{sunstein2001republic}
\begin{bbook}
\bauthor{\bsnm{Sunstein}, \binits{C.}}:
\bbtitle{Republic.com}.
\bpublisher{Princeton University Press},
\blocation{Princeton, N.J}
(\byear{2001})
\end{bbook}
\endbibitem

\bibitem[\protect\citeauthoryear{Cinelli et~al.}{2020}]{Cinelli.2020}
\begin{botherref}
\oauthor{\bsnm{Cinelli}, \binits{M.}},
\oauthor{\bsnm{Brugnoli}, \binits{E.}},
\oauthor{\bsnm{Schmidt}, \binits{A.L.}},
\oauthor{\bsnm{Zollo}, \binits{F.}},
\oauthor{\bsnm{Quattrociocchi}, \binits{W.}},
\oauthor{\bsnm{Scala}, \binits{A.}}:
Selective exposure shapes the facebook news diet.
PLOS ONE
\textbf{15}(3)
(2020)
\end{botherref}
\endbibitem

\bibitem[\protect\citeauthoryear{Pariser}{2011}]{pariser_filter_2011}
\begin{bbook}
\bauthor{\bsnm{Pariser}, \binits{E.}}:
\bbtitle{The Filter Bubble: What the Internet Is Hiding from You}.
\bpublisher{The Penguin Group},
\blocation{London}
(\byear{2011})
\end{bbook}
\endbibitem

\bibitem[\protect\citeauthoryear{Sunstein}{2017}]{sunstein2017republic}
\begin{bbook}
\bauthor{\bsnm{Sunstein}, \binits{C.}}:
\bbtitle{Republic : Divided Democracy in the Age of Social Media}.
\bpublisher{Princeton University Press},
\blocation{Princeton}
(\byear{2017})
\end{bbook}
\endbibitem

\bibitem[\protect\citeauthoryear{Van~Aelst et~al.}{2017}]{van_aelst_political_2017}
\begin{botherref}
\oauthor{\bsnm{Van~Aelst}, \binits{P.}},
\oauthor{\bsnm{Strömbäck}, \binits{J.}},
\oauthor{\bsnm{Aalberg}, \binits{T.}},
\oauthor{\bsnm{Esser}, \binits{F.}},
\oauthor{\bsnm{De~Vreese}, \binits{C.}},
\oauthor{\bsnm{Matthes}, \binits{J.}},
\oauthor{\bsnm{Hopmann}, \binits{D.}},
\oauthor{\bsnm{Salgado}, \binits{S.}},
\oauthor{\bsnm{Hubé}, \binits{N.}},
\oauthor{\bsnm{Stępińska}, \binits{A.}}:
Political communication in a high-choice media environment: a challenge for democracy?
\textbf{41}(1),
3--27
(2017).
{ISBN}: 2380-8985 Publisher: Taylor \& Francis
\end{botherref}
\endbibitem

\bibitem[\protect\citeauthoryear{Dubois and Blank}{2018}]{Dubois.2018}
\begin{barticle}
\bauthor{\bsnm{Dubois}, \binits{E.}},
\bauthor{\bsnm{Blank}, \binits{G.}}:
\batitle{The echo chamber is overstated: the moderating effect of political interest and diverse media}.
\bjtitle{Information Communication {\&} Society}
\bvolume{21}(\bissue{5, SI}),
\bfpage{729}--\blpage{745}
(\byear{2018})
\doiurl{10.1080/1369118X.2018.1428656}
\end{barticle}
\endbibitem

\bibitem[\protect\citeauthoryear{Habermas}{2010}]{habermas}
\begin{bchapter}
\bauthor{\bsnm{Habermas}, \binits{J.}}:
\bctitle{Hat die demokratie noch eine epistemische dimension?}
In: \bbtitle{Kleine Politische Schriften 11. Ach Europa},
pp. \bfpage{138}--\blpage{191}.
\bpublisher{Suhrkamp Verlag AG},
\blocation{Frankfurt}
(\byear{2010})
\end{bchapter}
\endbibitem

\bibitem[\protect\citeauthoryear{Dahlgren}{2005}]{dahlgreen}
\begin{barticle}
\bauthor{\bsnm{Dahlgren}, \binits{P.}}:
\batitle{The internet, public spheres, and political communication: Dispersion and deliberation}.
\bjtitle{Political Communication}
\bvolume{22}(\bissue{2}),
\bfpage{147}--\blpage{162}
(\byear{2005})
\end{barticle}
\endbibitem

\bibitem[\protect\citeauthoryear{Terren and Borge}{2021}]{Terren.2021}
\begin{barticle}
\bauthor{\bsnm{Terren}, \binits{L.}},
\bauthor{\bsnm{Borge}, \binits{R.}}:
\batitle{Echo chambers on social media: A systematic review of the literature}.
\bjtitle{Review of Communication Research}
\bvolume{9},
\bfpage{99}--\blpage{118}
(\byear{2021})
\doiurl{10.12840/ISSN.2255-4165.028}
\end{barticle}
\endbibitem

\bibitem[\protect\citeauthoryear{Boulianne et~al.}{2020}]{Boulianne.2020}
\begin{barticle}
\bauthor{\bsnm{Boulianne}, \binits{S.}},
\bauthor{\bsnm{Koc-Michalska}, \binits{K.}},
\bauthor{\bsnm{Bimber}, \binits{B.}}:
\batitle{Right-wing populism, social media and echo chambers in western democracies}.
\bjtitle{New Media {\&} Society}
\bvolume{22}(\bissue{4}),
\bfpage{683}--\blpage{699}
(\byear{2020})
\doiurl{10.1177/1461444819893983}
\end{barticle}
\endbibitem

\bibitem[\protect\citeauthoryear{Cinelli et~al.}{2021}]{Cinelli.2021}
\begin{botherref}
\oauthor{\bsnm{Cinelli}, \binits{M.}},
\oauthor{\bsnm{Morales}, \binits{G.D.F.}},
\oauthor{\bsnm{Galeazzi}, \binits{A.}},
\oauthor{\bsnm{Quattrociocchi}, \binits{W.}},
\oauthor{\bsnm{Starnini}, \binits{M.}}:
The echo chamber effect on social media.
Proceedings of the National Academy of Sciences of the United States of America
\textbf{118}(9)
(2021)
\end{botherref}
\endbibitem

\bibitem[\protect\citeauthoryear{Benkler et~al.}{2018}]{benkler_network_2018}
\begin{bbook}
\bauthor{\bsnm{Benkler}, \binits{Y.}},
\bauthor{\bsnm{Faris}, \binits{R.}},
\bauthor{\bsnm{Roberts}, \binits{H.}}:
\bbtitle{Network Propaganda: Manipulation, Disinformation, and Radicalization in American Politics}.
\bpublisher{Oxford University Press},
\blocation{Oxford}
(\byear{2018})
\end{bbook}
\endbibitem

\bibitem[\protect\citeauthoryear{Jacobs and Wallach}{2021}]{jacobs}
\begin{bchapter}
\bauthor{\bsnm{Jacobs}, \binits{A.Z.}},
\bauthor{\bsnm{Wallach}, \binits{H.}}:
\bctitle{Measurement and fairness}.
In: \bbtitle{Proceedings of the 2021 ACM Conference on Fairness, Accountability, and Transparency}.
\bsertitle{FAccT '21},
pp. \bfpage{375}--\blpage{385}.
\bpublisher{Association for Computing Machinery},
\blocation{New York, NY, USA}
(\byear{2021}).
\burl{https://doi.org/10.1145/3442188.3445901}
\end{bchapter}
\endbibitem

\bibitem[\protect\citeauthoryear{Jackman}{2008}]{measurement}
\begin{bchapter}
\bauthor{\bsnm{Jackman}, \binits{S.}}:
\bctitle{Measurement}.
In: \bbtitle{The Oxford Handbook of Political Methodology}.
\bpublisher{Oxford University Press},
\blocation{Oxford}
(\byear{2008}).
\burl{https://doi.org/10.1093/oxfordhb/9780199286546.003.0006}
\end{bchapter}
\endbibitem

\bibitem[\protect\citeauthoryear{Arguedas et~al.}{2022}]{arguedas_echo_2022}
\begin{botherref}
\oauthor{\bsnm{Arguedas}, \binits{A.R.}},
\oauthor{\bsnm{Robertson}, \binits{C.}},
\oauthor{\bsnm{Fletcher}, \binits{R.}},
\oauthor{\bsnm{Nielsen}, \binits{R.K.}}:
Echo chambers, filter bubbles, and polarisation: a literature review.
Reuter Institute for the Study of Journalism
(2022).
Accessed 2022-03-10
\end{botherref}
\endbibitem

\bibitem[\protect\citeauthoryear{Mahmoudi et~al.}{2024}]{mahmoudi}
\begin{barticle}
\bauthor{\bsnm{Mahmoudi}, \binits{A.}},
\bauthor{\bsnm{Jemielniak}, \binits{D.}},
\bauthor{\bsnm{Ciechanowski}, \binits{L.}}:
\batitle{Echo chambers in online social networks: A systematic literature review}.
\bjtitle{IEEE Access}
\bvolume{12},
\bfpage{9594}--\blpage{9620}
(\byear{2024})
\doiurl{10.1109/ACCESS.2024.3353054}
\end{barticle}
\endbibitem

\bibitem[\protect\citeauthoryear{Michiels et~al.}{}]{michielsWhatAreFilter2022}
\begin{botherref}
\oauthor{\bsnm{Michiels}, \binits{L.}},
\oauthor{\bsnm{Leysen}, \binits{J.}},
\oauthor{\bsnm{Smets}, \binits{A.}},
\oauthor{\bsnm{Goethals}, \binits{B.}}:
What are filter bubbles really? a review of the conceptual and empirical work.
In: Adjunct Proceedings of the 30th ACM Conference on User Modeling, Adaptation and Personalization,
pp. 274--279.
ACM.
\doiurl{10.1145/3511047.3538028} .
\url{https://dl.acm.org/doi/10.1145/3511047.3538028}
Accessed 2025-01-29
\end{botherref}
\endbibitem

\bibitem[\protect\citeauthoryear{Kitchens et~al.}{2020}]{Kitchens.2020}
\begin{barticle}
\bauthor{\bsnm{Kitchens}, \binits{B.}},
\bauthor{\bsnm{Johnson}, \binits{S.L.}},
\bauthor{\bsnm{Gray}, \binits{P.}}:
\batitle{Understanding echo chambers and filter bubbles: The impact of social media on diversification and partisan shifts in news consumption}.
\bjtitle{MIS Quarterly}
\bvolume{44}(\bissue{4}),
\bfpage{1619}--\blpage{1649}
(\byear{2020})
\end{barticle}
\endbibitem

\bibitem[\protect\citeauthoryear{Interian et~al.}{}]{Interian}
\begin{botherref}
\oauthor{\bsnm{Interian}, \binits{R.}},
\oauthor{\bsnm{Marzo}, \binits{R.G.}},
\oauthor{\bsnm{Mendoza}, \binits{I.}},
\oauthor{\bsnm{Ribeiro}, \binits{C.C.}}:
Network polarization, filter bubbles, and echo chambers: an annotated review of measures and reduction methods.
INTERNATIONAL TRANSACTIONS IN OPERATIONAL RESEARCH
\doiurl{10.1111/itor.13224}
\end{botherref}
\endbibitem

\bibitem[\protect\citeauthoryear{Page et~al.}{2021}]{page_prisma_2021}
\begin{botherref}
\oauthor{\bsnm{Page}, \binits{M.J.}},
\oauthor{\bsnm{{McKenzie}}, \binits{J.E.}},
\oauthor{\bsnm{Bossuyt}, \binits{P.M.}},
\oauthor{\bsnm{Boutron}, \binits{I.}},
\oauthor{\bsnm{Hoffmann}, \binits{T.C.}},
\oauthor{\bsnm{Mulrow}, \binits{C.D.}},
\oauthor{\bsnm{Shamseer}, \binits{L.}},
\oauthor{\bsnm{Tetzlaff}, \binits{J.M.}},
\oauthor{\bsnm{Akl}, \binits{E.A.}},
\oauthor{\bsnm{Brennan}, \binits{S.E.}},
\oauthor{\bsnm{Chou}, \binits{R.}},
\oauthor{\bsnm{Glanville}, \binits{J.}},
\oauthor{\bsnm{Grimshaw}, \binits{J.M.}},
\oauthor{\bsnm{Hróbjartsson}, \binits{A.}},
\oauthor{\bsnm{Lalu}, \binits{M.M.}},
\oauthor{\bsnm{Li}, \binits{T.}},
\oauthor{\bsnm{Loder}, \binits{E.W.}},
\oauthor{\bsnm{Mayo-Wilson}, \binits{E.}},
\oauthor{\bsnm{{McDonald}}, \binits{S.}},
\oauthor{\bsnm{{McGuinness}}, \binits{L.A.}},
\oauthor{\bsnm{Stewart}, \binits{L.A.}},
\oauthor{\bsnm{Thomas}, \binits{J.}},
\oauthor{\bsnm{Tricco}, \binits{A.C.}},
\oauthor{\bsnm{Welch}, \binits{V.A.}},
\oauthor{\bsnm{Whiting}, \binits{P.}},
\oauthor{\bsnm{Moher}, \binits{D.}}:
The {PRISMA} 2020 statement: An updated guideline for reporting systematic reviews
\textbf{18}(3),
1003583
(2021).
Publisher: Public Library of Science.
Accessed 2022-01-07
\end{botherref}
\endbibitem

\bibitem[\protect\citeauthoryear{Gough}{2007}]{gough_weight_2007}
\begin{botherref}
\oauthor{\bsnm{Gough}, \binits{D.}}:
Weight of evidence: a framework for the appraisal of the quality and relevance of evidence
\textbf{22}(2),
213--228
(2007).
Accessed 2021-11-04
\end{botherref}
\endbibitem

\bibitem[\protect\citeauthoryear{Grant and Booth}{2009}]{grant_typology_2009}
\begin{botherref}
\oauthor{\bsnm{Grant}, \binits{M.J.}},
\oauthor{\bsnm{Booth}, \binits{A.}}:
A typology of reviews: an analysis of 14 review types and associated methodologies
\textbf{26}(2),
91--108
(2009)
\end{botherref}
\endbibitem

\bibitem[\protect\citeauthoryear{Booth et~al.}{2012}]{booth_systematic_2012}
\begin{bbook}
\bauthor{\bsnm{Booth}, \binits{A.}},
\bauthor{\bsnm{Papaioannou}, \binits{D.}},
\bauthor{\bsnm{Sutton}, \binits{A.}}:
\bbtitle{Systematic Approaches to a Successful Literature Review},
(\byear{2012})
\end{bbook}
\endbibitem

\bibitem[\protect\citeauthoryear{Biondi-Zoccai et~al.}{2011}]{biondi-zoccai_rough_2011}
\begin{botherref}
\oauthor{\bsnm{Biondi-Zoccai}, \binits{G.}},
\oauthor{\bsnm{Lotrionte}, \binits{M.}},
\oauthor{\bsnm{Landoni}, \binits{G.}},
\oauthor{\bsnm{Modena}, \binits{M.G.}}:
The rough guide to systematic reviews and meta-analyses
\textbf{3}(3),
161--173
(2011).
Accessed 2022-01-07
\end{botherref}
\endbibitem

\bibitem[\protect\citeauthoryear{{We Are Social} et~al.}{2024}]{wearesocial_2024}
\begin{botherref}
\oauthor{\bsnm{{We Are Social}}},
\oauthor{\bsnm{{DataReportal}}},
\oauthor{\bsnm{{Meltwater}}}:
Most Popular Social Networks Worldwide as of April 2024, Ranked by Number of Monthly Active Users (in millions) [Graph].
In Statista.
Retrieved April 25, 2024, from \url{https://www.statista.com/statistics/272014/global-social-networks-ranked-by-number-of-users/}
(2024)
\end{botherref}
\endbibitem

\bibitem[\protect\citeauthoryear{Silva and Neiva}{2016}]{silva}
\begin{botherref}
\oauthor{\bsnm{Silva}, \binits{R.}},
\oauthor{\bsnm{Neiva}, \binits{F.}}:
Systematic literature review in computer science - a practical guide
(2016)
\doiurl{10.13140/RG.2.2.35453.87524}
\end{botherref}
\endbibitem

\bibitem[\protect\citeauthoryear{Mahood et~al.}{2014}]{mahood2014searching}
\begin{barticle}
\bauthor{\bsnm{Mahood}, \binits{Q.}},
\bauthor{\bsnm{Van~Eerd}, \binits{D.}},
\bauthor{\bsnm{Irvin}, \binits{E.}}:
\batitle{Searching for grey literature for systematic reviews: challenges and benefits}.
\bjtitle{Research synthesis methods}
\bvolume{5}(\bissue{3}),
\bfpage{221}--\blpage{234}
(\byear{2014})
\end{barticle}
\endbibitem

\bibitem[\protect\citeauthoryear{Haddaway et~al.}{2015}]{haddaway_role_2015}
\begin{botherref}
\oauthor{\bsnm{Haddaway}, \binits{N.R.}},
\oauthor{\bsnm{Collins}, \binits{A.M.}},
\oauthor{\bsnm{Coughlin}, \binits{D.}},
\oauthor{\bsnm{Kirk}, \binits{S.}}:
The role of google scholar in evidence reviews and its applicability to grey literature searching
\textbf{10}(9),
0138237
(2015).
Accessed 2022-01-07
\end{botherref}
\endbibitem

\bibitem[\protect\citeauthoryear{Boeker et~al.}{2013}]{Boeker.2013}
\begin{barticle}
\bauthor{\bsnm{Boeker}, \binits{M.}},
\bauthor{\bsnm{Vach}, \binits{W.}},
\bauthor{\bsnm{Motschall}, \binits{E.}}:
\batitle{Google scholar as replacement for systematic literature searches: good relative recall and precision are not enough}.
\bjtitle{BMC medical research methodology}
\bvolume{13}(\bissue{1}),
\bfpage{131}
(\byear{2013})
\doiurl{10.1186/1471-2288-13-131}
\end{barticle}
\endbibitem

\bibitem[\protect\citeauthoryear{Dickersin}{1990}]{publication}
\begin{barticle}
\bauthor{\bsnm{Dickersin}, \binits{K.}}:
\batitle{The existence of publication bias and risk factors for its occurrence}.
\bjtitle{JAMA : the journal of the American Medical Association}
\bvolume{263},
\bfpage{1385}--\blpage{9}
(\byear{1990})
\doiurl{10.1001/jama.263.10.1385}
\end{barticle}
\endbibitem

\bibitem[\protect\citeauthoryear{Drucker et~al.}{2016}]{DRUCKER2016e109}
\begin{barticle}
\bauthor{\bsnm{Drucker}, \binits{A.M.}},
\bauthor{\bsnm{Fleming}, \binits{P.}},
\bauthor{\bsnm{Chan}, \binits{A.-W.}}:
\batitle{Research techniques made simple: Assessing risk of bias in systematic reviews}.
\bjtitle{Journal of Investigative Dermatology}
\bvolume{136}(\bissue{11}),
\bfpage{109}--\blpage{114}
(\byear{2016})
\doiurl{10.1016/j.jid.2016.08.021}
\end{barticle}
\endbibitem

\bibitem[\protect\citeauthoryear{Lombard et~al.}{2006}]{lombard_intercoder_2006}
\begin{barticle}
\bauthor{\bsnm{Lombard}, \binits{M.}},
\bauthor{\bsnm{Snyder-Duch}, \binits{J.}},
\bauthor{\bsnm{Bracken}, \binits{C.C.}}:
\batitle{Content analysis in mass communication: Assessment and reporting of intercoder reliability}.
\bjtitle{Human Communication Research}
\bvolume{28}(\bissue{4}),
\bfpage{587}--\blpage{604}
(\byear{2006})
\doiurl{10.1111/j.1468-2958.2002.tb00826.x}
{\href{https://arxiv.org/abs/https://academic.oup.com/hcr/article-pdf/28/4/587/22337849/jhumcom0587.pdf}{{https://academic.oup.com/hcr/article-pdf/28/4/587/22337849/jhumcom0587.pdf}}}
\end{barticle}
\endbibitem

\bibitem[\protect\citeauthoryear{O’Connor and Joffe}{2020}]{oconnor_jaffe_2020}
\begin{barticle}
\bauthor{\bsnm{O’Connor}, \binits{C.}},
\bauthor{\bsnm{Joffe}, \binits{H.}}:
\batitle{Intercoder reliability in qualitative research: Debates and practical guidelines}.
\bjtitle{International Journal of Qualitative Methods}
\bvolume{19},
\bfpage{1609406919899220}
(\byear{2020})
\doiurl{10.1177/1609406919899220}
{\href{https://arxiv.org/abs/https://doi.org/10.1177/1609406919899220}{{https://doi.org/10.1177/1609406919899220}}}
\end{barticle}
\endbibitem

\bibitem[\protect\citeauthoryear{Cohen}{1960}]{cohen1960coefficient}
\begin{barticle}
\bauthor{\bsnm{Cohen}, \binits{J.}}:
\batitle{A coefficient of agreement for nominal scales}.
\bjtitle{Educational and psychological measurement}
\bvolume{20}(\bissue{1}),
\bfpage{37}--\blpage{46}
(\byear{1960})
\end{barticle}
\endbibitem

\bibitem[\protect\citeauthoryear{Elo and Kyng{\"a}s}{2008}]{elo2008qualitative}
\begin{barticle}
\bauthor{\bsnm{Elo}, \binits{S.}},
\bauthor{\bsnm{Kyng{\"a}s}, \binits{H.}}:
\batitle{The qualitative content analysis process}.
\bjtitle{Journal of advanced nursing}
\bvolume{62}(\bissue{1}),
\bfpage{107}--\blpage{115}
(\byear{2008})
\end{barticle}
\endbibitem

\bibitem[\protect\citeauthoryear{Hsieh and Shannon}{2005}]{hsieh2005three}
\begin{barticle}
\bauthor{\bsnm{Hsieh}, \binits{H.-F.}},
\bauthor{\bsnm{Shannon}, \binits{S.E.}}:
\batitle{Three approaches to qualitative content analysis}.
\bjtitle{Qualitative health research}
\bvolume{15}(\bissue{9}),
\bfpage{1277}--\blpage{1288}
(\byear{2005})
\end{barticle}
\endbibitem

\bibitem[\protect\citeauthoryear{Olteanu et~al.}{2019}]{olteanu2019social}
\begin{barticle}
\bauthor{\bsnm{Olteanu}, \binits{A.}},
\bauthor{\bsnm{Castillo}, \binits{C.}},
\bauthor{\bsnm{Diaz}, \binits{F.}},
\bauthor{\bsnm{K{\i}c{\i}man}, \binits{E.}}:
\batitle{Social data: Biases, methodological pitfalls, and ethical boundaries}.
\bjtitle{Frontiers in big data}
\bvolume{2},
\bfpage{13}
(\byear{2019})
\end{barticle}
\endbibitem

\bibitem[\protect\citeauthoryear{Siddaway et~al.}{2019}]{siddaway2019systematic}
\begin{barticle}
\bauthor{\bsnm{Siddaway}, \binits{A.P.}},
\bauthor{\bsnm{Wood}, \binits{A.M.}},
\bauthor{\bsnm{Hedges}, \binits{L.V.}}:
\batitle{How to do a systematic review: a best practice guide for conducting and reporting narrative reviews, meta-analyses, and meta-syntheses}.
\bjtitle{Annual review of psychology}
\bvolume{70}(\bissue{1}),
\bfpage{747}--\blpage{770}
(\byear{2019})
\end{barticle}
\endbibitem

\bibitem[\protect\citeauthoryear{Popay et~al.}{2006}]{popay2006guidance}
\begin{barticle}
\bauthor{\bsnm{Popay}, \binits{J.}},
\bauthor{\bsnm{Roberts}, \binits{H.}},
\bauthor{\bsnm{Sowden}, \binits{A.}},
\bauthor{\bsnm{Petticrew}, \binits{M.}},
\bauthor{\bsnm{Arai}, \binits{L.}},
\bauthor{\bsnm{Rodgers}, \binits{M.}},
\bauthor{\bsnm{Britten}, \binits{N.}},
\bauthor{\bsnm{Roen}, \binits{K.}},
\bauthor{\bsnm{Duffy}, \binits{S.}}, \betal:
\batitle{Guidance on the conduct of narrative synthesis in systematic reviews}.
\bjtitle{A product from the ESRC methods programme Version}
\bvolume{1}(\bissue{1}),
\bfpage{92}
(\byear{2006})
\end{barticle}
\endbibitem

\bibitem[\protect\citeauthoryear{{We Are Social, Hootsuite, DataReportal.}}{2024}]{most}
\begin{botherref}
\oauthor{\bsnm{{We Are Social, Hootsuite, DataReportal.}}}:
Most popular social networks worldwide as of {November} 2024, ranked by number of monthly active users (in millions).
publisher: Statista
(2024).
\url{https://www.statista.com/statistics/272014/global-social-networks-ranked-by-number-of-users/}
Accessed 2022-04-06
\end{botherref}
\endbibitem

\bibitem[\protect\citeauthoryear{Bessi et~al.}{2016}]{Bessi.2016}
\begin{botherref}
\oauthor{\bsnm{Bessi}, \binits{A.}},
\oauthor{\bsnm{Zollo}, \binits{F.}},
\oauthor{\bsnm{{Del Vicario}}, \binits{M.}},
\oauthor{\bsnm{Puliga}, \binits{M.}},
\oauthor{\bsnm{Scala}, \binits{A.}},
\oauthor{\bsnm{Caldarelli}, \binits{G.}},
\oauthor{\bsnm{Uzzi}, \binits{B.}},
\oauthor{\bsnm{Quattrociocchi}, \binits{W.}}:
Users polarization on facebook and youtube.
PLOS ONE
\textbf{11}(8)
(2016)
\end{botherref}
\endbibitem

\bibitem[\protect\citeauthoryear{{Del Vicario} et~al.}{2016}]{DelVicario.2016}
\begin{barticle}
\bauthor{\bsnm{{Del Vicario}}, \binits{M.}},
\bauthor{\bsnm{Bessi}, \binits{A.}},
\bauthor{\bsnm{Zollo}, \binits{F.}},
\bauthor{\bsnm{Petroni}, \binits{F.}},
\bauthor{\bsnm{Scala}, \binits{A.}},
\bauthor{\bsnm{Caldarelli}, \binits{G.}},
\bauthor{\bsnm{Stanley}, \binits{H.E.}},
\bauthor{\bsnm{Quattrociocchi}, \binits{W.}}:
\batitle{The spreading of misinformation online}.
\bjtitle{Proceedings of the National Academy of Sciences of the United States of America}
\bvolume{113}(\bissue{3}),
\bfpage{554}--\blpage{559}
(\byear{2016})
\end{barticle}
\endbibitem

\bibitem[\protect\citeauthoryear{Schmidt et~al.}{2018}]{Schmidt.2018}
\begin{barticle}
\bauthor{\bsnm{Schmidt}, \binits{A.L.}},
\bauthor{\bsnm{Zollo}, \binits{F.}},
\bauthor{\bsnm{Scala}, \binits{A.}},
\bauthor{\bsnm{Betsch}, \binits{C.}},
\bauthor{\bsnm{Quattrociocchi}, \binits{W.}}:
\batitle{Polarization of the vaccination debate on facebook}.
\bjtitle{Vaccine}
\bvolume{36}(\bissue{25}),
\bfpage{3606}--\blpage{3612}
(\byear{2018})
\doiurl{10.1016/j.vaccine.2018.05.040}
\end{barticle}
\endbibitem

\bibitem[\protect\citeauthoryear{Zollo et~al.}{2017}]{Zollo.2017}
\begin{botherref}
\oauthor{\bsnm{Zollo}, \binits{F.}},
\oauthor{\bsnm{Bessi}, \binits{A.}},
\oauthor{\bsnm{{Del Vicario}}, \binits{M.}},
\oauthor{\bsnm{Scala}, \binits{A.}},
\oauthor{\bsnm{Caldarelli}, \binits{G.}},
\oauthor{\bsnm{Shekhtman}, \binits{L.}},
\oauthor{\bsnm{Havlin}, \binits{S.}},
\oauthor{\bsnm{Quattrociocchi}, \binits{W.}}:
Debunking in a world of tribes.
PLOS ONE
\textbf{12}(7)
(2017)
\end{botherref}
\endbibitem

\bibitem[\protect\citeauthoryear{Batorski and Grzywinska}{2018}]{Batorski.2018}
\begin{barticle}
\bauthor{\bsnm{Batorski}, \binits{D.}},
\bauthor{\bsnm{Grzywinska}, \binits{I.}}:
\batitle{Three dimensions of the public sphere on facebook}.
\bjtitle{Information, Communication {\&} Society}
\bvolume{21}(\bissue{3}),
\bfpage{356}--\blpage{374}
(\byear{2018})
\doiurl{10.1080/1369118X.2017.1281329}
\end{barticle}
\endbibitem

\bibitem[\protect\citeauthoryear{Cota et~al.}{2019}]{Cota.2019}
\begin{botherref}
\oauthor{\bsnm{Cota}, \binits{W.}},
\oauthor{\bsnm{Ferreira}, \binits{S.C.}},
\oauthor{\bsnm{Pastor-Satorras}, \binits{R.}},
\oauthor{\bsnm{Starnini}, \binits{M.}}:
Quantifying echo chamber effects in information spreading over political communication networks.
EPJ Data Science
\textbf{8}(1)
(2019)
\end{botherref}
\endbibitem

\bibitem[\protect\citeauthoryear{Guo et~al.}{2020}]{Guo.2020}
\begin{barticle}
\bauthor{\bsnm{Guo}, \binits{L.}},
\bauthor{\bsnm{Rohde}, \binits{J.A.}},
\bauthor{\bsnm{Wu}, \binits{H.D.}}:
\batitle{Who is responsible for twitter's echo chamber problem? evidence from 2016 us election networks}.
\bjtitle{Information, Communication {\&} Society}
\bvolume{23}(\bissue{2}),
\bfpage{234}--\blpage{251}
(\byear{2020})
\doiurl{10.1080/1369118X.2018.1499793}
\end{barticle}
\endbibitem

\bibitem[\protect\citeauthoryear{Nikolov et~al.}{2019}]{Nikolov.2019}
\begin{barticle}
\bauthor{\bsnm{Nikolov}, \binits{D.}},
\bauthor{\bsnm{Lalmas}, \binits{M.}},
\bauthor{\bsnm{Flammini}, \binits{A.}},
\bauthor{\bsnm{Menczer}, \binits{F.}}:
\batitle{Quantifying biases in online information exposure}.
\bjtitle{Journal of the Association for Information Science and Technology}
\bvolume{70}(\bissue{3}),
\bfpage{218}--\blpage{229}
(\byear{2019})
\doiurl{10.1002/asi.24121}
\end{barticle}
\endbibitem

\bibitem[\protect\citeauthoryear{Masip et~al.}{2020}]{Masip.2020}
\begin{barticle}
\bauthor{\bsnm{Masip}, \binits{P.}},
\bauthor{\bsnm{Suau}, \binits{J.}},
\bauthor{\bsnm{Ruiz-Caballero}, \binits{C.}}:
\batitle{Incidental exposure to non-like-minded news through social media: Opposing voices in echo-chambers' news feeds}.
\bjtitle{Media and Communication}
\bvolume{8}(\bissue{4}),
\bfpage{53}--\blpage{62}
(\byear{2020})
\end{barticle}
\endbibitem

\bibitem[\protect\citeauthoryear{Guarino et~al.}{2020}]{Guarino.2020}
\begin{botherref}
\oauthor{\bsnm{Guarino}, \binits{S.}},
\oauthor{\bsnm{Trino}, \binits{N.}},
\oauthor{\bsnm{Celestini}, \binits{A.}},
\oauthor{\bsnm{Chessa}, \binits{A.}},
\oauthor{\bsnm{Riotta}, \binits{G.}}:
Characterizing networks of propaganda on twitter: a case study.
Applied Network Science
\textbf{5}(1)
(2020)
\end{botherref}
\endbibitem

\bibitem[\protect\citeauthoryear{{Del Valle} and {Borge Bravo}}{2018}]{DelValle.2018}
\begin{barticle}
\bauthor{\bsnm{{Del Valle}}, \binits{M.E.}},
\bauthor{\bsnm{{Borge Bravo}}, \binits{R.}}:
\batitle{Echo chambers in parliamentary twitter networks: The catalan case}.
\bjtitle{International Journal of Communication}
\bvolume{12},
\bfpage{1715}--\blpage{1735}
(\byear{2018})
\end{barticle}
\endbibitem

\bibitem[\protect\citeauthoryear{Matuszewski}{2019}]{Matuszewski.2019b}
\begin{botherref}
\oauthor{\bsnm{Matuszewski}, \binits{P.}}:
Selective exposure on polish political and news media facebook pages.
Polish Sociological Review
(206),
177--197
(2019)
\doiurl{10.26412/psr206.04}
\end{botherref}
\endbibitem

\bibitem[\protect\citeauthoryear{Rafail and Freitas}{2019}]{Rafail.2019}
\begin{botherref}
\oauthor{\bsnm{Rafail}, \binits{P.}},
\oauthor{\bsnm{Freitas}, \binits{I.}}:
Grievance articulation and community reactions in the men's rights movement online.
Social Media + Society
\textbf{5}(2)
(2019)
\doiurl{10.1177/2056305119841387}
\end{botherref}
\endbibitem

\bibitem[\protect\citeauthoryear{Whittaker et~al.}{2021}]{Whittaker.2021}
\begin{botherref}
\oauthor{\bsnm{Whittaker}, \binits{J.}},
\oauthor{\bsnm{Looney}, \binits{S.}},
\oauthor{\bsnm{Reed}, \binits{A.}},
\oauthor{\bsnm{Votta}, \binits{F.}}:
Recommender systems and the amplification of extremist content.
Internet Policy Review
\textbf{10}(3)
(2021)
\end{botherref}
\endbibitem

\bibitem[\protect\citeauthoryear{Monti et~al.}{2023}]{monti2023}
\begin{bchapter}
\bauthor{\bsnm{Monti}, \binits{C.}},
\bauthor{\bsnm{D’Ignazi}, \binits{J.}},
\bauthor{\bsnm{Starnini}, \binits{M.}},
\bauthor{\bsnm{Morales}, \binits{G.D.F.}}:
\bctitle{Evidence of demographic rather than ideological segregation in news discussion on reddit}.
In: \bbtitle{Proceedings of the ACM Web Conference 2023 (WWW '23)},
pp. \bfpage{1}--\blpage{10}.
\bpublisher{ACM},
\blocation{New York, NY, USA}
(\byear{2023}).
\doiurl{10.1145/3543507.3583468} .
\burl{https://doi.org/10.1145/3543507.3583468}
\end{bchapter}
\endbibitem

\bibitem[\protect\citeauthoryear{Treen et~al.}{2022}]{treen2022}
\begin{barticle}
\bauthor{\bsnm{Treen}, \binits{K.}},
\bauthor{\bsnm{Williams}, \binits{H.}},
\bauthor{\bsnm{O’Neill}, \binits{S.}},
\bauthor{\bsnm{Coan}, \binits{T.G.}}:
\batitle{Discussion of climate change on reddit: Polarized discourse or deliberative debate?}
\bjtitle{Environmental Communication}
\bvolume{16}(\bissue{5}),
\bfpage{680}--\blpage{698}
(\byear{2022})
{\href{https://arxiv.org/abs/https://doi.org/10.1080/17524032.2022.2050776}{{https://doi.org/10.1080/17524032.2022.2050776}}}
\end{barticle}
\endbibitem

\bibitem[\protect\citeauthoryear{Morales et~al.}{2021}]{Morales.2021}
\begin{botherref}
\oauthor{\bsnm{Morales}, \binits{G.D.F.}},
\oauthor{\bsnm{Monti}, \binits{C.}},
\oauthor{\bsnm{Starnini}, \binits{M.}}:
No echo in the chambers of political interactions on reddit.
Scientific Reports
\textbf{11}(1)
(2021)
\end{botherref}
\endbibitem

\bibitem[\protect\citeauthoryear{Gao et~al.}{2023}]{Gao2023}
\begin{barticle}
\bauthor{\bsnm{Gao}, \binits{Y.}},
\bauthor{\bsnm{Liu}, \binits{F.}},
\bauthor{\bsnm{Gao}, \binits{L.}}:
\batitle{Echo chamber effects on short video platforms}.
\bjtitle{Scientific Reports}
\bvolume{13},
\bfpage{6282}
(\byear{2023})
\doiurl{10.1038/s41598-023-33370-1}
\end{barticle}
\endbibitem

\bibitem[\protect\citeauthoryear{Dubois et~al.}{2020}]{Dubois.2020}
\begin{botherref}
\oauthor{\bsnm{Dubois}, \binits{E.}},
\oauthor{\bsnm{Minaeian}, \binits{S.}},
\oauthor{\bsnm{Paquet-Labelle}, \binits{A.}},
\oauthor{\bsnm{Beaudry}, \binits{S.}}:
Who to trust on social media: How opinion leaders and seekers avoid disinformation and echo chambers.
Social Media + Society
\textbf{6}(2)
(2020)
\doiurl{10.1177/2056305120913993}
\end{botherref}
\endbibitem

\bibitem[\protect\citeauthoryear{Schmidt et~al.}{2020}]{Schmidt.2020}
\begin{botherref}
\oauthor{\bsnm{Schmidt}, \binits{A.L.}},
\oauthor{\bsnm{Peruzzi}, \binits{A.}},
\oauthor{\bsnm{Scala}, \binits{A.}},
\oauthor{\bsnm{Cinelli}, \binits{M.}},
\oauthor{\bsnm{Pomerantsev}, \binits{P.}},
\oauthor{\bsnm{Applebaum}, \binits{A.}},
\oauthor{\bsnm{Gaston}, \binits{S.}},
\oauthor{\bsnm{Fusi}, \binits{N.}},
\oauthor{\bsnm{Peterson}, \binits{Z.}},
\oauthor{\bsnm{Severgnini}, \binits{G.}},
\oauthor{\bsnm{Cesco}, \binits{A.F.}},
\oauthor{\bsnm{Casati}, \binits{D.}},
\oauthor{\bsnm{Novak}, \binits{P.K.}},
\oauthor{\bsnm{Stanley}, \binits{H.E.}},
\oauthor{\bsnm{Zollo}, \binits{F.}},
\oauthor{\bsnm{Quattrociocchi}, \binits{W.}}:
Measuring social response to different journalistic techniques on facebook.
Humanities {\&} Social Sciences Communications
\textbf{7}(1)
(2020)
\end{botherref}
\endbibitem

\bibitem[\protect\citeauthoryear{Koivula et~al.}{2019}]{Koivula.2019}
\begin{barticle}
\bauthor{\bsnm{Koivula}, \binits{A.}},
\bauthor{\bsnm{Kaakinen}, \binits{M.}},
\bauthor{\bsnm{Oksanen}, \binits{A.}},
\bauthor{\bsnm{Rasanen}, \binits{P.}}:
\batitle{The role of political activity in the formation of online identity bubbles}.
\bjtitle{Policy and Internet}
\bvolume{11}(\bissue{4}),
\bfpage{396}--\blpage{417}
(\byear{2019})
\end{barticle}
\endbibitem

\bibitem[\protect\citeauthoryear{Jones-Jang and Chung}{2022}]{jones2022can}
\begin{botherref}
\oauthor{\bsnm{Jones-Jang}, \binits{S.M.}},
\oauthor{\bsnm{Chung}, \binits{M.}}:
Can we blame social media for polarization? counter-evidence against filter bubble claims during the covid-19 pandemic.
new media \& society,
14614448221099591
(2022)
\end{botherref}
\endbibitem

\bibitem[\protect\citeauthoryear{Burnett et~al.}{2022}]{burnett2022selfcensoring}
\begin{botherref}
\oauthor{\bsnm{Burnett}, \binits{A.}},
\oauthor{\bsnm{Knighton}, \binits{D.}},
\oauthor{\bsnm{Wilson}, \binits{C.}}:
The self-censoring majority: How political identity and ideology impacts willingness to self-censor and fear of isolation in the united states.
Social Media + Society
\textbf{8}(3)
(2022)
\end{botherref}
\endbibitem

\bibitem[\protect\citeauthoryear{Justwan et~al.}{2018}]{Justwan.2018}
\begin{barticle}
\bauthor{\bsnm{Justwan}, \binits{F.}},
\bauthor{\bsnm{Baumgaertner}, \binits{B.}},
\bauthor{\bsnm{Carlisle}, \binits{J.E.}},
\bauthor{\bsnm{Clark}, \binits{A.K.}},
\bauthor{\bsnm{Clark}, \binits{M.}}:
\batitle{Social media echo chambers and satisfaction with democracy among democrats and republicans in the aftermath of the 2016 us elections}.
\bjtitle{Journal of Elections Public Opinion and Parties}
\bvolume{28}(\bissue{4}),
\bfpage{424}--\blpage{442}
(\byear{2018})
\end{barticle}
\endbibitem

\bibitem[\protect\citeauthoryear{Neely}{2021}]{Neely.2021}
\begin{botherref}
\oauthor{\bsnm{Neely}, \binits{S.R.}}:
Politically motivated avoidance in social networks: A study of facebook and the 2020 presidential election.
Social Media + Society
\textbf{7}(4)
(2021)
\doiurl{10.1177/20563051211055438}
\end{botherref}
\endbibitem

\bibitem[\protect\citeauthoryear{Eady et~al.}{2019}]{Eady.2019}
\begin{botherref}
\oauthor{\bsnm{Eady}, \binits{G.}},
\oauthor{\bsnm{Nagler}, \binits{J.}},
\oauthor{\bsnm{Guess}, \binits{A.}},
\oauthor{\bsnm{Zilinsky}, \binits{J.}},
\oauthor{\bsnm{Tucker}, \binits{J.A.}}:
How many people live in political bubbles on social media? evidence from linked survey and twitter data.
SAGE OPEN
\textbf{9}(1)
(2019)
\doiurl{10.1177/2158244019832705}
\end{botherref}
\endbibitem

\bibitem[\protect\citeauthoryear{Boutyline and Willer}{2017}]{Boutyline.2017}
\begin{barticle}
\bauthor{\bsnm{Boutyline}, \binits{A.}},
\bauthor{\bsnm{Willer}, \binits{R.}}:
\batitle{The social structure of political echo chambers: Variation in ideological homophily in online networks}.
\bjtitle{Political Psychology}
\bvolume{38}(\bissue{3}),
\bfpage{551}--\blpage{569}
(\byear{2017})
\doiurl{10.1111/pops.12337}
\end{barticle}
\endbibitem

\bibitem[\protect\citeauthoryear{Beam et~al.}{2018}]{Beam.2018}
\begin{barticle}
\bauthor{\bsnm{Beam}, \binits{M.A.}},
\bauthor{\bsnm{Hutchens}, \binits{M.J.}},
\bauthor{\bsnm{Hmielowski}, \binits{J.D.}}:
\batitle{Facebook news and (de)polarization: reinforcing spirals in the 2016 us election}.
\bjtitle{Information, Communication {\&} Society}
\bvolume{21}(\bissue{7}),
\bfpage{940}--\blpage{958}
(\byear{2018})
\doiurl{10.1080/1369118X.2018.1444783}
\end{barticle}
\endbibitem

\bibitem[\protect\citeauthoryear{Barbera et~al.}{2015}]{Barbera.2015}
\begin{barticle}
\bauthor{\bsnm{Barbera}, \binits{P.}},
\bauthor{\bsnm{Jost}, \binits{J.T.}},
\bauthor{\bsnm{Nagler}, \binits{J.}},
\bauthor{\bsnm{Tucker}, \binits{J.A.}},
\bauthor{\bsnm{Bonneau}, \binits{R.}}:
\batitle{Tweeting from left to right: Is online political communication more than an echo chamber?}
\bjtitle{Psychological Science}
\bvolume{26}(\bissue{10}),
\bfpage{1531}--\blpage{1542}
(\byear{2015})
\doiurl{10.1177/0956797615594620}
\end{barticle}
\endbibitem

\bibitem[\protect\citeauthoryear{Nordbrandt}{2019}]{Nordbrandt.}
\begin{barticle}
\bauthor{\bsnm{Nordbrandt}, \binits{M.}}:
\batitle{Affective polarization in the digital age: Testing the direction of the relationship between social media and users' feelings for out-group parties}.
\bjtitle{New Media {\&} Society}
(\byear{2019})
\doiurl{10.1177/14614448211044393}
\end{barticle}
\endbibitem

\bibitem[\protect\citeauthoryear{Nyhan et~al.}{2023}]{Nyhan2023}
\begin{barticle}
\bauthor{\bsnm{Nyhan}, \binits{B.}},
\bauthor{\bsnm{Settle}, \binits{J.}},
\bauthor{\bsnm{Thorson}, \binits{E.}}, \betal:
\batitle{Like-minded sources on facebook are prevalent but not polarizing}.
\bjtitle{Nature}
\bvolume{620},
\bfpage{137}--\blpage{144}
(\byear{2023})
\doiurl{10.1038/s41586-023-06297-w}
\end{barticle}
\endbibitem

\bibitem[\protect\citeauthoryear{Wang et~al.}{2020}]{Wang.2020}
\begin{botherref}
\oauthor{\bsnm{Wang}, \binits{X.}},
\oauthor{\bsnm{Sirianni}, \binits{A.D.}},
\oauthor{\bsnm{Tang}, \binits{S.}},
\oauthor{\bsnm{Zheng}, \binits{Z.}},
\oauthor{\bsnm{Fu}, \binits{F.}}:
Public discourse and social network echo chambers driven by socio-cognitive biases.
Physical Review X
\textbf{10}(4)
(2020)
\doiurl{10.1103/PhysRevX.10.041042}
\end{botherref}
\endbibitem

\bibitem[\protect\citeauthoryear{Lin et~al.}{2023}]{Lin2023}
\begin{barticle}
\bauthor{\bsnm{Lin}, \binits{H.}},
\bauthor{\bsnm{Wang}, \binits{Y.}},
\bauthor{\bsnm{Lee}, \binits{J.}},
\bauthor{\bsnm{Kim}, \binits{Y.}}:
\batitle{The effects of disagreement and unfriending on political polarization: a moderated-mediation model of cross-cutting discussion on affective polarization via unfriending contingent upon exposure to incivility}.
\bjtitle{Journal of Computer-Mediated Communication}
\bvolume{28}(\bissue{4}),
\bfpage{022}
(\byear{2023})
\doiurl{10.1093/jcmc/zmad022}
\end{barticle}
\endbibitem

\bibitem[\protect\citeauthoryear{Powers et~al.}{2019}]{Powers.2019}
\begin{barticle}
\bauthor{\bsnm{Powers}, \binits{E.}},
\bauthor{\bsnm{Koliska}, \binits{M.}},
\bauthor{\bsnm{Guha}, \binits{P.}}:
\batitle{{\textquotedbl}shouting matches and echo chambers{\textquotedbl}: Perceived identity threats and political self-censorship on social media}.
\bjtitle{International Journal of Communication}
\bvolume{13},
\bfpage{3630}--\blpage{3649}
(\byear{2019})
\end{barticle}
\endbibitem

\bibitem[\protect\citeauthoryear{Bail et~al.}{2018}]{bail2018exposure}
\begin{barticle}
\bauthor{\bsnm{Bail}, \binits{C.A.}},
\bauthor{\bsnm{Argyle}, \binits{L.P.}},
\bauthor{\bsnm{Brown}, \binits{T.W.}},
\bauthor{\bsnm{Bumpus}, \binits{J.P.}},
\bauthor{\bsnm{Chen}, \binits{H.}},
\bauthor{\bsnm{Hunzaker}, \binits{M.F.}},
\bauthor{\bsnm{Lee}, \binits{J.}},
\bauthor{\bsnm{Mann}, \binits{M.}},
\bauthor{\bsnm{Merhout}, \binits{F.}},
\bauthor{\bsnm{Volfovsky}, \binits{A.}}:
\batitle{Exposure to opposing views on social media can increase political polarization}.
\bjtitle{Proceedings of the National Academy of Sciences}
\bvolume{115}(\bissue{37}),
\bfpage{9216}--\blpage{9221}
(\byear{2018})
\end{barticle}
\endbibitem

\bibitem[\protect\citeauthoryear{Bodo et~al.}{2019}]{Bodo.2019}
\begin{barticle}
\bauthor{\bsnm{Bodo}, \binits{B.}},
\bauthor{\bsnm{Helberger}, \binits{N.}},
\bauthor{\bsnm{Eskens}, \binits{S.}},
\bauthor{\bsnm{Moller}, \binits{J.}}:
\batitle{Interested in diversity: The role of user attitudes, algorithmic feedback loops, and policy in news personalization}.
\bjtitle{Digital Jorunalism}
\bvolume{7}(\bissue{2}),
\bfpage{206}--\blpage{229}
(\byear{2019})
\doiurl{10.1080/21670811.2018.1521292}
\end{barticle}
\endbibitem

\bibitem[\protect\citeauthoryear{Chan et~al.}{2019}]{Chan.2019}
\begin{barticle}
\bauthor{\bsnm{Chan}, \binits{M.}},
\bauthor{\bsnm{Chen}, \binits{H.-T.}},
\bauthor{\bsnm{Lee}, \binits{F.L.F.}}:
\batitle{Examining the roles of political social network and internal efficacy on social media news engagement: A comparative study of six asian countries}.
\bjtitle{International Journal of Press-Politics}
\bvolume{24}(\bissue{2}),
\bfpage{127}--\blpage{145}
(\byear{2019})
\doiurl{10.1177/1940161218814480}
\end{barticle}
\endbibitem

\bibitem[\protect\citeauthoryear{Matz}{2021}]{matz2021personal}
\begin{barticle}
\bauthor{\bsnm{Matz}, \binits{S.C.}}:
\batitle{Personal echo chambers: Openness-to-experience is linked to higher levels of psychological interest diversity in large-scale behavioral data}.
\bjtitle{Journal of Personality and Social Psychology}
\bvolume{121}(\bissue{6}),
\bfpage{1284}--\blpage{1300}
(\byear{2021})
\end{barticle}
\endbibitem

\bibitem[\protect\citeauthoryear{Boulianne and Koc-Michalska}{2022}]{Boulianne.2022}
\begin{barticle}
\bauthor{\bsnm{Boulianne}, \binits{S.}},
\bauthor{\bsnm{Koc-Michalska}, \binits{K.}}:
\batitle{The role of personality in political talk and like-minded discussion}.
\bjtitle{International Journal of Press-Politics}
\bvolume{27}(\bissue{1}),
\bfpage{285}--\blpage{310}
(\byear{2022})
\doiurl{10.1177/1940161221994096}
\end{barticle}
\endbibitem

\bibitem[\protect\citeauthoryear{Kaiser and Rauchfleisch}{2020}]{Kaiser.2020}
\begin{botherref}
\oauthor{\bsnm{Kaiser}, \binits{J.}},
\oauthor{\bsnm{Rauchfleisch}, \binits{A.}}:
Birds of a feather get recommended together: Algorithmic homophily in youtube's channel recommendations in the united states and germany.
Social Media + Society
\textbf{6}(4)
(2020)
\doiurl{10.1177/2056305120969914}
\end{botherref}
\endbibitem

\bibitem[\protect\citeauthoryear{Hilbert et~al.}{2018}]{Hilbert.2018}
\begin{barticle}
\bauthor{\bsnm{Hilbert}, \binits{M.}},
\bauthor{\bsnm{Ahmed}, \binits{S.}},
\bauthor{\bsnm{Cho}, \binits{J.}},
\bauthor{\bsnm{Liu}, \binits{B.}},
\bauthor{\bsnm{Luu}, \binits{J.}}:
\batitle{Communicating with algorithms: A transfer entropy analysis of emotions-based escapes from online echo chambers}.
\bjtitle{COMMUNICATION METHODS AND MEASURES}
\bvolume{12}(\bissue{4}),
\bfpage{260}--\blpage{275}
(\byear{2018})
\doiurl{10.1080/19312458.2018.1479843}
\end{barticle}
\endbibitem

\bibitem[\protect\citeauthoryear{Hartmann et~al.}{2024}]{hartmann2024addressing}
\begin{botherref}
\oauthor{\bsnm{Hartmann}, \binits{D.}},
\oauthor{\bsnm{Pereira}, \binits{J.R.L.}},
\oauthor{\bsnm{Streitbörger}, \binits{C.}},
\oauthor{\bsnm{Berendt}, \binits{B.}}:
Addressing the Regulatory Gap: Moving Towards an EU AI Audit Ecosystem Beyond the AIA by Including Civil Society
(2024)
\end{botherref}
\endbibitem

\bibitem[\protect\citeauthoryear{Bright}{2018}]{Bright.2018}
\begin{barticle}
\bauthor{\bsnm{Bright}, \binits{J.}}:
\batitle{Explaining the emergence of political fragmentation on social media: The role of ideology and extremism}.
\bjtitle{Journal of Computer-Mediated Communication}
\bvolume{23}(\bissue{1}),
\bfpage{17}--\blpage{33}
(\byear{2018})
\end{barticle}
\endbibitem

\bibitem[\protect\citeauthoryear{Bastos et~al.}{2018}]{Bastos.2018}
\begin{botherref}
\oauthor{\bsnm{Bastos}, \binits{M.}},
\oauthor{\bsnm{Mercea}, \binits{D.}},
\oauthor{\bsnm{Baronchelli}, \binits{A.}}:
The geographic embedding of online echo chambers: Evidence from the brexit campaign.
PLOS ONE
\textbf{13}(11)
(2018)
\end{botherref}
\endbibitem

\bibitem[\protect\citeauthoryear{Mahrt}{2019}]{Mahrt2019}
\begin{bbook}
\bauthor{\bsnm{Mahrt}, \binits{M.}}:
\bbtitle{Beyond Filter Bubbles and Echo Chambers: The Integrative Potential of the Internet}.
\bsertitle{Digital Communication Research},
vol. \bseriesno{5},
p. \bfpage{246}.
\bpublisher{Digital Communications Research},
\blocation{Berlin}
(\byear{2019}).
\doiurl{10.17174/dcr.v5.0}
\end{bbook}
\endbibitem

\bibitem[\protect\citeauthoryear{Grusauskaite et~al.}{2024}]{Grusauskaite2023}
\begin{barticle}
\bauthor{\bsnm{Grusauskaite}, \binits{K.}},
\bauthor{\bsnm{Carbone}, \binits{L.}},
\bauthor{\bsnm{Harambam}, \binits{J.}},
\bauthor{\bsnm{Aupers}, \binits{S.}}:
\batitle{Debating (in) echo chambers: How culture shapes communication in conspiracy theory networks on youtube}.
\bjtitle{New Media \& Society}
\bvolume{26}(\bissue{12}),
\bfpage{7037}--\blpage{7057}
(\byear{2024})
\doiurl{10.1177/14614448231162585}
\end{barticle}
\endbibitem

\bibitem[\protect\citeauthoryear{Mosleh et~al.}{2021}]{Mosleh.2021}
\begin{botherref}
\oauthor{\bsnm{Mosleh}, \binits{M.}},
\oauthor{\bsnm{Pennycook}, \binits{G.}},
\oauthor{\bsnm{Arechar}, \binits{A.A.}},
\oauthor{\bsnm{Rand}, \binits{D.G.}}:
Cognitive reflection correlates with behavior on twitter.
Nature Communications
\textbf{12}(1)
(2021)
\end{botherref}
\endbibitem

\bibitem[\protect\citeauthoryear{Cann et~al.}{2021}]{Cann.2021}
\begin{botherref}
\oauthor{\bsnm{Cann}, \binits{T.J.B.}},
\oauthor{\bsnm{Weaver}, \binits{I.S.}},
\oauthor{\bsnm{Williams}, \binits{H.T.P.}}:
Ideological biases in social sharing of online information about climate change.
PLOS ONE
\textbf{16}(4)
(2021)
\end{botherref}
\endbibitem

\bibitem[\protect\citeauthoryear{Guerrero-Sole}{2018}]{GuerreroSole.2018}
\begin{botherref}
\oauthor{\bsnm{Guerrero-Sole}, \binits{F.}}:
Interactive behavior in political discussions on twitter: Politicians, media, and citizens' patterns of interaction in the 2015 and 2016 electoral campaigns in spain.
Social Media + Society
\textbf{4}(4)
(2018)
\doiurl{10.1177/2056305118808776}
\end{botherref}
\endbibitem

\bibitem[\protect\citeauthoryear{{Del Vicario} et~al.}{2017}]{DelVicario.2017}
\begin{barticle}
\bauthor{\bsnm{{Del Vicario}}, \binits{M.}},
\bauthor{\bsnm{Zollo}, \binits{F.}},
\bauthor{\bsnm{Caldarelli}, \binits{G.}},
\bauthor{\bsnm{Scala}, \binits{A.}},
\bauthor{\bsnm{Quattrociocchi}, \binits{W.}}:
\batitle{Mapping social dynamics on facebook: The brexit debate}.
\bjtitle{Social Networks}
\bvolume{50},
\bfpage{6}--\blpage{16}
(\byear{2017})
\end{barticle}
\endbibitem

\bibitem[\protect\citeauthoryear{Brugnoli et~al.}{2019}]{Brugnoli.2019}
\begin{botherref}
\oauthor{\bsnm{Brugnoli}, \binits{E.}},
\oauthor{\bsnm{Cinelli}, \binits{M.}},
\oauthor{\bsnm{Quattrociocchi}, \binits{W.}},
\oauthor{\bsnm{Scala}, \binits{A.}}:
Recursive patterns in online echo chambers.
Scientific Reports
\textbf{9}
(2019)
\end{botherref}
\endbibitem

\bibitem[\protect\citeauthoryear{Del~Vicario et~al.}{2016}]{delvicarion2016emotional}
\begin{botherref}
\oauthor{\bsnm{Del~Vicario}, \binits{M.}},
\oauthor{\bsnm{Vivaldo}, \binits{G.}},
\oauthor{\bsnm{Bessi}, \binits{A.}},
\oauthor{\bsnm{Zollo}, \binits{F.}},
\oauthor{\bsnm{Scala}, \binits{A.}},
\oauthor{\bsnm{Caldarelli}, \binits{G.}},
\oauthor{\bsnm{Quattrociocchi}, \binits{W.}}:
Echo chambers: Emotional contagion and group polarization on facebook.
Scientific Reports
\textbf{6}
(2016)
\doiurl{10.1038/srep37825}
\end{botherref}
\endbibitem

\bibitem[\protect\citeauthoryear{Asatani et~al.}{2021}]{Asatani.2021}
\begin{botherref}
\oauthor{\bsnm{Asatani}, \binits{K.}},
\oauthor{\bsnm{Yamano}, \binits{H.}},
\oauthor{\bsnm{Sakaki}, \binits{T.}},
\oauthor{\bsnm{Sakata}, \binits{I.}}:
Dense and influential core promotion of daily viral information spread in political echo chambers.
Scientific Reports
\textbf{11}(1)
(2021)
\end{botherref}
\endbibitem

\bibitem[\protect\citeauthoryear{Choi et~al.}{2020}]{choi2020rumor}
\begin{barticle}
\bauthor{\bsnm{Choi}, \binits{D.}},
\bauthor{\bsnm{Chun}, \binits{S.}},
\bauthor{\bsnm{Oh}, \binits{H.}},
\bauthor{\bsnm{Han}, \binits{J.}},
\bauthor{\bsnm{Kwon}, \binits{T.}}:
\batitle{Rumor propagation is amplified by echo chambers in social media}.
\bjtitle{Scientific reports}
\bvolume{10}(\bissue{1}),
\bfpage{310}
(\byear{2020})
\end{barticle}
\endbibitem

\bibitem[\protect\citeauthoryear{Wolfowicz et~al.}{2023}]{Wolfowicz.}
\begin{botherref}
\oauthor{\bsnm{Wolfowicz}, \binits{M.}},
\oauthor{\bsnm{Weisburd}, \binits{D.}},
\oauthor{\bsnm{Hasisi}, \binits{B.}}:
Examining the interactive effects of the filter bubble and the echo chamber on radicalization.
Journal of Experimental Criminology
(2023)
\end{botherref}
\endbibitem

\bibitem[\protect\citeauthoryear{Nguyen}{2020}]{nguyen_2020}
\begin{barticle}
\bauthor{\bsnm{Nguyen}, \binits{C.T.}}:
\batitle{Echo chambers and epistemic bubbles}.
\bjtitle{Episteme}
\bvolume{17}(\bissue{2}),
\bfpage{141}--\blpage{161}
(\byear{2020})
\end{barticle}
\endbibitem

\bibitem[\protect\citeauthoryear{McPherson et~al.}{2001}]{mcpherson}
\begin{barticle}
\bauthor{\bsnm{McPherson}, \binits{M.}},
\bauthor{\bsnm{Smith-Lovin}, \binits{L.}},
\bauthor{\bsnm{Cook}, \binits{J.M.}}:
\batitle{Birds of a feather: Homophily in social networks}.
\bjtitle{Annual Review of Sociology}
\bvolume{27}(\bissue{1}),
\bfpage{415}--\blpage{444}
(\byear{2001})
\end{barticle}
\endbibitem

\bibitem[\protect\citeauthoryear{Wollebaek et~al.}{2019}]{Wollebaek.2019}
\begin{botherref}
\oauthor{\bsnm{Wollebaek}, \binits{D.}},
\oauthor{\bsnm{Karlsen}, \binits{R.}},
\oauthor{\bsnm{Steen-Johnsen}, \binits{K.}},
\oauthor{\bsnm{Enjolras}, \binits{B.}}:
Anger, fear, and echo chambers: The emotional basis for online behavior.
Social Media + Society
\textbf{5}(2)
(2019)
\doiurl{10.1177/2056305119829859}
\end{botherref}
\endbibitem

\bibitem[\protect\citeauthoryear{Jamieson and Hall}{2008}]{jamieson2008echo}
\begin{bbook}
\bauthor{\bsnm{Jamieson}, \binits{K.}},
\bauthor{\bsnm{Hall}, \binits{C.}}:
\bbtitle{Echo Chamber : Rush Limbaugh and the Conservative Media Establishment}.
\bpublisher{Oxford University Press},
\blocation{Oxford New York}
(\byear{2008})
\end{bbook}
\endbibitem

\bibitem[\protect\citeauthoryear{Sandvig et~al.}{2014}]{sandvig_auditing_2014}
\begin{botherref}
\oauthor{\bsnm{Sandvig}, \binits{C.}},
\oauthor{\bsnm{Hamilton}, \binits{K.}},
\oauthor{\bsnm{Karahalios}, \binits{K.}},
\oauthor{\bsnm{Langbort}, \binits{C.}}:
Auditing {Algorithms}: {Research} {Methods} for {Detecting} {Discrimination} on {Internet} {Platforms}.
Data and Discrimination: Converting Critical Concerns into Productive Inquiry
(2014)
\end{botherref}
\endbibitem

\bibitem[\protect\citeauthoryear{Ohme et~al.}{2023}]{ohme}
\begin{barticle}
\bauthor{\bsnm{Ohme}, \binits{J.}},
\bauthor{\bsnm{Araujo}, \binits{T.}},
\bauthor{\bsnm{Boeschoten}, \binits{L.}},
\bauthor{\bsnm{Freelon}, \binits{D.}},
\bauthor{\bsnm{Ram}, \binits{N.}},
\bauthor{\bsnm{Reeves}, \binits{B.B.}},
\bauthor{\bsnm{Robinson}, \binits{T.N.}}:
\batitle{Digital trace data collection for social media effects research: Apis, data donation, and (screen) tracking}.
\bjtitle{Communication Methods and Measures}
\bvolume{0}(\bissue{0}),
\bfpage{1}--\blpage{18}
(\byear{2023})
\end{barticle}
\endbibitem

\bibitem[\protect\citeauthoryear{Aruguete et~al.}{2023}]{Aruguete.2023}
\begin{barticle}
\bauthor{\bsnm{Aruguete}, \binits{N.}},
\bauthor{\bsnm{Calvo}, \binits{E.}},
\bauthor{\bsnm{Ventura}, \binits{T.}}:
\batitle{News by popular demand: Ideological congruence, issue salience, and media reputation in news sharing( 1 )}.
\bjtitle{International Journal of Press-Politics}
(\byear{2023})
\doiurl{10.1177/19401612211057068}
\end{barticle}
\endbibitem

\bibitem[\protect\citeauthoryear{Wieringa et~al.}{2018}]{Wieringa.2018}
\begin{botherref}
\oauthor{\bsnm{Wieringa}, \binits{M.}},
\oauthor{\bsnm{{van Geenen}}, \binits{D.}},
\oauthor{\bsnm{Schafer}, \binits{M.T.}},
\oauthor{\bsnm{Gorzeman}, \binits{L.}}:
Political topic-communities and their framing practices in the dutch twittersphere.
Internet Policy Review
\textbf{7}(2)
(2018)
\end{botherref}
\endbibitem

\bibitem[\protect\citeauthoryear{Rusche}{2022}]{rusche2022}
\begin{botherref}
\oauthor{\bsnm{Rusche}, \binits{F.}}:
Few voices, strong echo: Measuring follower homogeneity of politicians’ twitter accounts.
New Media \& Society,
14614448221099860
(2022)
{\href{https://arxiv.org/abs/https://doi.org/10.1177/14614448221099860}{{https://doi.org/10.1177/14614448221099860}}}
\end{botherref}
\endbibitem

\bibitem[\protect\citeauthoryear{Kratzke}{2023}]{Kratzke}
\begin{botherref}
\oauthor{\bsnm{Kratzke}, \binits{N.}}:
How to find orchestrated trolls? a case study on identifying polarized twitter echo chambers.
Computers
\textbf{12}(3)
(2023)
\doiurl{10.3390/computers12030057}
\end{botherref}
\endbibitem

\bibitem[\protect\citeauthoryear{Hagen et~al.}{2022}]{Hagen.2022}
\begin{barticle}
\bauthor{\bsnm{Hagen}, \binits{L.}},
\bauthor{\bsnm{Fox}, \binits{A.}},
\bauthor{\bsnm{O'Leary}, \binits{H.}},
\bauthor{\bsnm{Dyson}, \binits{D.}},
\bauthor{\bsnm{Walker}, \binits{K.}},
\bauthor{\bsnm{Lengacher}, \binits{C.A.}},
\bauthor{\bsnm{Hernandez}, \binits{R.}}:
\batitle{The role of influential actors in fostering the polarized covid-19 vaccine discourse on twitter: Mixed methods of machine learning and inductive coding}.
\bjtitle{JMIR infodemiology}
\bvolume{2}(\bissue{1}),
\bfpage{34231}--\blpage{34231}
(\byear{2022})
\doiurl{10.2196/34231}
\end{barticle}
\endbibitem

\bibitem[\protect\citeauthoryear{Colleoni et~al.}{2014}]{Colleoni.2014}
\begin{barticle}
\bauthor{\bsnm{Colleoni}, \binits{E.}},
\bauthor{\bsnm{Rozza}, \binits{A.}},
\bauthor{\bsnm{Arvidsson}, \binits{A.}}:
\batitle{Echo chamber or public sphere? predicting political orientation and measuring political homophily in twitter using big data}.
\bjtitle{Journal of Communication}
\bvolume{64}(\bissue{2}),
\bfpage{317}--\blpage{332}
(\byear{2014})
\end{barticle}
\endbibitem

\bibitem[\protect\citeauthoryear{Bruns}{2017}]{bruns2017echo}
\begin{bchapter}
\bauthor{\bsnm{Bruns}, \binits{A.}}:
\bctitle{Echo chamber? what echo chamber? reviewing the evidence}.
In: \bbtitle{6th Biennial Future of Journalism Conference (FOJ17)}
(\byear{2017})
\end{bchapter}
\endbibitem

\bibitem[\protect\citeauthoryear{Matuszewski and Szabo}{2019}]{Matuszewski.2019}
\begin{botherref}
\oauthor{\bsnm{Matuszewski}, \binits{P.}},
\oauthor{\bsnm{Szabo}, \binits{G.}}:
Are echo chambers based on partisanship? twitter and political polarity in poland and hungary.
Social Media + Society
\textbf{5}(2)
(2019)
\doiurl{10.1177/2056305119837671}
\end{botherref}
\endbibitem

\bibitem[\protect\citeauthoryear{{Del Valle} et~al.}{2022}]{DelValle.2022}
\begin{barticle}
\bauthor{\bsnm{{Del Valle}}, \binits{M.E.}},
\bauthor{\bsnm{Broersma}, \binits{M.}},
\bauthor{\bsnm{Ponsioen}, \binits{A.}}:
\batitle{Political interaction beyond party lines: Communication ties and party polarization in parliamentary twitter networks}.
\bjtitle{Social Science Computer Review}
(\byear{2022})
\doiurl{10.1177/0894439320987569}
\end{barticle}
\endbibitem

\bibitem[\protect\citeauthoryear{Enjolras and Salway}{2022}]{Enjolras.2022}
\begin{barticle}
\bauthor{\bsnm{Enjolras}, \binits{B.}},
\bauthor{\bsnm{Salway}, \binits{A.}}:
\batitle{Homophily and polarization on political twitter during the 2017 norwegian election}.
\bjtitle{Social Network Analysis and Mining}
\bvolume{13}(\bissue{1}),
\bfpage{10}
(\byear{2022})
\doiurl{10.1007/s13278-022-01018-z}
\end{barticle}
\endbibitem

\bibitem[\protect\citeauthoryear{Furman and Tunc}{2020}]{Furman.2020}
\begin{barticle}
\bauthor{\bsnm{Furman}, \binits{I.}},
\bauthor{\bsnm{Tunc}, \binits{A.}}:
\batitle{The end of the habermassian ideal? political communication on twitter during the 2017 turkish constitutional referendum}.
\bjtitle{Policy and Internet}
\bvolume{12}(\bissue{3}),
\bfpage{311}--\blpage{331}
(\byear{2020})
\doiurl{10.1002/poi3.218}
\end{barticle}
\endbibitem

\bibitem[\protect\citeauthoryear{Radicioni et~al.}{2021}]{Radicioni.2021b}
\begin{botherref}
\oauthor{\bsnm{Radicioni}, \binits{T.}},
\oauthor{\bsnm{Saracco}, \binits{F.}},
\oauthor{\bsnm{Pavan}, \binits{E.}},
\oauthor{\bsnm{Squartini}, \binits{T.}}:
Analysing twitter semantic networks: the case of 2018 italian elections.
Scientific Reports
\textbf{11}(1)
(2021)
\end{botherref}
\endbibitem

\bibitem[\protect\citeauthoryear{Gaumont et~al.}{2018}]{Gaumont.2018}
\begin{botherref}
\oauthor{\bsnm{Gaumont}, \binits{N.}},
\oauthor{\bsnm{Panahi}, \binits{M.}},
\oauthor{\bsnm{Chavalarias}, \binits{D.}}:
Reconstruction of the socio-semantic dynamics of political activist twitter networks-method and application to the 2017 french presidential election.
PLOS ONE
\textbf{13}(9)
(2018)
\end{botherref}
\endbibitem

\bibitem[\protect\citeauthoryear{Radicioni et~al.}{2021}]{Radicioni.2021}
\begin{botherref}
\oauthor{\bsnm{Radicioni}, \binits{T.}},
\oauthor{\bsnm{Squartini}, \binits{T.}},
\oauthor{\bsnm{Pavan}, \binits{E.}},
\oauthor{\bsnm{Saracco}, \binits{F.}}:
Networked partisanship and framing: A socio-semantic network analysis of the italian debate on migration.
PLOS ONE
\textbf{16}(8)
(2021)
\end{botherref}
\endbibitem

\bibitem[\protect\citeauthoryear{Ceron and Splendore}{2019}]{Ceron.2019}
\begin{barticle}
\bauthor{\bsnm{Ceron}, \binits{A.}},
\bauthor{\bsnm{Splendore}, \binits{S.}}:
\batitle{'cheap talk'? second screening and the irrelevance of tv political debates}.
\bjtitle{Journalism}
\bvolume{20}(\bissue{8}),
\bfpage{1108}--\blpage{1123}
(\byear{2019})
\doiurl{10.1177/1464884919845443}
\end{barticle}
\endbibitem

\bibitem[\protect\citeauthoryear{Samantray and Pin}{2019}]{Samantray.2019}
\begin{botherref}
\oauthor{\bsnm{Samantray}, \binits{A.}},
\oauthor{\bsnm{Pin}, \binits{P.}}:
Credibility of climate change denial in social media.
Palgrave Communications
\textbf{5}
(2019)
\end{botherref}
\endbibitem

\bibitem[\protect\citeauthoryear{Shore et~al.}{2018}]{Shore.2018}
\begin{barticle}
\bauthor{\bsnm{Shore}, \binits{J.}},
\bauthor{\bsnm{Baek}, \binits{J.}},
\bauthor{\bsnm{Dellarocas}, \binits{C.}}:
\batitle{Network structure and patterns of information diversity on twitter}.
\bjtitle{Management Information Systems Quarterly}
\bvolume{42}(\bissue{3}),
\bfpage{849}--\blpage{972}
(\byear{2018})
\end{barticle}
\endbibitem

\bibitem[\protect\citeauthoryear{Flamino et~al.}{2023}]{Flamino.2023}
\begin{barticle}
\bauthor{\bsnm{Flamino}, \binits{J.}},
\bauthor{\bsnm{Galeazzi}, \binits{A.}},
\bauthor{\bsnm{Feldman}, \binits{S.}},
\bauthor{\bsnm{Macy}, \binits{M.W.}},
\bauthor{\bsnm{Cross}, \binits{B.}},
\bauthor{\bsnm{Zhou}, \binits{Z.}},
\bauthor{\bsnm{Serafino}, \binits{M.}},
\bauthor{\bsnm{Bovet}, \binits{A.}},
\bauthor{\bsnm{Makse}, \binits{H.A.}},
\bauthor{\bsnm{Szymanski}, \binits{B.K.}}:
\batitle{Political polarization of news media and influencers on twitter in the 2016 and 2020 us presidential elections}.
\bjtitle{Nature Human Behaviour}
(\byear{2023})
\doiurl{10.1038/s41562-023-01550-8}
\end{barticle}
\endbibitem

\bibitem[\protect\citeauthoryear{Tyagi et~al.}{2021}]{Tyagi.2021}
\begin{botherref}
\oauthor{\bsnm{Tyagi}, \binits{A.}},
\oauthor{\bsnm{Uyheng}, \binits{J.}},
\oauthor{\bsnm{Carley}, \binits{K.M.}}:
Heated conversations in a warming world: affective polarization in online climate change discourse follows real-world climate anomalies.
Social Network Analysis and Mining
\textbf{11}(1)
(2021)
\end{botherref}
\endbibitem

\bibitem[\protect\citeauthoryear{Bessi et~al.}{2015}]{Bessi.2015}
\begin{botherref}
\oauthor{\bsnm{Bessi}, \binits{A.}},
\oauthor{\bsnm{Zollo}, \binits{F.}},
\oauthor{\bsnm{{Del Vicario}}, \binits{M.}},
\oauthor{\bsnm{Scala}, \binits{A.}},
\oauthor{\bsnm{Caldarelli}, \binits{G.}},
\oauthor{\bsnm{Quattrociocchi}, \binits{W.}}:
Trend of narratives in the age of misinformation.
PLOS ONE
\textbf{10}(8)
(2015)
\end{botherref}
\endbibitem

\bibitem[\protect\citeauthoryear{Zollo et~al.}{2015}]{Zollo.2015}
\begin{botherref}
\oauthor{\bsnm{Zollo}, \binits{F.}},
\oauthor{\bsnm{Novak}, \binits{P.K.}},
\oauthor{\bsnm{{Del Vicario}}, \binits{M.}},
\oauthor{\bsnm{Bessi}, \binits{A.}},
\oauthor{\bsnm{Mozetic}, \binits{I.}},
\oauthor{\bsnm{Scala}, \binits{A.}},
\oauthor{\bsnm{Caldarelli}, \binits{G.}},
\oauthor{\bsnm{Quattrociocchi}, \binits{W.}}:
Emotional dynamics in the age of misinformation.
PLOS ONE
\textbf{10}(9)
(2015)
\end{botherref}
\endbibitem

\bibitem[\protect\citeauthoryear{Bakshy et~al.}{2015}]{Bakshy.2015}
\begin{barticle}
\bauthor{\bsnm{Bakshy}, \binits{E.}},
\bauthor{\bsnm{Messing}, \binits{S.}},
\bauthor{\bsnm{Adamic}, \binits{L.A.}}:
\batitle{Political science. exposure to ideologically diverse news and opinion on facebook}.
\bjtitle{Science (New York, N.Y.)}
\bvolume{348}(\bissue{6239}),
\bfpage{1130}--\blpage{1132}
(\byear{2015})
\doiurl{10.1126/science.aaa1160}
\end{barticle}
\endbibitem

\bibitem[\protect\citeauthoryear{Etta et~al.}{2022}]{etta}
\begin{botherref}
\oauthor{\bsnm{Etta}, \binits{G.}},
\oauthor{\bsnm{Cinelli}, \binits{M.}},
\oauthor{\bsnm{Galeazzi}, \binits{A.}},
\oauthor{\bsnm{Valensise}, \binits{C.M.}},
\oauthor{\bsnm{Quattrociocchi}, \binits{W.}},
\oauthor{\bsnm{Conti}, \binits{M.}}:
Comparing the impact of social media regulations on news consumption.
IEEE Transactions on Computational Social Systems,
1--11
(2022)
\doiurl{10.1109/TCSS.2022.3171391}
\end{botherref}
\endbibitem

\bibitem[\protect\citeauthoryear{Bessi}{2016}]{Bessi.2016b}
\begin{barticle}
\bauthor{\bsnm{Bessi}, \binits{A.}}:
\batitle{Personality traits and echo chambers on facebook}.
\bjtitle{Computers in Human Behavior}
\bvolume{65},
\bfpage{319}--\blpage{324}
(\byear{2016})
\doiurl{10.1016/j.chb.2016.08.016}
\end{barticle}
\endbibitem

\bibitem[\protect\citeauthoryear{Goel et~al.}{2023}]{goel}
\begin{botherref}
\oauthor{\bsnm{Goel}, \binits{V.}},
\oauthor{\bsnm{Sahnan}, \binits{D.}},
\oauthor{\bsnm{Dutta}, \binits{S.}},
\oauthor{\bsnm{Bandhakavi}, \binits{A.}},
\oauthor{\bsnm{Chakraborty}, \binits{T.}}:
Hatemongers ride on echo chambers to escalate hate speech diffusion.
PNAS Nexus
\textbf{2}(3)
(2023)
{\href{https://arxiv.org/abs/https://academic.oup.com/pnasnexus/article-pdf/2/3/pgad041/49703225/pgad041.pdf}{{https://academic.oup.com/pnasnexus/article-pdf/2/3/pgad041/49703225/pgad041.pdf}}}.
pgad041
\end{botherref}
\endbibitem

\bibitem[\protect\citeauthoryear{Fletcher et~al.}{2021}]{fletcher}
\begin{barticle}
\bauthor{\bsnm{Fletcher}, \binits{R.}},
\bauthor{\bsnm{Kalogeropoulos}, \binits{A.}},
\bauthor{\bsnm{Nielsen}, \binits{R.K.}}:
\batitle{More diverse, more politically varied: How social media, search engines and aggregators shape news repertoires in the united kingdom}.
\bjtitle{New Media \& Society}
\bvolume{0}(\bissue{0}),
\bfpage{14614448211027393}
(\byear{2021})
{\href{https://arxiv.org/abs/https://doi.org/10.1177/14614448211027393}{{https://doi.org/10.1177/14614448211027393}}}
\end{barticle}
\endbibitem

\bibitem[\protect\citeauthoryear{Flaxman et~al.}{2016}]{Flaxman.2016}
\begin{barticle}
\bauthor{\bsnm{Flaxman}, \binits{S.}},
\bauthor{\bsnm{Goel}, \binits{S.}},
\bauthor{\bsnm{Rao}, \binits{J.M.}}:
\batitle{Filter bubbles, echo chambers, and online news consumption}.
\bjtitle{Public Opinion Quarterly}
\bvolume{80},
\bfpage{298}--\blpage{320}
(\byear{2016})
\end{barticle}
\endbibitem

\bibitem[\protect\citeauthoryear{Muise et~al.}{2022}]{muise2022}
\begin{barticle}
\bauthor{\bsnm{Muise}, \binits{D.}},
\bauthor{\bsnm{Hosseinmardi}, \binits{H.}},
\bauthor{\bsnm{Howland}, \binits{B.}},
\bauthor{\bsnm{Mobius}, \binits{M.}},
\bauthor{\bsnm{Rothschild}, \binits{D.}},
\bauthor{\bsnm{Watts}, \binits{D.J.}}:
\batitle{Quantifying partisan news diets in web and tv audiences}.
\bjtitle{Science Advances}
\bvolume{8}(\bissue{28}),
\bfpage{0083}
(\byear{2022})
{\href{https://arxiv.org/abs/https://www.science.org/doi/pdf/10.1126/sciadv.abn0083}{{https://www.science.org/doi/pdf/10.1126/sciadv.abn0083}}}
\end{barticle}
\endbibitem

\bibitem[\protect\citeauthoryear{Urman}{2019}]{Urman.2019}
\begin{barticle}
\bauthor{\bsnm{Urman}, \binits{A.}}:
\batitle{News consumption of russian vkontakte users: Polarization and news avoidance}.
\bjtitle{International Journal of Communication}
\bvolume{13},
\bfpage{5158}--\blpage{5182}
(\byear{2019})
\end{barticle}
\endbibitem

\bibitem[\protect\citeauthoryear{Bond and Sweitzer}{2022}]{bond2022}
\begin{barticle}
\bauthor{\bsnm{Bond}, \binits{R.M.}},
\bauthor{\bsnm{Sweitzer}, \binits{M.D.}}:
\batitle{Political homophily in a large-scale online communication network}.
\bjtitle{Communication Research}
\bvolume{49}(\bissue{1}),
\bfpage{93}--\blpage{115}
(\byear{2022})
{\href{https://arxiv.org/abs/https://doi.org/10.1177/0093650218813655}{{https://doi.org/10.1177/0093650218813655}}}
\end{barticle}
\endbibitem

\bibitem[\protect\citeauthoryear{Roth et~al.}{2020}]{Roth.2020}
\begin{botherref}
\oauthor{\bsnm{Roth}, \binits{C.}},
\oauthor{\bsnm{Mazieres}, \binits{A.}},
\oauthor{\bsnm{Menezes}, \binits{T.}}:
Tubes and bubbles topological confinement of youtube recommendations.
PLOS ONE
\textbf{15}(4)
(2020)
\end{botherref}
\endbibitem

\bibitem[\protect\citeauthoryear{Lima et~al.}{2018}]{Lima.2018}
\begin{bchapter}
\bauthor{\bsnm{Lima}, \binits{L.}},
\bauthor{\bsnm{Reis}, \binits{J.C.S.}},
\bauthor{\bsnm{Melo}, \binits{P.}},
\bauthor{\bsnm{Murai}, \binits{F.}},
\bauthor{\bsnm{Ara{\'u}jo}, \binits{L.}},
\bauthor{\bsnm{Vikatos}, \binits{P.}},
\bauthor{\bsnm{Benevenuto}, \binits{F.}}:
\bctitle{Inside the right-leaning echo chambers: Characterizing gab, an unmoderated social system}.
In: \bbtitle{Proceedings of the 2018 IEEE/ACM International Conference on Advances in Social Networks Analysis and Mining}.
\bsertitle{ASONAM '18},
pp. \bfpage{515}--\blpage{522}.
\bpublisher{{IEEE Press}}, \blocation{???}
(\byear{2018})
\end{bchapter}
\endbibitem

\bibitem[\protect\citeauthoryear{Wang and Qian}{2021}]{Wang.2021}
\begin{botherref}
\oauthor{\bsnm{Wang}, \binits{D.}},
\oauthor{\bsnm{Qian}, \binits{Y.}}:
Echo chamber effect in rumor rebuttal discussions about covid-19 in china: Social media content and network analysis study.
Journal of Medical Internet Research
\textbf{23}(3)
(2021)
\doiurl{10.2196/27009}
\end{botherref}
\endbibitem

\bibitem[\protect\citeauthoryear{Wang et~al.}{2022}]{wang2022}
\begin{barticle}
\bauthor{\bsnm{Wang}, \binits{D.}},
\bauthor{\bsnm{Zhou}, \binits{Y.}},
\bauthor{\bsnm{Ma}, \binits{F.}}:
\batitle{Opinion leaders and structural hole spanners influencing echo chambers in discussions about covid-19 vaccines on social media in china: Network analysis}.
\bjtitle{J Med Internet Res}
\bvolume{24}(\bissue{11}),
\bfpage{40701}
(\byear{2022})
\doiurl{10.2196/40701}
\end{barticle}
\endbibitem

\bibitem[\protect\citeauthoryear{Light and Moody}{2021}]{oxford1}
\begin{bchapter}
\bauthor{\bsnm{Light}, \binits{R.}},
\bauthor{\bsnm{Moody}, \binits{J.}}:
\bctitle{16network basics: Points, lines, and positions}.
In: \bbtitle{The Oxford Handbook of Social Networks}.
\bpublisher{Oxford University Press},
\blocation{Oxford}
(\byear{2021}).
\burl{https://doi.org/10.1093/oxfordhb/9780190251765.013.2}
\end{bchapter}
\endbibitem

\bibitem[\protect\citeauthoryear{Cinelli et~al.}{2021}]{Cinelli.2021b}
\begin{botherref}
\oauthor{\bsnm{Cinelli}, \binits{M.}},
\oauthor{\bsnm{Pelicon}, \binits{A.}},
\oauthor{\bsnm{Mozetic}, \binits{I.}},
\oauthor{\bsnm{Quattrociocchi}, \binits{W.}},
\oauthor{\bsnm{Novak}, \binits{P.K.}},
\oauthor{\bsnm{Zollo}, \binits{F.}}:
Dynamics of online hate and misinformation.
Scientific Reports
\textbf{11}(1)
(2021)
\end{botherref}
\endbibitem

\bibitem[\protect\citeauthoryear{Traag et~al.}{2019}]{traag_louvain_2019}
\begin{botherref}
\oauthor{\bsnm{Traag}, \binits{V.A.}},
\oauthor{\bsnm{Waltman}, \binits{L.}},
\oauthor{\bsnm{Eck}, \binits{N.J.}}:
From louvain to leiden: guaranteeing well-connected communities
\textbf{9}(1),
5233
(2019).
Number: 1 Publisher: Nature Publishing Group.
Accessed 2022-06-29
\end{botherref}
\endbibitem

\bibitem[\protect\citeauthoryear{Kim et~al.}{2017}]{latentspace}
\begin{botherref}
\oauthor{\bsnm{Kim}, \binits{B.}},
\oauthor{\bsnm{Lee}, \binits{K.}},
\oauthor{\bsnm{Xue}, \binits{L.}},
\oauthor{\bsnm{Niu}, \binits{X.}}:
A review of dynamic network models with latent variables.
Statistics Surveys
\textbf{12}
(2017)
\doiurl{10.1214/18-SS121}
\end{botherref}
\endbibitem

\bibitem[\protect\citeauthoryear{Bovet and Grindrod}{2022}]{bovet2022}
\begin{botherref}
\oauthor{\bsnm{Bovet}, \binits{A.}},
\oauthor{\bsnm{Grindrod}, \binits{P.}}:
Organization and evolution of the uk far-right network on telegram.
APPLIED NETWORK SCIENCE
\textbf{7}(1)
(2022)
\doiurl{10.1007/s41109-022-00513-8}
\end{botherref}
\endbibitem

\bibitem[\protect\citeauthoryear{Srba et~al.}{2023}]{srba2023}
\begin{botherref}
\oauthor{\bsnm{Srba}, \binits{I.}},
\oauthor{\bsnm{Moro}, \binits{R.}},
\oauthor{\bsnm{Tomlein}, \binits{M.}},
\oauthor{\bsnm{Pecher}, \binits{B.}},
\oauthor{\bsnm{Simko}, \binits{J.}},
\oauthor{\bsnm{Stefancova}, \binits{E.}},
\oauthor{\bsnm{Kompan}, \binits{M.}},
\oauthor{\bsnm{Hrckova}, \binits{A.}},
\oauthor{\bsnm{Podrouzek}, \binits{J.}},
\oauthor{\bsnm{Gavornik}, \binits{A.}},
\oauthor{\bsnm{Bielikova}, \binits{M.}}:
Auditing youtube’s recommendation algorithm\&nbsp;for misinformation filter bubbles
\textbf{1}(1)
(2023)
\end{botherref}
\endbibitem

\bibitem[\protect\citeauthoryear{Zerback and Kobilke}{2022}]{zerback}
\begin{botherref}
\oauthor{\bsnm{Zerback}, \binits{T.}},
\oauthor{\bsnm{Kobilke}, \binits{L.}}:
The role of affective and cognitive attitude extremity in perceived viewpoint diversity exposure.
New Media \& Society,
14614448221117484
(2022)
{\href{https://arxiv.org/abs/https://doi.org/10.1177/14614448221117484}{{https://doi.org/10.1177/14614448221117484}}}
\end{botherref}
\endbibitem

\bibitem[\protect\citeauthoryear{Mosleh et~al.}{2021}]{Mosleh.2021b}
\begin{botherref}
\oauthor{\bsnm{Mosleh}, \binits{M.}},
\oauthor{\bsnm{Martel}, \binits{C.}},
\oauthor{\bsnm{Eckles}, \binits{D.}},
\oauthor{\bsnm{Rand}, \binits{D.G.}}:
Shared partisanship dramatically increases social tie formation in a twitter field experiment.
Proceedings of the National Academy of Sciences of the United States of America
\textbf{118}(7)
(2021)
\end{botherref}
\endbibitem

\bibitem[\protect\citeauthoryear{Suelflow et~al.}{2019}]{Suelflow.2019}
\begin{barticle}
\bauthor{\bsnm{Suelflow}, \binits{M.}},
\bauthor{\bsnm{Schaefer}, \binits{S.}},
\bauthor{\bsnm{Winter}, \binits{S.}}:
\batitle{Selective attention in the news feed: An eye-tracking study on the perception and selection of political news posts on facebook}.
\bjtitle{New Media {\&} Society}
\bvolume{21}(\bissue{1}),
\bfpage{168}--\blpage{190}
(\byear{2019})
\doiurl{10.1177/1461444818791520}
\end{barticle}
\endbibitem

\bibitem[\protect\citeauthoryear{Mellon and Prosser}{2017}]{Mellon.2017}
\begin{barticle}
\bauthor{\bsnm{Mellon}, \binits{J.}},
\bauthor{\bsnm{Prosser}, \binits{C.}}:
\batitle{Twitter and facebook are not representative of the general population: Political attitudes and demographics of british social media users}.
\bjtitle{Research {\&} Politics}
\bvolume{4}(\bissue{3}),
\bfpage{205316801772000}
(\byear{2017})
\end{barticle}
\endbibitem

\bibitem[\protect\citeauthoryear{Center}{2021}]{pewresearch}
\begin{botherref}
\oauthor{\bsnm{Center}, \binits{P.R.}}:
Social Media Fact Sheet
(2021).
\url{https://www.pewresearch.org/internet/fact-sheet/social-media/}
\end{botherref}
\endbibitem

\bibitem[\protect\citeauthoryear{Jasny et~al.}{2015}]{jasny_empirical_2015}
\begin{botherref}
\oauthor{\bsnm{Jasny}, \binits{L.}},
\oauthor{\bsnm{Waggle}, \binits{J.}},
\oauthor{\bsnm{Fisher}, \binits{D.R.}}:
An empirical examination of echo chambers in {US} climate policy networks
\textbf{5}(8),
782--786
(2015)
\end{botherref}
\endbibitem

\bibitem[\protect\citeauthoryear{Guess et~al.}{2018}]{avoiding}
\begin{botherref}
\oauthor{\bsnm{Guess}, \binits{A.}},
\oauthor{\bsnm{Lyons}, \binits{B.}},
\oauthor{\bsnm{Nyhan}, \binits{B.}},
\oauthor{\bsnm{Reifler}, \binits{J.}}:
Avoiding the echo chamber about echo chambers: Why selective exposure to like-minded political news is less prevalent than you think
(2018)
\end{botherref}
\endbibitem

\bibitem[\protect\citeauthoryear{Barbera}{2013}]{barbera2013}
\begin{botherref}
\oauthor{\bsnm{Barbera}, \binits{P.}}:
Birds of the same feather tweet together: Bayesian ideal point estimation using twitter data.
SSRN Electronic Journal
\textbf{23}
(2013)
\doiurl{10.2139/ssrn.2108098}
\end{botherref}
\endbibitem

\bibitem[\protect\citeauthoryear{Gallwitz and Kreil}{2021}]{Gallwitz.2021}
\begin{botherref}
\oauthor{\bsnm{Gallwitz}, \binits{F.}},
\oauthor{\bsnm{Kreil}, \binits{M.}}:
The rise and fall of 'social bot' research.
SSRN
(2021)
\end{botherref}
\endbibitem

\bibitem[\protect\citeauthoryear{Neudert et~al.}{2017}]{neudert2017junk}
\begin{botherref}
\oauthor{\bsnm{Neudert}, \binits{L.}},
\oauthor{\bsnm{Kollanyi}, \binits{B.}},
\oauthor{\bsnm{Howard}, \binits{P.N.}}:
Junk news and bots during the german parliamentary election: What are german voters sharing over twitter?
(2017)
\end{botherref}
\endbibitem

\bibitem[\protect\citeauthoryear{Silva and Proksch}{2021}]{silva_proksch_2021}
\begin{barticle}
\bauthor{\bsnm{Silva}, \binits{B.C.}},
\bauthor{\bsnm{Proksch}, \binits{S.-O.}}:
\batitle{Fake it ‘til you make it: A natural experiment to identify european politicians’ benefit from twitter bots}.
\bjtitle{American Political Science Review}
\bvolume{115}(\bissue{1}),
\bfpage{316}--\blpage{322}
(\byear{2021})
\end{barticle}
\endbibitem

\bibitem[\protect\citeauthoryear{VanderWeele and Weihua}{2014}]{Morgan.2014}
\begin{bchapter}
\bauthor{\bsnm{VanderWeele}, \binits{T.J.}},
\bauthor{\bsnm{Weihua}, \binits{A.}}:
\bctitle{Social networks and causal inference}.
In: \bbtitle{Handbook of Causal Analysis for Social Research}.
\bsertitle{Handbooks of Sociology and Social Research},
pp. \bfpage{353}--\blpage{374}.
\bpublisher{Springer},
\blocation{Dordrecht}
(\byear{2014})
\end{bchapter}
\endbibitem

\bibitem[\protect\citeauthoryear{Frank and Xu}{2021}]{causal}
\begin{bchapter}
\bauthor{\bsnm{Frank}, \binits{K.A.}},
\bauthor{\bsnm{Xu}, \binits{R.}}:
\bctitle{Causal inference for social network analysis}.
In: \bbtitle{The Oxford Handbook of Social Networks}.
\bpublisher{Oxford University Press},
\blocation{Oxford}
(\byear{2021}).
\burl{https://doi.org/10.1093/oxfordhb/9780190251765.013.21}
\end{bchapter}
\endbibitem

\bibitem[\protect\citeauthoryear{Brashears and Gladstone}{2021}]{causalexp}
\begin{bchapter}
\bauthor{\bsnm{Brashears}, \binits{M.E.}},
\bauthor{\bsnm{Gladstone}, \binits{E.}}:
\bctitle{Social network experiments}.
In: \bbtitle{The Oxford Handbook of Social Networks}.
\bpublisher{Oxford University Press},
\blocation{Oxford}
(\byear{2021}).
\burl{https://doi.org/10.1093/oxfordhb/9780190251765.013.12}
\end{bchapter}
\endbibitem

\bibitem[\protect\citeauthoryear{Walther and Mccoy}{2021}]{telegram}
\begin{botherref}
\oauthor{\bsnm{Walther}, \binits{S.}},
\oauthor{\bsnm{Mccoy}, \binits{A.}}:
Us extremism on telegram: Fueling disinformation, conspiracy theories, and accelerationism.
Perspectives on Terrorism
\textbf{15}
(2021)
\end{botherref}
\endbibitem

\bibitem[\protect\citeauthoryear{Stokel-Walker}{2022}]{STOKELWALKER20228}
\begin{barticle}
\bauthor{\bsnm{Stokel-Walker}, \binits{C.}}:
\batitle{How to fight disinformation}.
\bjtitle{New Scientist}
\bvolume{253}(\bissue{3378}),
\bfpage{8}
(\byear{2022})
\doiurl{10.1016/S0262-4079(22)00458-4}
\end{barticle}
\endbibitem

\bibitem[\protect\citeauthoryear{Madrigal}{2012}]{madrigal2012dark}
\begin{botherref}
\oauthor{\bsnm{Madrigal}, \binits{A.}}:
Dark social: We have the whole history of the web wrong.
The Atlantic
\textbf{12}
(2012)
\end{botherref}
\endbibitem

\bibitem[\protect\citeauthoryear{Belur et~al.}{2021}]{belur}
\begin{barticle}
\bauthor{\bsnm{Belur}, \binits{J.}},
\bauthor{\bsnm{Tompson}, \binits{L.}},
\bauthor{\bsnm{Thornton}, \binits{A.}},
\bauthor{\bsnm{Simon}, \binits{M.}}:
\batitle{Interrater reliability in systematic review methodology: Exploring variation in coder decision-making}.
\bjtitle{Sociological Methods \& Research}
\bvolume{50}(\bissue{2}),
\bfpage{837}--\blpage{865}
(\byear{2021})
\end{barticle}
\endbibitem

\bibitem[\protect\citeauthoryear{Tomlein et~al.}{2021}]{Tomlein.2021}
\begin{bchapter}
\bauthor{\bsnm{Tomlein}, \binits{M.}},
\bauthor{\bsnm{Pecher}, \binits{B.}},
\bauthor{\bsnm{Simko}, \binits{J.}},
\bauthor{\bsnm{Srba}, \binits{I.}},
\bauthor{\bsnm{Moro}, \binits{R.}},
\bauthor{\bsnm{Stefancova}, \binits{E.}},
\bauthor{\bsnm{Kompan}, \binits{M.}},
\bauthor{\bsnm{Hrckova}, \binits{A.}},
\bauthor{\bsnm{Podrouzek}, \binits{J.}},
\bauthor{\bsnm{Bielikova}, \binits{M.}}:
\bctitle{An audit of misinformation filter bubbles on youtube: Bubble bursting and recent behavior changes}.
In: \bbtitle{Fifteenth ACM Conference on Recommender Systems},
pp. \bfpage{1}--\blpage{11}.
\bpublisher{{Association for Computing Machinery}},
\blocation{New York, NY, USA}
(\byear{2021}).
\doiurl{10.1145/3460231.3474241}
\end{bchapter}
\endbibitem

\bibitem[\protect\citeauthoryear{Ackermann and Stadelmann-Steffen}{2022}]{ackermann2022voting}
\begin{barticle}
\bauthor{\bsnm{Ackermann}, \binits{K.}},
\bauthor{\bsnm{Stadelmann-Steffen}, \binits{I.}}:
\batitle{Voting in the echo chamber? patterns of political online activities and voting behavior in switzerland}.
\bjtitle{Swiss Political Science Review}
\bvolume{28}(\bissue{2}),
\bfpage{377}--\blpage{400}
(\byear{2022})
\end{barticle}
\endbibitem

\bibitem[\protect\citeauthoryear{Auxier and Vitak}{2019}]{Auxier.2019}
\begin{botherref}
\oauthor{\bsnm{Auxier}, \binits{B.E.}},
\oauthor{\bsnm{Vitak}, \binits{J.}}:
Factors motivating customization and echo chamber creation within digital news environments.
Social Media + Society
\textbf{5}(2)
(2019)
\doiurl{10.1177/2056305119847506}
\end{botherref}
\endbibitem

\bibitem[\protect\citeauthoryear{Bodrunova et~al.}{2023}]{Bodrunova2023}
\begin{bchapter}
\bauthor{\bsnm{Bodrunova}, \binits{S.S.}},
\bauthor{\bsnm{Blekanov}, \binits{I.S.}},
\bauthor{\bsnm{Tarasov}, \binits{N.}}:
\bctitle{Echo chambers on russian twitter}.
In: \beditor{\bsnm{Meiselwitz}, \binits{G.}} (ed.)
\bbtitle{Social Computing and Social Media: Applications in Communication, Social Communities, and E-Services}.
\bsertitle{Lecture Notes in Computer Science},
vol. \bseriesno{14025},
pp. \bfpage{167}--\blpage{182}.
\bpublisher{Springer},
\blocation{Cham}
(\byear{2023}).
\doiurl{10.1007/978-3-031-35801-0_13} .
\bcomment{First Online: 09 July 2023}
\end{bchapter}
\endbibitem

\bibitem[\protect\citeauthoryear{Cargnino and Neubaum}{2022}]{cargnino2022better}
\begin{botherref}
\oauthor{\bsnm{Cargnino}, \binits{M.}},
\oauthor{\bsnm{Neubaum}, \binits{G.}}:
Is it better to strike a balance? how exposure to congruent and incongruent opinion climates on social networking sites impacts users’ processing and selection of information.
New Media \& Society,
14614448221083914
(2022)
\end{botherref}
\endbibitem

\bibitem[\protect\citeauthoryear{Cargnino et~al.}{2023}]{cargnino2023we}
\begin{barticle}
\bauthor{\bsnm{Cargnino}, \binits{M.}},
\bauthor{\bsnm{Neubaum}, \binits{G.}},
\bauthor{\bsnm{Winter}, \binits{S.}}:
\batitle{We’re a good match: Selective political friending on social networking sites}.
\bjtitle{Communications}
\bvolume{48}(\bissue{2}),
\bfpage{202}--\blpage{225}
(\byear{2023})
\end{barticle}
\endbibitem

\bibitem[\protect\citeauthoryear{Cheng et~al.}{2023}]{cheng2023birds}
\begin{botherref}
\oauthor{\bsnm{Cheng}, \binits{Z.}},
\oauthor{\bsnm{Marcos-Marne}, \binits{H.}},
\oauthor{\bsnm{Z{\'u}{\~n}iga}, \binits{H.}}:
Birds of a feather get angrier together: Social media news use and social media political homophily as antecedents of political anger.
Political Behavior,
1--17
(2023)
\end{botherref}
\endbibitem

\bibitem[\protect\citeauthoryear{De~Lima-Santos and Ceron}{2023}]{DeLimaSantos2023}
\begin{bchapter}
\bauthor{\bsnm{De~Lima-Santos}, \binits{M.F.}},
\bauthor{\bsnm{Ceron}, \binits{W.}}:
\bctitle{Disinformation echo chambers on facebook}.
In: \bbtitle{Fighting Fake Facts},
pp. \bfpage{61}--\blpage{90}.
\bpublisher{MDPI},
\blocation{Basel}
(\byear{2023}).
\bcomment{Chap. 4}.
\burl{https://doi.org/10.3390/books978-3-0365-1347-8-4}
\end{bchapter}
\endbibitem

\bibitem[\protect\citeauthoryear{Ebeling et~al.}{2023}]{Ebeling2023}
\begin{barticle}
\bauthor{\bsnm{Ebeling}, \binits{R.}},
\bauthor{\bsnm{Nobre}, \binits{J.}},
\bauthor{\bsnm{Becker}, \binits{K.}}:
\batitle{A multi-dimensional framework to analyze group behavior based on political polarization}.
\bjtitle{Expert Systems with Applications}
\bvolume{233},
\bfpage{120768}
(\byear{2023})
\doiurl{10.1016/j.eswa.2023.120768}
\end{barticle}
\endbibitem

\bibitem[\protect\citeauthoryear{Enjolras}{2023}]{Enjolras2023}
\begin{barticle}
\bauthor{\bsnm{Enjolras}, \binits{B.}}:
\batitle{Does relational polarization entail ideological polarization? the case of the 2017 norwegian election campaign on twitter}.
\bjtitle{International Journal of Communication}
\bvolume{17},
\bfpage{2394}--\blpage{2421}
(\byear{2023})
\end{barticle}
\endbibitem

\bibitem[\protect\citeauthoryear{Erickson et~al.}{2023}]{Erickson2023}
\begin{botherref}
\oauthor{\bsnm{Erickson}, \binits{J.}},
\oauthor{\bsnm{Yan}, \binits{B.}},
\oauthor{\bsnm{Huang}, \binits{J.}}:
Bridging echo chambers? understanding political partisanship through semantic network analysis.
Social Media + Society
\textbf{9}(3)
(2023)
\doiurl{10.1177/20563051231186368}
\end{botherref}
\endbibitem

\bibitem[\protect\citeauthoryear{Hada et~al.}{2023}]{Hada2023}
\begin{bchapter}
\bauthor{\bsnm{Hada}, \binits{R.}},
\bauthor{\bsnm{Ebrahimi~Fard}, \binits{A.}},
\bauthor{\bsnm{Shugars}, \binits{S.}},
\bauthor{\bsnm{Bianchi}, \binits{F.}},
\bauthor{\bsnm{Rossini}, \binits{P.}},
\bauthor{\bsnm{Hovy}, \binits{D.}},
\bauthor{\bsnm{Tromble}, \binits{R.}},
\bauthor{\bsnm{Tintarev}, \binits{N.}}:
\bctitle{Beyond digital "echo chambers": The role of viewpoint diversity in political discussion}.
In: \bbtitle{Proceedings of the Sixteenth ACM International Conference on Web Search and Data Mining}.
\bsertitle{WSDM '23},
pp. \bfpage{33}--\blpage{41}.
\bpublisher{Association for Computing Machinery},
\blocation{New York, NY, USA}
(\byear{2023}).
\doiurl{10.1145/3539597.3570487} .
\burl{https://doi.org/10.1145/3539597.3570487}
\end{bchapter}
\endbibitem

\bibitem[\protect\citeauthoryear{Hong and Kim}{2016}]{Hong2016}
\begin{barticle}
\bauthor{\bsnm{Hong}, \binits{S.}},
\bauthor{\bsnm{Kim}, \binits{S.H.}}:
\batitle{Political polarization on twitter: Implications for the use of social media in digital governments}.
\bjtitle{Government Information Quarterly}
\bvolume{33}(\bissue{4}),
\bfpage{777}--\blpage{782}
(\byear{2016})
\doiurl{10.1016/j.giq.2016.04.007}
\end{barticle}
\endbibitem

\bibitem[\protect\citeauthoryear{Liu et~al.}{2021}]{Liu.2021}
\begin{botherref}
\oauthor{\bsnm{Liu}, \binits{R.}},
\oauthor{\bsnm{Greene}, \binits{K.T.}},
\oauthor{\bsnm{Liu}, \binits{R.}},
\oauthor{\bsnm{Mandic}, \binits{M.}},
\oauthor{\bsnm{Valentino}, \binits{B.A.}},
\oauthor{\bsnm{Vosoughi}, \binits{S.}},
\oauthor{\bsnm{Subrahmanian}, \binits{V.S.}}:
Using impression data to improve models of online social influence.
Scientific Reports
\textbf{11}(1)
(2021)
\end{botherref}
\endbibitem

\bibitem[\protect\citeauthoryear{Ludwig et~al.}{2023}]{ludwig2023divided}
\begin{barticle}
\bauthor{\bsnm{Ludwig}, \binits{K.}},
\bauthor{\bsnm{Grote}, \binits{A.}},
\bauthor{\bsnm{Iana}, \binits{A.}},
\bauthor{\bsnm{Alam}, \binits{M.}},
\bauthor{\bsnm{Paulheim}, \binits{H.}},
\bauthor{\bsnm{Sack}, \binits{H.}},
\bauthor{\bsnm{Weinhardt}, \binits{C.}},
\bauthor{\bsnm{M{\"u}ller}, \binits{P.}}:
\batitle{Divided by the algorithm? the (limited) effects of content-and sentiment-based news recommendation on affective, ideological, and perceived polarization}.
\bjtitle{Social Science Computer Review}
\bvolume{41}(\bissue{6}),
\bfpage{2188}--\blpage{2210}
(\byear{2023})
\end{barticle}
\endbibitem

\bibitem[\protect\citeauthoryear{Min et~al.}{2019}]{Min.2019}
\begin{botherref}
\oauthor{\bsnm{Min}, \binits{Y.}},
\oauthor{\bsnm{Jiang}, \binits{T.}},
\oauthor{\bsnm{Jin}, \binits{C.}},
\oauthor{\bsnm{Li}, \binits{Q.}},
\oauthor{\bsnm{Jin}, \binits{X.}}:
Endogenetic structure of filter bubble in social networks.
Royal Society Open Science
\textbf{6}(11)
(2019)
\end{botherref}
\endbibitem

\bibitem[\protect\citeauthoryear{Mirlohi et~al.}{2022}]{mirlohi2022social}
\begin{botherref}
\oauthor{\bsnm{Mirlohi}, \binits{A.}},
\oauthor{\bsnm{Mahdavimoghaddam}, \binits{J.}},
\oauthor{\bsnm{Jovanovic}, \binits{J.}},
\oauthor{\bsnm{Al-Obeidat}, \binits{F.N.}},
\oauthor{\bsnm{Khani}, \binits{M.}},
\oauthor{\bsnm{Ghorbani}, \binits{A.A.}},
\oauthor{\bsnm{Bagheri}, \binits{E.}}:
Social alignment contagion in online social networks.
IEEE Transactions on Computational Social Systems
(2022)
\end{botherref}
\endbibitem

\bibitem[\protect\citeauthoryear{Shmargad and Klar}{2019}]{Shmargad.2019}
\begin{barticle}
\bauthor{\bsnm{Shmargad}, \binits{Y.}},
\bauthor{\bsnm{Klar}, \binits{S.}}:
\batitle{How partisan online environments shape communication with political outgroups}.
\bjtitle{International Journal of Communication}
\bvolume{13},
\bfpage{2287}--\blpage{2313}
(\byear{2019})
\end{barticle}
\endbibitem

\bibitem[\protect\citeauthoryear{Tsai et~al.}{2020}]{Tsai2020}
\begin{barticle}
\bauthor{\bsnm{Tsai}, \binits{W.-H.S.}},
\bauthor{\bsnm{Tao}, \binits{W.}},
\bauthor{\bsnm{Chuan}, \binits{C.-H.}},
\bauthor{\bsnm{Hong}, \binits{C.}}:
\batitle{Echo chambers and social mediators in public advocacy issue networks}.
\bjtitle{Public Relations Review}
\bvolume{46}(\bissue{1}),
\bfpage{101882}
(\byear{2020})
\doiurl{10.1016/j.pubrev.2020.101882}
\end{barticle}
\endbibitem

\bibitem[\protect\citeauthoryear{Turetsky and Riddle}{2018}]{Turetsky.2018}
\begin{barticle}
\bauthor{\bsnm{Turetsky}, \binits{K.M.}},
\bauthor{\bsnm{Riddle}, \binits{T.A.}}:
\batitle{Porous chambers, echoes of valence and stereotypes: A network analysis of online news coverage interconnectedness following a nationally polarizing race-related event}.
\bjtitle{Social Psychological and Personality Science}
\bvolume{9}(\bissue{2}),
\bfpage{163}--\blpage{175}
(\byear{2018})
\end{barticle}
\endbibitem

\bibitem[\protect\citeauthoryear{Vaccari et~al.}{2016}]{Vaccari.2016}
\begin{botherref}
\oauthor{\bsnm{Vaccari}, \binits{C.}},
\oauthor{\bsnm{Valeriani}, \binits{A.}},
\oauthor{\bsnm{Barbera}, \binits{P.}},
\oauthor{\bsnm{Jost}, \binits{J.T.}},
\oauthor{\bsnm{Nagler}, \binits{J.}},
\oauthor{\bsnm{Tucker}, \binits{J.A.}}:
Of echo chambers and contrarian clubs: Exposure to political disagreement among german and italian users of twitter.
Social Media + Society
\textbf{2}(3)
(2016)
\doiurl{10.1177/2056305116664221}
\end{botherref}
\endbibitem

\bibitem[\protect\citeauthoryear{Villa et~al.}{2021}]{Villa.2021}
\begin{botherref}
\oauthor{\bsnm{Villa}, \binits{G.}},
\oauthor{\bsnm{Pasi}, \binits{G.}},
\oauthor{\bsnm{Viviani}, \binits{M.}}:
Echo chamber detection and analysis a topology- and content-based approach in the covid-19 scenario.
Social Network Analysis and Mining
\textbf{11}(1)
(2021)
\end{botherref}
\endbibitem

\bibitem[\protect\citeauthoryear{Williams et~al.}{2015}]{Williams2015}
\begin{barticle}
\bauthor{\bsnm{Williams}, \binits{H.T.P.}},
\bauthor{\bsnm{McMurray}, \binits{J.R.}},
\bauthor{\bsnm{Kurz}, \binits{T.}},
\bauthor{\bsnm{{Hugo Lambert}}, \binits{F.}}:
\batitle{Network analysis reveals open forums and echo chambers in social media discussions of climate change}.
\bjtitle{Global Environmental Change}
\bvolume{32},
\bfpage{126}--\blpage{138}
(\byear{2015})
\doiurl{10.1016/j.gloenvcha.2015.03.006}
\end{barticle}
\endbibitem

\end{thebibliography}

\pagebreak
 \appendix
 \section*{Appendix}
 \label{sec:appendix}
\renewcommand{\thesubsection}{\Alph{subsection}}

\begin{appendices}
\subsection{Users and Data Points of Platforms in Corpus Literature}

\begin{figure}[h]
    \centering
    \begin{subfigure}{0.49 \textwidth}
        \includegraphics[width=\linewidth]{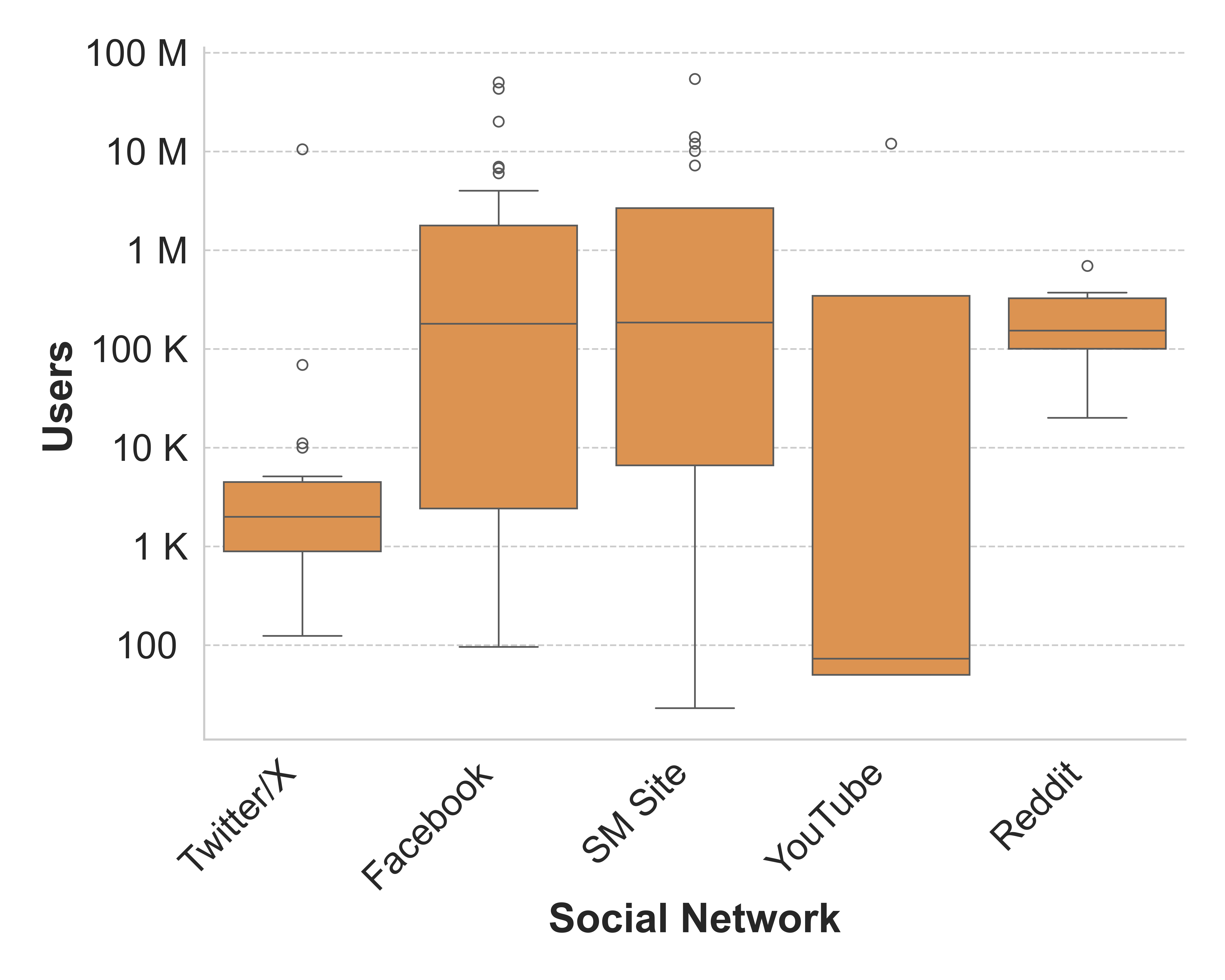}
        \caption{The amount of users included in trace data sets per platform (logarithmic scale).}
        \label{fig:users}
    \end{subfigure}
    \begin{subfigure}{0.49 \textwidth}
        \includegraphics[width=\linewidth]{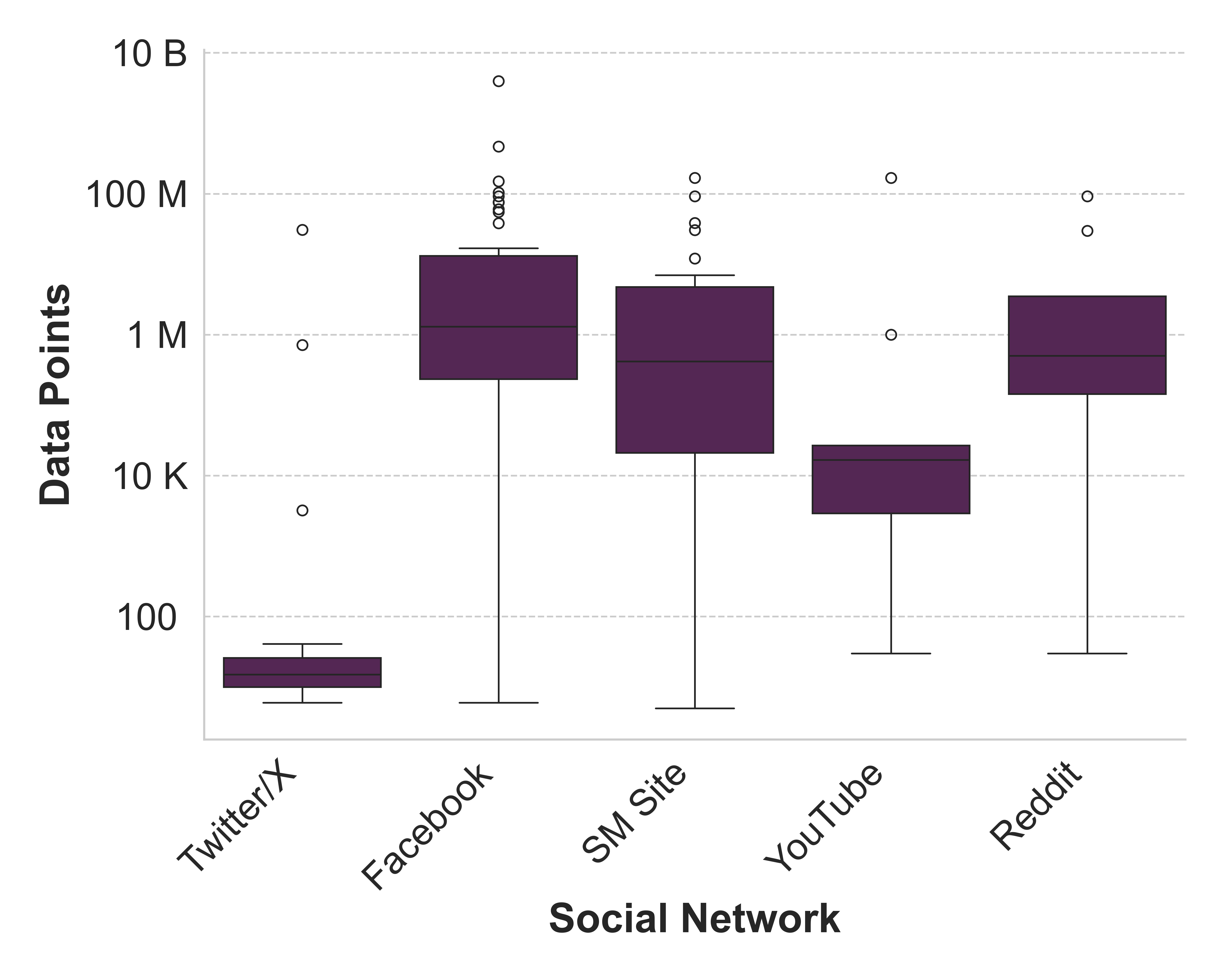}
        \caption{Data points (retweets, posts, likes,\dots) per platform (logarithmic scale).}
        \label{fig:data}
    \end{subfigure}
    \captionof{figure}{For social networks that are analyzed ten or more times in the review corpus the variation in the number of users and the data points between the different Networks.}
    \label{fig:usersAndData}
\end{figure} 
\newpage
\subsection{Publication Sources in Corpus Literature}
\begin{table}[ht]

\caption{Publication Source}
\centering
\small
\begin{tabular}{p{3cm}p{0.5cm}p{2cm}p{0.5cm}p{4cm}p{0.5cm}}
  
\multicolumn{4}{c}{\textit{a. Publication   Source}} & \multicolumn{2}{c}{\textit{d. Focus}}                                                                                                           \\ \hline
\multicolumn{3}{p{6cm}}{Source}                                                                                                                & \#  & Focus & \# \\ \hline
\multicolumn{3}{p{6cm}}{PLoS One}                                                                                                              & 11  & Echo chamber hypothesis & 48     \\
\multicolumn{3}{p{6cm}}{Social Media + Society}                                                                             & 9   & News sharing and consuming & 9     \\
\multicolumn{3}{p{6cm}}{Scientific Reports}                                                                                          & 9 & Recommender systems and echo chambers & 8       \\
\multicolumn{3}{p{6cm}}{Proceedings of the National Academy of Sciences} & 5  & Cognitive states, emotions, personality traits and EC & 8       \\      
\multicolumn{3}{p{6cm}}{New Media \& Society}                                                                               & 5 & Opinion dynamics on social media & 5       \\
\multicolumn{3}{p{6cm}}{International Journal of Communication}                                                            & 4 & COVID-19, vaccines and echo chambers &  5     \\
\multicolumn{3}{p{6cm}}{International Journal of Press-Politics}                                                      & 3     & Misinformation and echo chambers& 5   \\
\multicolumn{3}{p{6cm}}{Information, Communication \& Society}                                                          & 3   & Extremism, radicalization and echo chambers & 4     \\
\multicolumn{3}{p{6cm}}{Social Network Analysis and Mining}                                                          & 3     & Elections and echo chambers& 4   \\
\multicolumn{3}{p{6cm}}{Royal Society Open Science}                                                                      & 2   & Climate change and echo chambers  & 4    \\
\multicolumn{3}{p{6cm}}{Policy and Internet}                                                                               & 2   & Migitation of echo chambers & 3     \\
\multicolumn{3}{p{6cm}}{Internet Policy Review}                                                                       & 2   & Parliaments discussions and echo chambers & 2      \\

\multicolumn{4}{c}{\textit{b. Discipline}}                                                             & \multicolumn{2}{c}{\textit{c. Document Type}}                       \\ \hline
Area                                                & \#                             & & & Type                                         & \#  \\ \hline
Computer Science                                    &     31                                 &&& Journal                                      & 92       \\
Social Science                                      &       23                              & && Conference Paper                             & 11       \\
Communication Science                               &        21                              &                                              &          \\
Political Science                                  &       14                               &                                              &          \\
Psychology  and Medicine                                          &    9                                  &                                              &   \\ 
Economy                                          &  5                                    &                                              &   \\ 
\end{tabular}
\label{tab:pub}
\end{table}
\newpage
\subsection{Table of Results}

\begin{tabular}{p{2cm}p{1.5cm}p{1.5cm}p{2.5cm}p{2.5cm}p{1.5cm}}

\toprule
              Authors &               Country &            Method &                                              Focus &                                      Platform &                      Result \\
\midrule
\citet{ackermann2022voting} &                 Swiss &            Surveys &                                   Elections and EC &                            Social Media in General &    \cellcolor{gray!15}Mixed \\
 \citet{Aruguete.2023} & Argentina, Brasil, US & CSS &                              News sharing behavior &                                       Twitter & \cellcolor{Gray}Positive \\
  \citet{Asatani.2021} &                 Japan & CSS &                                      EC hypothesis &                                       Twitter & \cellcolor{Gray}Positive \\
   \citet{Auxier.2019}&                    US &            Surveys & Cognitive states, emotions, personality traits  & Social Media in General, mobile news applications  &    \cellcolor{gray!15}Mixed \\
     \citet{bail2018exposure}&                    US &        Experiment &                                      EC hypothesis &                                       Twitter &  \cellcolor{blue!15}Negative \\
   \citet{Bakshy.2015} &                    US & CSS &                                      EC hypothesis &                                      Facebook & \cellcolor{Gray}Positive \\
  \citet{Barbera.2015} &                    US & CSS &                                      EC hypothesis &                                       Twitter &    \cellcolor{gray!15}Mixed \\
   \citet{Bastos.2018} &                    UK & CSS &              Geographic embedding of Echo Chambers &                                       Twitter & \cellcolor{Gray}Positive \\
\citet{Batorski.2018} &                Poland & CSS &                                      EC hypothesis &                                      Facebook & \cellcolor{Gray}Positive \\
     \citet{Beam.2018} &                    US &            Surveys &                                      EC hypothesis &                                      Facebook &  \cellcolor{blue!15}Negative \\
           \citet{Bessi.2016b} &                    US & CSS &                     cognitive associations with EC &                                      Facebook & \cellcolor{Gray}Positive \\
    \citet{Bessi.2016} &                    US & CSS &                                      EC hypothesis &                             Facebook, Youtube & \cellcolor{Gray}Positive \\
    \citet{Bessi.2015} &                 Italy & CSS &                                     Misinformation &                                      Facebook & \cellcolor{Gray}Positive \\
     \ &           Netherlands &            Surveys &                         News sharing and consuming &                            Social Media in General & \cellcolor{Gray}Positive \\
     \citet{bond2022}&                       & CSS &                                      EC hypothesis &                                        bluedit & \cellcolor{Gray}Positive \\
\bottomrule
\end{tabular}
\newpage
\begin{tabular}{p{2cm}p{1.5cm}p{1.5cm}p{2.5cm}p{2.5cm}p{1.5cm}}
\toprule
 Authors &               Country &            Method &                                              Focus &                                      Platform &                      Result \\
\midrule
\citet{Boulianne.2022} &        France, UK, US &            Surveys &        Extremism, radicalization and echo chambers &                            Social Media in General &  \cellcolor{blue!15}Negative \\
\citet{Boulianne.2020}&        France, UK, US &            Surveys & Cognitive states, emotions, personality traites  &                            Social Media in General &  \cellcolor{blue!15}Negative \\
\citet{Boutyline.2017}&                    US & CSS &                                      EC hypothesis &                                       Twitter & \cellcolor{Gray}Positive \\
    \citet{bovet2022} &                    UK & CSS &                        Extremism and echo chambers &                                               &    \cellcolor{gray!15}Mixed \\
          \citet{Bright.2018} &                    EU & CSS &                                      EC hypothesis &                                       Twitter &    \cellcolor{gray!15}Mixed \\
       \citet{Brugnoli.2019} &                 Italy & CSS &                                      EC hypothesis &                                      Facebook & \cellcolor{Gray}Positive \\
           \citet{bruns2017echo} &             Australia & CSS &                                      EC hypothesis &                                       Twitter &  \cellcolor{blue!15}Negative \\
\citet{Bodrunova2023} &                Russia & CSS &                              EC Hypothesis &                                       Twitter & \cellcolor{Gray}Positive \\
  \citet{burnett2022selfcensoring} &                    US &            Surveys &                         News sharing and consuming &                            Social Media in General &  \cellcolor{blue!15}Negative \\
     \citet{Cann.2021}&                US, UK & CSS &                   Climate change and echo chambers &                                       Twitter & \cellcolor{Gray}Positive \\
     \citet{cargnino2022better} &              Germany &        Experiment & Cognitive states, emotions, personality traites  &                      Twitter & \cellcolor{blue!15}Negative \\
   \citet{cargnino2023we}&                      &        Experiment &                                                Echo chamber hypothesis &                              & \cellcolor{Gray}Positive \\
      \citet{Ceron.2019}&                Italy & CSS &                 TV, social media and echo chambers &                      Twitter & \cellcolor{Gray}Positive \\
       \citet{Chan.2019} & Taiwan, Japan, Korea &            Surveys &                         News sharing and consuming &           Social Media in General &    \cellcolor{gray!15}Mixed \\
\citet{cheng2023birds} &                      &            Surveys & Cognitive states, emotions, personality traites  &                              & \cellcolor{blue!15}Negative \\
     
     \bottomrule
\end{tabular}

\begin{tabular}{p{2cm}p{1.5cm}p{1.5cm}p{2.5cm}p{2.5cm}p{1.5cm}}

\toprule
              Authors &               Country &            Method &                                              Focus &                                      Platform &                      Result \\
\midrule
       \citet{choi2020rumor}&                   US & CSS &                                     Misinformation &                      Twitter & \cellcolor{Gray}Positive \\
    \citet{Cinelli.2020} &                   EU & CSS &                                      EC hypothesis &                     Facebook & \cellcolor{Gray}Positive \\
    \citet{Cinelli.2021} &                   US & CSS &                          recommendation algorithms &    Twitter, Facebook, Reddit &    \cellcolor{gray!15}Mixed \\
   \citet{Cinelli.2021b} &                Italy & CSS &                                     Misinformation &                      YouTube & \cellcolor{Gray}Positive \\
   \citet{Colleoni.2014} &                   US & CSS &                                      EC hypothesis &                      Twitter &    \cellcolor{gray!15}Mixed \\
       \citet{Cota.2019}&               Brasli & CSS &                                      EC hypothesis &                      Twitter & \cellcolor{Gray}Positive \\

 \citet{DelValle.2018} &                Spain & CSS &       Parliamentary discussions and echo chambers  &                      Twitter & \cellcolor{Gray}Positive \\
      \citet{DelValle.2022}  &          Netherlands & CSS &       Parliamentary discussions and echo chambers  &                      Twitter & \cellcolor{blue!15}Negative \\
\citet{DeLimaSantos2023} &              Brazil & CSS &                Covid 19 vaccines and EC &                                     Facebook & \cellcolor{Gray}Positive \\
\citet{DelVicario.2016} &                   US & CSS &                                     Misinformation &                     Facebook & \cellcolor{Gray}Positive \\
\citet{delvicarion2016emotional} &                   US & CSS &                         Emotions and echo chambers &                     Facebook & \cellcolor{Gray}Positive \\
\citet{DelVicario.2017} &                   UK & CSS &                                      EC hypothesis &                     Facebook & \cellcolor{Gray}Positive \\
     \citet{Dubois.2018}&                   UK &            Surveys &                                      EC hypothesis &           Social Media in General & \cellcolor{blue!15}Negative \\
     \citet{Dubois.2020} &               France &            Surveys &                                      EC hypothesis &           Social Media in General &    \cellcolor{gray!15}Mixed \\
      \citet{Ebeling2023} &              Brazil & Case Study &                Covid 19 vaccines and EC &                                     Twitter & \cellcolor{Gray}Positive \\
       \citet{Eady.2019} &                   US &     Mixed Methods &                                      EC hypothesis &                      Twitter & \cellcolor{blue!15}Negative \\
   \citet{Enjolras2023} &              Norway & CSS &                      Elections and EC &                                     Twitter & \cellcolor{gray!15}Mixed \\
   \citet{Enjolras.2022} &               Norway & CSS &                                                Echo chamber hypothesis &                      Twitter &    \cellcolor{gray!15}Mixed \\
   \citet{Erickson2023} &               USA & CSS &                      Elections and EC &                                      Reddit & \cellcolor{Gray}Positive \\
       \citet{etta}&                   US & CSS &                                      EC hypothesis &                 Twitter, Gab & \cellcolor{Gray}Positive \\
\bottomrule
\end{tabular}

\begin{tabular}{p{2cm}p{1.5cm}p{1.5cm}p{2.5cm}p{2.5cm}p{1.5cm}}
\toprule
 Authors &               Country &            Method &                                              Focus &                                      Platform &                      Result \\
\midrule
        \citet{Flamino.2023}&                   US & CSS &                                      EC hypothesis &                      Twitter &    \cellcolor{gray!15}Mixed \\
    \citet{Flaxman.2016} &                      & CSS &                                      EC hypothesis &                              &    \cellcolor{gray!15}Mixed \\
   \citet{fletcher} &                   UK & CSS &                              News sharing behavior &                      Browser &    \cellcolor{gray!15}Mixed \\
     \citet{Furman.2020}&                      & CSS &                                      EC hypothesis &                      Twitter & \cellcolor{blue!15}Negative \\
     \citet{Gao2023} &               China & CSS &                          EC Hypothesis &                          Douyin, TikTok, Bilibili & \cellcolor{Gray}Positive \\

    \citet{Gaumont.2018} &               France & CSS &                                   Opinion dynamics &                      Twitter & \cellcolor{Gray}Positive \\
    \citet{goel} &                   US & CSS &                                                Echo chamber hypothesis &         Reddit, Twitter, Gab &    \cellcolor{gray!15}Mixed \\
      \citet{Grusauskaite2023} &                 - & Mixed-Methods &                Conspiracy and EC &                                     YouTube & \cellcolor{Gray}Positive \\
    \citet{Guarino.2020} &                Italy & CSS &                                      EC hypothesis &                      Twitter & \cellcolor{Gray}Positive \\
\citet{GuerreroSole.2018} &                Spain & CSS &                                      EC hypothesis &                      Twitter & \cellcolor{Gray}Positive \\
     \citet{Guo.2020} &                   US & CSS &                                      EC hypothesis &                      Twitter &    \cellcolor{gray!15}Mixed \\
     \citet{Hada2023} &                 - & CSS &       Extremism, radicalization, and EC &                                     Twitter & \cellcolor{Gray}Positive \\
     \citet{Hagen.2022} &                    US &     Mixed Methods &                                                Vaccines, misinformation, and echo chambers &                      Twitter &    \cellcolor{gray!15}Mixed \\
    \citet{Hilbert.2018} &                   US &     Mixed Methods &                         Recommender systems and EC &                      YouTube & \cellcolor{Gray}Positive \\
    \citet{Hong2016} &                 USA & CSS &                      Parliament representations and EC &                                     Twitter & \cellcolor{Gray}Positive \\
 \citet{jones2022can} &                   US &            Surveys &         Vaccines, misinformation and echo chambers &           Social Media in General & \cellcolor{blue!15}Negative \\
    \citet{Justwan.2018} &                   US &            Surveys & Cognitive states, emotions, personality traites  &           Social Media in General & \cellcolor{blue!15}Negative \\
\citet{Kaiser.2020}&          Germany, US & CSS &                         Recommender systems and EC &                      YouTube & \cellcolor{Gray}Positive \\
\bottomrule
\end{tabular}
\newpage
\begin{tabular}{p{2cm}p{1.5cm}p{1.5cm}p{2.5cm}p{2.5cm}p{1.5cm}}
\toprule
 Authors &               Country &            Method &                                              Focus &                                      Platform &                      Result \\
\midrule
   \citet{Kitchens.2020} &                      &     Mixed Methods &                                      sda: tracking &    Facebook, Twitter, Reddit &    \cellcolor{gray!15}Mixed \\
    \citet{Koivula.2019} &              Finland &            Surveys &                         News sharing and consuming &           Social Media in General & \cellcolor{Gray}Positive \\
           \citet{Kratzke} &              Germany & CSS &                                      EC hypothesis &                      Twitter & \cellcolor{Gray}Positive \\
       \citet{Lima.2018} &                   US & CSS &                                                Echo chamber hypothesis &                          Gab & \cellcolor{Gray}Positive \\
        \citet{Liu.2021} &                   US &        Experiment &                         Recommender systems and EC &    self-constructed platform & \cellcolor{Gray}Positive \\
    \citet{ludwig2023divided} &              Germany &        Experiment &                         Recommender systems and EC &                              & \cellcolor{blue!15}Negative \\
      \citet{Masip.2020} &                Spain &            Surveys &                                      EC hypothesis & Facebook, Twitter, Instagram & \cellcolor{blue!15}Negative \\
       \citet{Matuszewski.2019} &               Poland & CSS &                                      EC hypothesis &                     Facebook & \cellcolor{Gray}Positive \\
\citet{Matuszewski.2019b} &      Poland, Hungary & CSS &                                      EC hypothesis &                      Twitter & \cellcolor{blue!15}Negative \\
              \citet{matz2021personal}&                   US &     Mixed Methods & Cognitive states, emotions, personality traites  &                     Facebook &    \cellcolor{gray!15}Mixed \\
           \citet{Min.2019} &                China &        Experiment &                                      EC hypothesis &                        Weibo & \cellcolor{Gray}Positive \\
        \citet{mirlohi2022social} &                      & CSS &                                                Contagion and echo chambers &                           Twitter, Forsquare   &    \cellcolor{gray!15}Mixed \\
      \citet{monti2023} &                   US & CSS &                                      EC hypothesis &                       Reddit & \cellcolor{blue!15}Negative \\
    \citet{Morales.2021}&                   US & CSS &                                      EC hypothesis &                       Reddit & \cellcolor{blue!15}Negative \\
     \citet{Mosleh.2021}&                   US &        Experiment &                                      EC hypothesis &                      Twitter & \cellcolor{Gray}Positive \\
     \citet{Mosleh.2021b} &                   US &     mixed Methods & Cognitive states, emotions, personality traites  &                      Twitter &    \cellcolor{gray!15}Mixed \\
      \citet{muise2022} &                   US & CSS &                 TV, social media and echo chambers &                      Browser &    \cellcolor{gray!15}Mixed \\
             \citet{Neely.2021} &                   US &            Surveys &                                   Elections and EC &                     Facebook &    \cellcolor{gray!15}Mixed \\
\bottomrule
\end{tabular}

\begin{tabular}{p{2cm}p{1.5cm}p{1.5cm}p{2.5cm}p{2.5cm}p{1.5cm}}
\toprule
 Authors &               Country &            Method &                                              Focus &                                      Platform &                      Result \\
\midrule
    \citet{Nikolov.2019} &                   US & CSS &                                   Elections and EC &           Social Media in General & \cellcolor{Gray}Positive \\
            \citet{Nordbrandt.}  &          Netherlands &            Surveys &                                      EC hypothesis &           Social Media in General & \cellcolor{blue!15}Negative \\
            \citet{Nyhan2023} & US & experiment& Facebook & \cellcolor{white}Negative \\
     \citet{Powers.2019} &                   US &            Surveys &                                   Elections and EC &           Social Media in General & \cellcolor{blue!15}Negative \\
 \citet{Radicioni.2021b} &                Italy & CSS &                 TV, social media and echo chambers &                      Twitter &    \cellcolor{gray!15}Mixed \\
   \citet{Radicioni.2021} &                Italy & CSS &                                      EC hypothesis &                      Twitter &    \cellcolor{gray!15}Mixed \\
     \citet{Rafail.2019} &                   US & CSS &                                      EC hypothesis &                       Reddit & \cellcolor{Gray}Positive \\
\citet{Roth.2020} &                   EU & CSS &                                                EC hypothesis &                      YouTube &    \cellcolor{gray!15}Mixed \\
            \citet{rusche2022} &              Germany & CSS &                                      EC hypothesis &                      Twitter &    \cellcolor{gray!15}Mixed \\
  \citet{Samantray.2019} &                      & CSS &                                     Climate change &                      Twitter &    \cellcolor{gray!15}Mixed \\
    Schmidt et al. 2020 &                Italy & CSS &                                      EC hypothesis &                     Facebook & \cellcolor{Gray}Positive \\
    \citet{Schmidt.2018} &                Italy & CSS &               COVID-19, vaccines and echo chambers &                     Facebook & \cellcolor{Gray}Positive \\
   \citet{Shmargad.2019} &                   US &        Experiment &                         News sharing and consuming &           Social Media in General & \cellcolor{Gray}Positive \\
      \citet{Shore.2018}&                   US & CSS &                                      EC hypothesis &                      Twitter &    \cellcolor{gray!15}Mixed \\
       \citet{srba2023} &                      & CSS &                                     Misinformation &                      YouTube &    \cellcolor{gray!15}Mixed \\
   \citet{Suelflow.2019} &                   US &        Experiment &                         Recommender systems and EC &                     Facebook & \cellcolor{Gray}Positive \\
        Sun et al. 2022 &                   US & CSS &                                           Covid-19 &                      YouTube &    \cellcolor{gray!15}Mixed \\
        \citet{Tomlein.2021}&                      & CSS &                                     Misinformation &                      YouTube &    \cellcolor{gray!15}Mixed \\
 \citet{Torregrosa.2020}&                   US & CSS &                        Extremism and echo chambers &                      Twitter & \cellcolor{Gray}Positive \\
      Treen et al. 2022 &                   US & CSS &                              Climate change and EC &                       Reddit &    \cellcolor{gray!15}Mixed \\
      \citet{Tsai2020} & - & CSS & Boycotts and EC & Twitter & \cellcolor{Gray}Positive \\
\bottomrule
\end{tabular}

\begin{tabular}{p{2cm}p{1.5cm}p{1.5cm}p{2.5cm}p{2.5cm}p{1.5cm}}
\toprule
 Authors &               Country &            Method &                                              Focus &                                      Platform &                      Result \\
\midrule
   \citet{Turetsky.2018} &                   US & CSS &                              News sharing behavior &                              & \cellcolor{blue!15}Negative \\
      \citet{Tyagi.2021} &                   US & CSS &                   climate chance and echo chambers &                      Twitter & \cellcolor{Gray}Positive \\
            \citet{Urman.2019} &               Russia & CSS &                                      EC hypothesis &                    Vkontakte & \cellcolor{Gray}Positive \\
   \citet{Vaccari.2016} &       Italy, Germany &            Surveys &                                      EC hypothesis &                      Twitter & \cellcolor{Gray}Positive \\
      \citet{Villa.2021} &                   EU & CSS &                         Covid-19 and Echo chambers &                      Twitter & \cellcolor{Gray}Positive \\
       \citet{Wang.2021} &                China & CSS &         Vaccines, misinformation and echo chambers &                        Weibo & \cellcolor{Gray}Positive \\
       \citet{wang2022} &                China & CSS &                                      EC hypothesis &                        Weibo & \cellcolor{Gray}Positive \\
       \citet{Wang.2020}&                China & CSS &                                     Misinformation &                        Weibo & \cellcolor{blue!15}Negative \\
       \citet{Williams2015} & - & CSS & Climate Change and EC & Twitter & \cellcolor{Gray}Positive \\
 \citet{Whittaker.2021} &                      &        Experiment &                                          extremism &         YouTube, Gab, Reddit &    \cellcolor{gray!15}Mixed \\
   \citet{Wieringa.2018} &          Netherlands & CSS &                                      EC hypothesis &                      Twitter & \cellcolor{Gray}Positive \\
      \citet{Wolfowicz.} &    Israel, Palestine &     Mixed Methods &                                          extremism &                      Twitter &    \cellcolor{gray!15}Mixed \\
  \citet{Wollebaek.2019} &               Norway &            Surveys & Cognitive states, emotions, personality traites  &           Social Media in General &    \cellcolor{gray!15}Mixed \\
    \citet{zerback} &              Germany &            Surveys &        Extremism, radicalization and echo chambers &           Social Media in General &    \cellcolor{gray!15}Mixed \\
      \citet{Zollo.2017} &                   US & CSS &                                      EC hypothesis &                     Facebook & \cellcolor{Gray}Positive \\
      \citet{Zollo.2015}&                   US & CSS &                         Emotions and echo chambers &                     Facebook & \cellcolor{Gray}Positive \\
\bottomrule
\end{tabular}
\newpage
\subsection{Platform-Specific Aggregated Tables}

\begin{table}[h!]
    \centering
    \caption{Corpus studies grouped by operationalization and conceptualization}
    \label{tab:granular_results}

    \begin{tabular}{l||p{0.6cm}|p{0.8cm}|p{0.8cm}|p{1.15cm}||p{0.8cm}|p{1.25cm}|p{1.15cm}|p{1.15cm}||c}
    \toprule
    & \multicolumn{4}{c||}{\textbf{Operationalization}} & \multicolumn{4}{c||}{\textbf{Conceptualization}} \\
    \midrule 
    Platform & CSS & Survey & Experi- ment & Mixed Methods & Homo- phily & Content Exposure & User Behaviour & Group Behaviour & Total \\ \midrule \midrule
Browser & 2 & 0 & 0 & 0 & 0 & 2 & 0 & 0 & 2 \\
Douyin & 2 & 0 & 0 & 0 & 1 & 1 & 0 & 0 & 2 \\
Facebook & 25 & 6 & 3 & 5 & 14 & 16 & 7 & 2 & 39 \\
Gab & 5 & 0 & 2 & 0 & 3 & 3 & 1 & 0 & 7 \\
Reddit & 13 & 0 & 2 & 2 & 8 & 5 & 3 & 1 & 17 \\
Twitter/X & 61 & 3 & 5 & 7 & 42 & 20 & 11 & 3 & 76 \\
Vkontakte & 2 & 0 & 0 & 0 & 1 & 1 & 0 & 0 & 2 \\
Weibo & 5 & 0 & 2 & 0 & 3 & 3 & 0 & 1 & 7 \\
YouTube & 7 & 0 & 2 & 2 & 6 & 4 & 1 & 0 & 11 \\
Self-Constr. & 0 & 0 & 2 & 0 & 0 & 1 & 1 & 0 & 2 \\
SM-Sites & 1 & 35 & 3 & 0 & 7 & 20 & 12 & 0 & 39 \\
\bottomrule
    \end{tabular}
   \end{table} 

\begin{table}[h!]
    \centering
     \caption{Results grouped by platform}
    \label{tab:granular_results2}
\begin{tabular}[h!]{l||c|c|c}
\toprule
 & pos. findings & neg. findings & mix. findings \\
Platform &  &  &  \\
\midrule \midrule
Browser & 0 & 2 & 1 \\
Douyin & 0 & 0 & 1 \\
Facebook & 17 & 2 & 4 \\
Gab & 1 & 0 & 0 \\
Reddit & 5 & 4 & 0 \\
Twitter & 34 & 9 & 9 \\
Vkontakte & 1 & 0 & 0 \\
Weibo & 3 & 1 & 0 \\
YouTube & 5 & 0 & 4 \\
Self-constr. & 1 & 0 & 0 \\
SM-Sites & 6 & 8 & 6 \\
\bottomrule
\end{tabular}

\end{table}
\newpage
\begin{table}[h!]
\caption{Corpus studies grouped by countries and platforms}
    \label{tab:granular_results3}
 \centering
\small
\begin{tabular}{l||p{0.7cm}|p{0.6cm}|p{0.7cm}|p{0.6cm}|p{0.8cm}|p{0.9cm}|p{0.7cm}|p{0.8cm}|p{0.6cm}|p{0.6cm}|p{0.6cm}||p{0.6cm}}
\toprule
  & Brow-ser & Dou-yin & Face-book & Gab & Reddit & Twitter /X & Vkon-takte & Weibo & You-Tube & Self-con.& SM-Sites & SUM\\
\midrule \midrule
Argentina & 0 & 0 & 0 & 0 & 0 & 1 & 0 & 0 & 0 & 0 & 0 & 1 \\
Australia & 0 & 0 & 0 & 0 & 0 & 1 & 0 & 0 & 0 & 0 & 0 & 1 \\
Brasil & 1 & 1 & 1 & 1 & 1 & 2 & 1 & 1 & 1 & 1 & 1 & 2 \\
Brazil & 0 & 0 & 1 & 0 & 0 & 1 & 0 & 0 & 0 & 0 & 0 & 2 \\
China & 0 & 1 & 0 & 0 & 0 & 0 & 0 & 4 & 0 & 0 & 1 & 6 \\
EU & 0 & 0 & 1 & 0 & 0 & 2 & 0 & 0 & 1 & 0 & 0 & 4 \\
Finland & 0 & 0 & 0 & 0 & 0 & 0 & 0 & 0 & 0 & 0 & 1 & 1 \\
France & 0 & 0 & 0 & 0 & 0 & 1 & 0 & 0 & 0 & 0 & 3 & 4 \\
Germany & 1 & 1 & 1 & 1 & 1 & 4 & 1 & 1 & 2 & 1 & 2 & 7 \\
Hungary & 1 & 1 & 1 & 1 & 1 & 1 & 1 & 1 & 1 & 1 & 1 & 1 \\
Israel & 0 & 0 & 0 & 0 & 0 & 1 & 0 & 0 & 0 & 0 & 0 & 1 \\
Italy & 0 & 0 & 4 & 0 & 0 & 5 & 0 & 0 & 1 & 0 & 0 & 10 \\
Japan & 1 & 1 & 1 & 1 & 1 & 2 & 1 & 1 & 1 & 1 & 1 & 2 \\
Netherlands & 0 & 0 & 0 & 0 & 0 & 3 & 0 & 0 & 0 & 0 & 2 & 5 \\
Norway & 0 & 0 & 0 & 0 & 0 & 2 & 0 & 0 & 0 & 0 & 1 & 3 \\
Palestine & 1 & 1 & 1 & 1 & 1 & 1 & 1 & 1 & 1 & 1 & 1 & 1 \\
Poland & 0 & 0 & 2 & 0 & 0 & 1 & 0 & 0 & 0 & 0 & 0 & 3 \\
Russia & 0 & 0 & 0 & 0 & 0 & 1 & 1 & 0 & 0 & 0 & 0 & 2 \\
South Korea & 0 & 0 & 0 & 0 & 0 & 0 & 0 & 0 & 0 & 0 & 1 & 1 \\
Spain & 0 & 0 & 1 & 0 & 0 & 3 & 0 & 0 & 0 & 0 & 0 & 3 \\
Swiss & 0 & 0 & 0 & 0 & 0 & 0 & 0 & 0 & 0 & 0 & 1 & 1 \\
Taiwan & 0 & 0 & 0 & 0 & 0 & 0 & 0 & 0 & 0 & 0 & 1 & 1 \\
Turkey & 1 & 1 & 1 & 1 & 1 & 1 & 1 & 1 & 1 & 1 & 1 & 1 \\
UK & 4 & 3 & 4 & 3 & 3 & 4 & 3 & 3 & 3 & 3 & 4 & 8 \\
US & 2 & 1 & 14 & 4 & 8 & 19 & 1 & 1 & 4 & 2 & 9 & 50 \\
\bottomrule

\end{tabular}

\end{table}
\newpage

\end{appendices}


\end{document}